\documentclass[reprint,amsmath,amssymb,amsfonts,aps,floatfix]{revtex4-2}

\usepackage{natbib}
\usepackage{graphicx,color,rotating}
\usepackage[latin1]{inputenc}
\usepackage{textcomp}
\usepackage{dcolumn}
\usepackage{bm}     
\usepackage{upgreek}
\usepackage{subdepth}  			
\usepackage{epstopdf}   		

\usepackage{array}
\newcolumntype{L}[1]{>{\raggedright\let\newline\\\arraybackslash\hspace{0pt}}m{#1}}
\newcolumntype{C}[1]{>{\centering\let\newline\\\arraybackslash\hspace{0pt}}m{#1}}
\newcolumntype{R}[1]{>{\raggedleft\let\newline\\\arraybackslash\hspace{0pt}}m{#1}}

\usepackage{hyperref}					
\usepackage{xcolor}						
\hypersetup{			     			
    colorlinks,
    allcolors={blue}
}
\newcommand{\qmarks}[1]{{``#1''}}

\newcommand{\mr}[1]{\ensuremath{\mathrm{#1}}}
\newcommand{\myvec}[1]{\bm{#1}}
\newcommand{\ee}{\mathrm{e}}
\newcommand{\ii}{\mathrm{i}}
\newcommand{\dm}{\mathrm{d}}

\newcommand{\avr}[1]{\big\langle #1 \big\rangle}

\DeclareMathOperator{\re}{Re}

\newcommand{\ve}{\varepsilon}

\newcommand{\pp}{\partial}

\newcommand{\nablabf}{\boldsymbol{\nabla}}

\newcommand{\Lapl}{\nabla^2}

\newcommand{\rot}{\nablabf\times}
\renewcommand{\div}{\nablabf\cdot}

\newcommand{\etal}{\textit{et~al.\ }}

\newcommand{\scap}{\!\cdot\!}

\newcommand{\dpst}{\displaystyle}

\newcommand{\eee}{\myvec{e}}
\newcommand{\een}{\myvec{e}}

\newcommand{\FFFrad}{\myvec{F}^\mathrm{rad}}

\newcommand{\FFFradII}{\FFFrad_{11}}
\newcommand{\FFFradin}{\FFFrad_{2,\mr{in}}}

\newcommand{\Imat}{\textbf{\textsf{I}}}

\newcommand{\kc}{k_\mathrm{c}}
\newcommand{\kcO}{k_\mathrm{c0}}
\newcommand{\kt}{k_\mathrm{t}}
\newcommand{\ks}{k_\mathrm{s}}

\newcommand{\MMM}{\myvec{M}}

\newcommand{\NNN}{\myvec{N}}
\newcommand{\nnn}{\myvec{n}}

\newcommand{\qqq}{\myvec{q}}

\newcommand{\rrr}{\myvec{r}}
\newcommand{\rrrO}{\rrr_0}

\newcommand{\rhat}{{\hat{r}}}

\newcommand{\sss}{\myvec{s}}

\newcommand{\uuu}{\myvec{u}}

\newcommand{\vvv}{\myvec{v}}

\newcommand{\XXX}{\myvec{X}}

\newcommand{\zerovec}{\boldsymbol{0}}

\newcommand{\calO}{\mathcal{O}}

\newcommand{\ansl}{a_n^\mr{sl}}
\newcommand{\anfl}{a_n^\mr{fl}}
\newcommand{\anflp}{a_n^\mr{fl \prime}}

\newcommand{\bnsl}{b_n^\mr{sl}}
\newcommand{\bnfl}{b_n^\mr{fl}}

\newcommand{\Bc}{B_\mr{c}}
\newcommand{\Bt}{B_\mr{t}}
\newcommand{\Bbc}{B_\mr{c}^\mr{b}}
\newcommand{\Bbt}{B_\mr{t}^\mr{b}}

\newcommand{\dnfl}{d_n^\mr{fl}}

\newcommand{\cO}{c_0}

\newcommand{\cpOTi}{\tilde{c}_{p0}}

\renewcommand{\cp}{c_p}
\newcommand{\cpO}{c_{p0}}

\newcommand{\cV}{c_V}

\newcommand{\Dth}{D^\mathrm{th}}
\newcommand{\DthO}{D^\mathrm{th}_0}

\newcommand{\DthOTi}{\tilde{D}_0^\mathrm{th}}

\newcommand{\Eac}{E_\mathrm{ac}}

\newcommand{\kth}{k^\mathrm{th}}

\newcommand{\kthO}{k^\mathrm{th}_0}
\newcommand{\kthOp}{k^\mathrm{th \prime}_0}
\newcommand{\kthOTi}{\tilde{k}^\mathrm{th}_0}

\newcommand{\kapT}{\kappa_T}

\newcommand{\kapSOTi}{\tilde{\kappa}_{s0}}
\newcommand{\kapTO}{\kappa_{T0}}

\newcommand{\kapS}{\kappa_s}
\newcommand{\kapSO}{\kappa_{s0}}

\newcommand{\kapsO}{\kappa_{s0}}

\newcommand{\Smn}[2]{S_{#1,#2}}

\newcommand{\ain}{\alpha_{i,n}}

\newcommand{\aknI}{\alpha_{k,n+1}}
\newcommand{\acn}{\alpha_{\mr{c},n}}
\newcommand{\acnsc}{\acn^\mr{sc}}
\newcommand{\acsc}[1]{\alpha^\mr{sc}_{\mr{c},#1}}
\newcommand{\acnp}{\alpha_{\mr{c},n}^\prime}
\newcommand{\acp}[1]{\alpha_{\mr{c},#1}^\prime}

\newcommand{\asn}{\alpha_{\mr{s},n}}
\newcommand{\asnsc}{\asn^\mr{sc}}
\newcommand{\assc}[1]{\alpha^\mr{sc}_{\mr{s},#1}}
\newcommand{\asnp}{\alpha_{\mr{s},n}^\prime}
\newcommand{\asp}[1]{\alpha_{\mr{s},#1}^\prime}

\newcommand{\atn}{\alpha_{\mr{t},n}}
\newcommand{\atnsc}{\atn^\mr{sc}}
\newcommand{\atsc}[1]{\alpha^\mr{sc}_{\mr{t},#1}}
\newcommand{\atnp}{\alpha_{\mr{t},n}^\prime}
\newcommand{\atp}[1]{\alpha_{\mr{t},#1}^\prime}

\newcommand{\alphap}{\alpha_p}

\newcommand{\alphapOTi}{\tilde{\alpha}_{p0}}
\newcommand{\alphapO}{{\alpha_{p0}}}

\newcommand{\delt}{\delta_t}

\newcommand{\dels}{\delta_s}

\newcommand{\etaB}{\eta^\mathrm{b}}
\newcommand{\etaOB}{\eta^\mathrm{b}_0}
\newcommand{\etaIB}{\eta^\mathrm{b}_1}

\newcommand{\etaO}{\eta_0}

\newcommand{\etaI}{\eta_1}

\newcommand{\etaOTi}{\tilde{\eta}_0}

\newcommand{\Gamt}{\Gamma_\mathrm{t}}
\newcommand{\Gams}{\Gamma_\mathrm{s}}

\newcommand{\gamST}{\gamma_\mr{surf}}

\newcommand{\gamO}{\gamma_0}

\newcommand{\nuI}{\nu_1}

\newcommand{\nuOB}{\nu_0^\mathrm{b}}
\newcommand{\nuIB}{\nu_1^\mathrm{b}}
\newcommand{\nuO}{\nu_0}

\newcommand{\surfP}{\partial\Omega_0}

\newcommand{\phiIc}{\phi_\mathrm{1c}}
\newcommand{\phiIt}{\phi_\mathrm{1t}}

\newcommand{\phiIcnp}{\phi_{\mr{1c},n}^{\prime}}
\newcommand{\phiIcp}{\phi_{\mr{1c}}^{\prime}}
\newcommand{\phiIcin}{\phi_\mr{1c}^\mr{in}}
\newcommand{\phiIcinC}{\phi_\mr{1c}^\mr{in *}}
\newcommand{\phiIcinn}{\phi_{\mr{1c},n}^\mr{in}}
\newcommand{\phiIcsc}{\phi_\mr{1c}^\mr{sc}}
\newcommand{\phiIcscn}{\phi_{\mr{1c},n}^\mr{sc}}
\newcommand{\phiItsc}{\phi_\mr{1t}^\mr{sc}}
\newcommand{\phiItscn}{\phi_{\mr{1t},n}^\mr{sc}}
\newcommand{\phiItnp}{\phi_{\mr{1t},n}^\prime}
\newcommand{\phiItp}{\phi_\mr{1t}^{\prime}}

\newcommand{\psiInp}{\psi_{1,n}^\prime}
\newcommand{\psiIscn}{\psi_{1,n}^\mr{sc}}
\newcommand{\psiIIn}{\psi_{2,n}}

\newcommand{\psiIInsc}{\psiIIn^\mr{sc}}
\newcommand{\psiIInp}{\psiIIn^\prime}
\newcommand{\psibf}{\bm{\psi}}
\newcommand{\psibfI}{\bm{\psi}_1}
\newcommand{\psibfII}{\bm{\psi}_2}
\newcommand{\psibfIsc}{\psibf^\mr{sc}_1}
\newcommand{\psibfIp}{\psibf^\mr{\prime}_1}
\newcommand{\psibfIIsc}{\psibfII^\mr{sc}}

\newcommand{\cP}{c_p}

\newcommand{\kO}{k_0}

\newcommand{\pIin}{p_1^\mr{in}}

\newcommand{\pST}{p_\mr{surf}}

\newcommand{\pIIin}{p_2^\mr{in}}
\newcommand{\pIIsc}{p_2^\mr{sc}}
\newcommand{\pIIp}{p'_2}

\newcommand{\vIr}{v_{1r}}
\newcommand{\vIrn}{v_{1r}^n}
\newcommand{\vIrp}{v_{1r}^{\prime}}
\newcommand{\vIrnm}{v_{1r}^{n\prime}}
\newcommand{\vIth}{v_{1\theta}}
\newcommand{\vIthn}{v_{1\theta}^n}
\newcommand{\vIthnm}{v_{1\theta}^{n\prime}}
\newcommand{\vIthp}{v_{1\theta}^{\prime}}
\newcommand{\vvvIin}{\myvec{v}_1^\mathrm{in}}
\newcommand{\vvvIinC}{\myvec{v}_1^\mathrm{in*}}

\newcommand{\vvvI}{\vvv_1}

\newcommand{\vvvIIin}{\vvv_2^\mr{in}}
\newcommand{\vvvIIsc}{\vvv_2^\mr{sc}}
\newcommand{\vvvIIp}{\vvv'_2}
\newcommand{\vvvIIrin}{v_{2r}^\mr{in}}
\newcommand{\vvvIIrsc}{v_{2r}^\mr{sc}}
\newcommand{\vvvIItin}{v_{2\theta}^\mr{in}}
\newcommand{\vvvIItsc}{v_{2\theta}^\mr{sc}}
\newcommand{\vvvIIrm}{v_{2r}'}
\newcommand{\vvvIItm}{v_{2\theta}'}

\newcommand{\xO}{x_0}

\newcommand{\rhoO}{\rho_0}

\newcommand{\rhoI}{\rho_1}

\newcommand{\rhoTi}{\tilde{\rho}}

\newcommand{\rhoOTi}{\tilde{\rho}_0}

\newcommand{\SIMHz}{\textrm{MHz}}
\newcommand{\SIkHz}{\textrm{kHz}}

\newcommand{\SIK}{\textrm{K}}

\newcommand{\SIkg}{\textrm{kg}}
\newcommand{\SIkgm}{\textrm{kg}\:\textrm{m$^{-3}$}}

\newcommand{\SIm}{\textrm{m}}

\newcommand{\SImum}{\textrm{\textmu{}m}}

\newcommand{\nn}{\nonumber}
\newcommand{\beq}[1]{\begin{equation} \eqlab{#1}}
\newcommand{\eeq}{\end{equation}}
\newcommand{\bsub}{\begin{subequations}}
\newcommand{\esub}{\end{subequations}}
\def\bal#1\eal{\begin{align}#1\end{align}}
\def\balat#1#2\ealat{\begin{alignat}{#1} #2 \end{alignat}}
\def\bsubal#1 #2\esubal{\bsuba{#1}\begin{align}#2\end{align} \esuba}     
\def\bsubalat#1#2#3\esubalat{\bsuba{#1} \begin{alignat}{#2} #3 \end{alignat} \esuba}
\newcommand{\bsuba}[1]{\bsub \eqlab{#1}}
\newcommand{\esuba}{\esub}

\newcommand{\eqlab}[1]{\label{eq:#1}}
\renewcommand{\eqref}[1]{Eq.~(\ref{eq:#1})}
\newcommand{\eqnoref}[1]{(\ref{eq:#1})}

\newcommand{\eqsref}[2]{Eqs.~(\ref{eq:#1}) and~(\ref{eq:#2})}
\newcommand{\eqsnoref}[2]{(\ref{eq:#1}) and~(\ref{eq:#2})}
\newcommand{\eqssref}[3]{Eqs.~(\ref{eq:#1}), (\ref{eq:#2}) and~(\ref{eq:#3})}
\newcommand{\eqssnoref}[3]{(\ref{eq:#1}), (\ref{eq:#2}) and~(\ref{eq:#3})}

\newcommand{\figref}[1]{Fig.~\ref{fig:#1}}

\newcommand{\figsref}[2]{Figs.~\ref{fig:#1} and~\ref{fig:#2}}
\newcommand{\figlab}[1]{\label{fig:#1}}
\newcommand{\appref}[1]{Appendix~\ref{sec:#1}}
\newcommand{\appnoref}[1]{\ref{sec:#1}}
\newcommand{\appsref}[2]{Appendices~\ref{sec:#1} and~\ref{sec:#2}}

\newcommand{\secref}[1]{Section~\ref{sec:#1}}
\newcommand{\secnoref}[1]{\ref{sec:#1}}
\newcommand{\secsref}[2]{Sections~\ref{sec:#1} and~\ref{sec:#2}}
\newcommand{\seclab}[1]{\label{sec:#1}}
\newcommand{\tabref}[1]{Table~\ref{tab:#1}}
\newcommand{\tabsref}[2]{Tables~\ref{tab:#1} and~\ref{tab:#2}}
\newcommand{\tablab}[1]{\label{tab:#1}}

\newcommand{\bt}{b_\mathrm{t}}
\newcommand{\bc}{b_\mathrm{c}}

\newcommand{\xtO}{x_\mr{t}}
\newcommand{\xc}{x_\mathrm{c}}

\newcommand{\xs}{x_\mathrm{s}}
\newcommand{\xt}{x_\mathrm{t}}

\newcommand{\xcp}{x_\mathrm{c}^{\prime}}
\newcommand{\xsp}{x_\mathrm{s}^{\prime}}
\newcommand{\xtp}{x_\mathrm{t}^{\prime}}

\newcommand{\Phiac}{\Phi_\mathrm{ac}}

\newcommand{\sigmabf}{\bm{\sigma}}

\newcommand{\cL}{c_\mathrm{lo}}
\newcommand{\cLO}{c_\mathrm{lo0}}
\newcommand{\cT}{c_\mathrm{tr}}
\newcommand{\cTO}{c_\mathrm{tr0}}

\newcommand{\phiII}{\phi_2}

\newcommand{\phiIIsc}{\phiII^\mr{sc}}

\newcommand{\phiIInsc}{\phi_{2,n}^\mr{sc}}
\newcommand{\phiIInp}{\phi_{2,n}^\prime}
\newcommand{\uuuI}{\myvec{u}_1}

\newcommand{\fl}{\mathrm{fl}}
\newcommand{\xl}{\mathrm{xl}}
\renewcommand{\sl}{\mathrm{sl}}

\definecolor{darkgreen}{rgb}{0.00, 0.50, 0.00}
\definecolor{DARKGREEN}{rgb}{0.00, 0.50, 0.00}
\definecolor{RED}{rgb}{1.00, 0.00, 0.00}
\definecolor{GREEN}{rgb}{0.00, 1.00, 0.00}
\definecolor{BLUE}{rgb}{0.00, 0.00, 1.00}
\definecolor{MAGENTA}{rgb}{1.00, 0.00, 1.00}

\begin{document}

\title{The acoustic radiation force on a spherical thermoviscous particle in a thermoviscous fluid including scattering and microstreaming}

\author{Bjørn G. Winckelmann}
\email{winckel@dtu.dk}
\affiliation{Department of Physics, Technical University of Denmark,\\
DTU Physics Building 309, DK-2800 Kongens Lyngby, Denmark}

\author{Henrik Bruus}
\email{bruus@fysik.dtu.dk}
\affiliation{Department of Physics, Technical University of Denmark,\\
DTU Physics Building 309, DK-2800 Kongens Lyngby, Denmark}

\date{18 December 2022}

\begin{abstract}
We derive general analytical expressions for the time-averaged acoustic radiation force on a small spherical particle suspended in a fluid and located in an axisymmetric incident acoustic wave. We treat the cases of the particle being either an elastic solid or a fluid particle. The effects of particle vibrations, acoustic scattering, acoustic microstreaming, heat conduction, and  temperature-dependent fluid viscosity are all included in the theory. Acoustic streaming inside the particle is also taken into account for the case of a fluid particle. No restrictions are placed on the widths of the viscous and thermal boundary layers relative to the particle radius. We compare the resulting acoustic radiation force with that obtained from previous theories in the literature, and we identify limits, where the theories agree, and specific cases of particle and fluid materials, where qualitative or significant quantitative deviations between the theories arise.
\end{abstract}

\maketitle


\section{Introduction}
A particle suspended in a fluid and subjected to an acoustic field experiences a steady time-averaged force due to the scattering of the wave, the so-called acoustic radiation force $\FFFrad$. The analytical theory of $\FFFrad$ has a long history, going back to King, who in 1934 studied an incompressible spherical particle in an ideal (non-thermoviscous) fluid \cite{King1934}, followed by Yosioka and Kawasima, who in 1955 included the compressibility of the particle \cite{Yosioka1955}, a result reformulated in terms of the acoustic potential by Gor'kov in 1962 \cite{Gorkov1962}. An important advance of the analytical theory of $\FFFrad$ was made by Doinikov, who in 1994 and 1997 replaced the ideal fluid with a viscous and heat conducting fluid and studied a rigid heat conducting solid particle and a viscous and heat conducting fluid particle and took both acoustic scattering and streaming into account \cite{Doinikov1994, Doinikov1994a, Doinikov1997, Doinikov1997a, Doinikov1997b}. More recently, further developments to the theory have been made by Settnes and Bruus in 2012 \cite{Settnes2012} and by Karlsen and Bruus in 2015 \cite{Karlsen2015}, who studied $\FFFrad$ for compressible particles in viscous and thermoviscous fluids, respectively, considering the acoustic scattering, but not the streaming, and by Doinikov, Fankhauser, and Dual in 2021, who studied an elastic solid particle in a viscoelastic fluid \cite{Doinikov2021}.

The increased interest in the theory of $\FFFrad$ in recent years is due to experimental advances and technological applications in microscale acoustofluidics and acoustic tweezers. In these technologies $\FFFrad$ is the primary working mechanism used for focusing \cite{Antfolk2014}, sorting \cite{Augustsson2016, Ai2017}, trapping \cite{Hammarstrom2012}, nanoscale separation \cite{Zhang2022}, and levitating particles \cite{Marzo2015, Riaud2017}. These advances have been supported by numerical simulations \cite{Muller2012, Skov2019}. The particular motivation for the present work is a combination of the numerical study by Baasch, Pavlic, and Dual, who in 2019 emphasized the importance of the acoustic streaming generated by the particle in a viscous fluid, the so-called microstreaming, for heavy microparticles \cite{Baasch2019}, and the numerical and experimental study of the importance of including temperature-dependent parameters in acoustofluidic systems \cite{Joergensen2021, Qui2021}.

In this work, we develop an extension of the analytical model of the acoustic radiation force $\FFFrad$ presented by  Doinikov in 1994 and 1997 \cite{Doinikov1994, Doinikov1994a, Doinikov1997a, Doinikov1997, Doinikov1997b} for either a  rigid or a fluid spherical particle suspended in a thermoviscous fluid. Our extension comprises the inclusion of (1) elastic instead of rigid solid particles, (2) temperature- and density-dependent material parameters, in particular the viscosity, (3) the tangential part of the Stokes drift in the boundary condition of the acoustic streaming on the particle surface, and (4) inner streaming in a fluid particle.

The structure of the paper is the following. The governing equations are presented in \secref{gov_eq} followed by the solution to the acoustic scattering and streaming problems in \secsref{FirstOrder}{SecondOrder}, respectively. The general results for $\FFFrad$ are derived in \secref{Evaluating_Frad}, with specific limiting cases presented and compared to the literature in \secref{D0_D1_limits}. In \secref{ResultsPlaneWave}, we study $\FFFrad$ versus particle size in a standing wave for eight selected combinations of particle and fluid materials. Finally, we conclude in \secref{conclusion}, and present mathematical details in Appendices~\secnoref{Legendre_integrals}--\secnoref{S_coeff_fl}. Some supporting \textsc{Matlab} scripts and lists of general analytical expressions for selected coefficients are provided in the Supplemental Material~\footnote{See Supplemental Material at \url{https://bruus-lab.dk/TMF/files/Supplemental_BGW_Frad.zip} for details on \textsc{Matlab} scripts for computing $D_0$ and $D_1$ in the long-wavelength limit, the general second-order coefficients $S_{ik,n}$, and how to compute them.}.

\section{Governing equations}
\seclab{gov_eq}

We consider a spherical particle, often called a 'sphere' for short, which can either be an isotropic elastic solid or a Newtonian fluid, in a time-harmonically perturbed surrounding fluid medium. The harmonic acoustic perturbation has the frequency $f$ and corresponding angular frequency $\omega=2\pi f$, and we use the following series expansions for all physical fields that describe the particle and the surrounding fluid,
\bal\eqlab{pertubation}
g(\rrr,t)&=g_0+\re\Big[g_1(\rrr)\,\ee^{-\ii\omega t}\Big]+g_2(\rrr,t).
\eal
The zero-order fields $g_0$ describe an initial unperturbed state of a surrounding quiescent and homogeneous fluid, and a stationary particle. The complex-valued first-order fields $g_1$ describe the first order acoustic response, which follows the actuation frequency $f$. The second-order fields $g_2$ describe a non-linear response containing small second-order harmonics and a steady time-averaged response. We only study the latter in this work.

All physical parameters $q$ characterizing the fluids and solids are temperature and density dependent. The first-order time-harmonic response includes a first-order density field $\rho_1$ and a first-order temperature field $T_1$, which in turn perturb the material parameters. Thus, we expand all parameters as,
 \bsubal{pertubation_params}
 \eqlab{pertubation_params_gen}
 q(\rrr,t)&=q_0+\re\Big[q_1(\rrr)\,\ee^{-\ii\omega t}\Big], \\
 \eqlab{pertubation_params_1}
 q_1(\rrr)&=\bigg(\frac{\pp q}{\pp T}\bigg)_{T=T_0} T_1(\rrr)+\bigg(\frac{\pp q}{\pp \rho}\bigg)_{\rho=\rho_0}\rho_1(\rrr).
 \esubal

Our objective is to evaluate the radiation force $\FFFrad$ on solid and fluid particles with a spherical equilibrium shape. When perturbed acoustically, the particle acquires the time-dependent volume $\Omega(t)$ with surface $\pp\Omega(t)$. $\FFFrad$ is the time average, an operation denoted by angled brackets $\avr{\dots}$, of the stress $\sigmabf$ integrated over the vibrating surface $\pp\Omega(t)$ with normal vector $\nnn$,
 \beq{Frad_fundamental}
 \FFFrad =\bigg\langle \oint_{\pp\Omega(t)} \sigmabf\cdot \nnn \: \dm S\bigg\rangle.
 \eeq
Assuming that the particle drifts very little during an acoustic period, $\FFFrad$ can be written as \cite{Doinikov1994a, Karlsen2015},
 \beq{Frad}
 \FFFrad = \oint_{\surfP} \langle \sigmabf_2-\rho_0\vvvI\vvvI \rangle \cdot \nnn \,\dm S,
 \eeq
where $\surfP$ is the equilibrium surface of the particle.

\vspace*{-5mm}
\subsection{Thermodynamic identities}

\vspace*{-3mm}
For both fluids and solids, we introduce the isothermal compressibility, $\kapT$, the isentropic compressibility $\kapS$, the isobaric thermal expansion coefficients $\alphap$, the specific heat capacity at constant pressure $\cp$ and at constant volume $\cV$, the ratio of specific heats $\gamma$, the thermal conductivity $\kth$, and the thermal diffusivity $\Dth$. These quantities are related by the expressions,
 \bsubal{thermodynamic_relations_general}
 \eqlab{thermodynamic_relation_kapT}
 \kapT&=\gamma\kapS &\text{(fluids and solids)},
 \\
 \eqlab{thermodynamic_relation_gamma}
 \gamma&=\frac{\cp}{\cV}=1+\frac{\alphap^2 T}{\rho \cp \kapS}  &\text{(fluids and solids)},
 \\
 \eqlab{Dth}
 \Dth &= \frac{\kth}{\rho\cp} &\text{(fluids and solids)},
 \esubal
where $\rho$ and $T$ is the mass density and the temperature field in the respective media.

\vspace*{-5 mm}

\subsection{Thermoviscous fluids}

\vspace*{-3 mm}

First, we present the governing equations for a thermoviscous Newtonian fluid. The fluid is described by the internal energy per mass $\ve$, the temperature $T$, pressure $p$, density $\rho$, and velocity field $\vvv$. We introduce the fluid stress tensor $\sigmabf$, which is expressed in terms of the pressure $p$, the fluid velocity field $\vvv$, the dynamic viscosity $\eta$, and the bulk viscosity $\etaB$,
 \bal
 \eqlab{sigma_fl}
 \sigmabf = \eta \Big[\nablabf \vvv +(\nablabf \vvv)^\textsf{T}\Big] +
 \big[(\etaB-\tfrac23 \eta)(\div \vvv) - p \big]\Imat .
 \eal
The governing equations for the fluids are the conservation of mass, momentum, and energy, without external body forces and heat sources \cite{Landau1993, Landau1980, Bruus2008},
 \bsubal{fluid_governing}
 \eqlab{continuity_eq}
 \pp_t\rho &=\div [-\rho \vvv], \\
 \eqlab{NS_eq}
 \pp_t(\rho\vvv)&=\div [\sigmabf -\rho \vvv\vvv], \\
 \eqlab{energy_eq}
 \pp_t(\rho\ve+\tfrac12 \rho v^2)&=\div[\vvv\cdot \sigmabf+\kth\nablabf T-\rho(\ve+\tfrac12 v^2)\vvv].
\esubal
To close the system of equations, we supplement with the first law of thermodynamics relating changes of the internal energy per mass $\ve$, the entropy per mass $s$, and the density $\rho$, and with the equation of state relating changes in density, pressure, and temperature,
 \bsubal{first_law}
 \eqlab{de}
 \dm \ve &= T \,\dm s - p\, \dm\bigg(\frac{1}{\rho}\bigg)=T\,\dm s+\frac{p}{\rho^2}\,\dm\rho,
 \\
 \eqlab{ds}
 \dm s & = \frac{\cP}{T} \,\dm T -\frac{\alphap}{\rho} \, \dm p,
 \\
 \eqlab{drho}
 \dm \rho & = \rho \kapT \,\dm p -\rho \alphap \, \dm T.
 \esubal
For fluids,  the isentropic sound speed $c$ is related to the isentropic compressibility through the relation
 \beq{isentropic_sound_speed_fluids}
 c= (\rho \kapS)^{-\frac12} \qquad \text{(fluids)}.
 \eeq
In summary, the fluid fields $g$ and fluid parameters $q$ that we expand according to \eqsref{pertubation}{pertubation_params}, respectively, are
 \bsubal{g_and_q_fluids}
 \eqlab{fluid_fields}
 g &= \{T,\rho, p,\vvv,\sigmabf, \ve, s \},
 \\
 \eqlab{fluid_parameters}
 q &= \{c, \eta, \etaB, \kth, \kapT, \kapS, \cp, \alphap, \gamma , \cV,\Dth \}.
 \esubal

\textit{\underline{First-order equations for fluids}}.
With the expansions~\eqsnoref{pertubation}{pertubation_params}, and using $\vvv_0 = \zerovec$, the first-order terms of \eqref{fluid_governing} become,
 \bsubal{1st_order_fluid_governing}
 \eqlab{1st_order_continuity}
 &-\ii \omega \rho_1 = -\rho_0 \div \vvv_1 ,
 \\
 \eqlab{1st_order_NS}
 &-\ii\omega \rho_0 \vvv_1 = \eta_0 \nabla^2 \vvv_1 +(\etaB_0+\tfrac13 \eta_0)\nablabf(\div \vvv_1) -\nablabf p_1,
 \\
 \eqlab{1st_order_energy}
 &-\ii \omega (\rho_0 \ve_1+\rho_1 \ve_0) = -p_0\div \vvv_1+\kthO \Lapl T_1-\rho_0\ve_0\div\vvv_1.
 \esubal
Similarly, the first-order terms of \eqref{first_law} become,
 \bsubal{eps1_rho1_s1}
 \eqlab{eps1}
 \ve_1 &= T_0s_1+\frac{p_0}{\rhoO^2}\rho_1, \\
 \eqlab{rho1}
 \rho_1 &= \rhoO\kapTO p_1-\rhoO \alphapO T_1, \\
 \eqlab{s1}
 s_1 &= \frac{\cpO}{T_0}T_1-\frac{\alphapO}{\rhoO}p_1.
\esubal
Inserting \eqref{eps1_rho1_s1} into \eqsref{1st_order_continuity}{1st_order_energy}, and eliminating \eqref{1st_order_continuity} from \eqref{1st_order_energy}, yields,
 \bsubal{1st_order_fluid_governing_2}
 \eqlab{1st_order_continuity_2}
 \div \vvv_1 &= \ii \omega \kapTO p_1 - \ii \omega \alphapO T_1, \\
 \eqlab{1st_order_energy_2}
 \DthO \nabla^2 T_1 &= -\ii \omega T_1 + \ii \omega \frac{\alphapO T_0}{\rhoO\cpO}p_1.
 \esubal
Equations~\eqsnoref{1st_order_NS}{1st_order_fluid_governing_2} determine the five components of the fields $\vvv_1$, $p_1$, and $T_1$, and they are conveniently solved by applying a Helmholtz decomposition of $\vvv_1$,
 \bal\eqlab{Helmholtz_decomp_v1}
 \vvv_1=\nablabf \phi_1 + \rot \psibfI.
 \eal
Inserting \eqref{Helmholtz_decomp_v1} into \eqsref{1st_order_NS}{1st_order_fluid_governing_2} leads to
 \bsubal{fluid_potential_governing}
 \eqlab{phi1_fl}
 -\omega^2 \phi_1 &= \Big(\frac{1}{\rhoO \kapTO}-\ii\omega \frac{\etaOB+\tfrac{4}{3}\etaO}{\rhoO}\Big)\nabla^2 \phi_1 + \ii\omega \frac{\alphapO}{\rhoO\kapTO}T_1, \\
 \eqlab{T1_fl}
 -\ii\omega T_1 &= \gamma \DthO \nabla^2 T_1-\frac{\gamO-1}{\alphapO}\nabla^2 \phi_1,
 \\
 \eqlab{psi1_fl}
 -\ii\omega\psibfI &=\nu_0\nabla^2 \psibfI,\; \text{ with }\;  \nu_0=\eta_0/\rho_0,
 \\
 \eqlab{p1}
 p_1 &= \ii\omega \rhoO \phi_1+(\etaOB+\tfrac{4}{3}\etaO)\nabla^2 \phi_1.
 \esubal
One can then combine \eqsref{phi1_fl}{T1_fl} to solve for $\phi_1$ and $T_1$ while $p_1$ is expressed in terms of $\phi_1$, and $\psibfI$ is found from \eqref{psi1_fl}.

\textit{\underline{Second-order time-averaged equations for fluids}}. In second order, we study only the steady part of the response obtained by time-averaging over an acoustic oscillation cycle. The fluid viscosities are expanded as in \eqref{pertubation_params}, so the second-order time-averaged terms of the stress tensor \eqref{sigma_fl}  become,
 \bal\eqlab{sigma2}
 &\avr{\sigmabf_2}=\eta_0 \Big[\nablabf \avr{\vvv_2} + (\nablabf \avr{\vvv_2})^\textsf{T}\Big]+(\etaB_0-\tfrac{2}{3}\eta_0)(\div \avr{\vvv_2}) \,\Imat
 \nn\\
 &\quad -\avr{p_2} \, \Imat + \avr{\eta_1 \Big[\nablabf \vvv_1 + (\nablabf \vvv_1)^\textsf{T}\Big]+(\etaB_1-\tfrac{2}{3}\eta_1)(\div \vvv_1)\,\Imat }.
 \eal
We introduce in \eqref{sigma2} the kinematic viscosities
$\nuO = \frac{1}{\rhoO}\etaO$,
$\nuOB = \frac{1}{\rhoO}\etaOB$,
$\nuI = \frac{1}{\rhoO}\etaI$, and
$\nuIB = \frac{1}{\rhoO}\etaIB$, so that the second-order terms in
\eqsref{continuity_eq}{NS_eq} become
 \bsubal{2nd_order_fluid_governing}
 \eqlab{2nd_order_continuity}
 \div &\avr{\vvv_2} = -\frac{1}{\rho_0}\div\avr{\rho_1\vvv_1} ,
 \\
 \eqlab{2nd_order_NS}
 \nuO & \nabla^2 \avr{\vvv_2}
 +
 (\nuOB+\tfrac{1}{3}\nuO)\nablabf (\div \avr{\vvv_2})
 - \frac{1}{\rhoO}\nablabf \avr{p_2}
 \\
 & =
 -\div\avr{\nuI\Big[\nablabf \vvv_1+(\nablabf \vvv_1)^\textsf{T}\Big]
 +[\nuIB-\tfrac{2}{3}\nuI](\div \vvv_1) \Imat }
 \nn\\\nn
 & \quad +\div \avr{\vvv_1\vvv_1} .
 \esubal
These equations make no reference to the second-order temperature field $T_2$, so the energy conservation equation \eqnoref{energy_eq} is not needed in second order.

\subsection{Thermoelastic solids}
A thermoelastic linear isotropic solid is described as in Refs.~\cite{Landau1986, Karlsen2015}. We introduce the mechanical displacement field $\uuu(\rrrO,t)$ of a solid element from its equilibrium location $\rrrO$, the temperature field $T$, and the mass density $\rho$. The solid stress tensor $\sigmabf$ is expressed in terms of $\uuu$ and the temperature $T$ relative to the equilibrium value $T_0$,
 \bal
 \eqlab{sigma_sl}
 \sigmabf &= -\frac{\alphap}{\kapT}(T-T_0)\Imat +\rho\cT^2\Big[(\nablabf\uuu)+(\nablabf\uuu)^\textsf{T}\Big] \nn\\
 &\quad +\rho (\cL^2 -2\cT^2)(\div \uuu) \Imat .
 \eal
Here, we have introduced the longitudinal $\cL$ and transverse $\cT$ speed of sound in the solid, which are related to the isentropic compressibility $\kapT$ and the density as,
 \bal \eqlab{kapT_solid}
 \frac{1}{\rho\kapT}=\cL^2-\frac{4}{3}\cT^2 \qquad \text{(Solids)}.
 \eal
We also introduce the adiabatic sound speed $c$ in solids,
 \bal\eqlab{c_sl}
 c^2=\cL^2+(\gamma-1)\frac{1}{\rho \kapT} \qquad \text{(Solids)}.
 \eal
The physical fields in solids are governed by the momentum equation and the heat diffusion equation,
 \bsubal{solid_governing}
 \eqlab{solid_displacement}
 \rho \pp_t^2 \uuu &= \div \sigmabf, \\
 \eqlab{solid_heat_flux}
\pp_t T +\frac{(\gamma-1)}{\alphap} \pp_t \div \uuu &=\frac{\gamma}{\rho \cp} \div [\kth \nablabf T] .
 \esubal
We express the velocity field $\vvv$ in the solid from $\uuu$ by,
 \bal\eqlab{solid_velocity}
 \vvv = \pp_t \uuu \qquad \text{(Solids)}.
 \eal
The solid fields $g$ and solid parameters $q$ that we expand according to \eqsref{pertubation}{pertubation_params}, respectively, are
 \bsubal{g_and_q_solid}
 \eqlab{solid_fields}
 g &= \{T,\rho, \uuu,\sigmabf, \vvv \},
 \\
 \eqlab{solid_parameters}
 q &= \{c, \cL, \cT, \kth, \kapT, \kapS, \cp, \alphap, \gamma , \cV,\Dth \}.
 \esubal

\textit{\underline{First-order equations in solids}}. Perturbation expansion of \eqref{solid_governing} gives the governing first-order equations,
 \bsubal{solid_governing_1st_order}
 \eqlab{solid_displacement_1st_order}
 -\rhoO \omega^2 \uuu_1 &= \div \sigmabf_1, \\
 \eqlab{solid_heat_flux_1st_order}
 -\ii\omega\frac{\gamO-1}{\alphapO}\div\uuu_1&=\gamO \DthO \nabla^2 T_1+\ii \omega T_1 .
 \esubal
The quantity $-\ii \omega \uuuI$ in the solid is the counterpart to the first-order velocity $\vvv_1$ in the fluid. To solve the system of equations~\eqnoref{solid_governing_1st_order}, we use the Helmholtz decomposition
 \beq{solid_Helmholtz}
 -\ii\omega \uuu_1 = \nablabf \phi_1 + \rot \psibfI.
 \eeq
Inserting \eqref{solid_Helmholtz} into \eqref{solid_governing_1st_order} we derive,
 \bsubal{solid_potential_governing}
 \eqlab{phi1_solid}
 -\omega^2 \phi_1 &= \cLO^2 \nabla^2 \phi_1+\ii\omega \frac{\alphapO}{\rhoO\kapTO}T_1 ,
 \\
 \eqlab{T1_solid}
 -\ii\omega T_1 &= \gamO \DthO \nabla^2 T_1-\frac{\gamO-1}{\alphapO}\nabla^2 \phi_1,
 \\
 \eqlab{psi1_solid}
 -\omega^2 \psibfI &=\cTO^2 \nabla^2 \psibfI.
 \esubal

\textit{\underline{The second-order response in the solid}} is not needed, as the time-averaged second-order velocity field is zero.

\begin{figure*}[t!]
\centering
\includegraphics[width=0.85\linewidth]{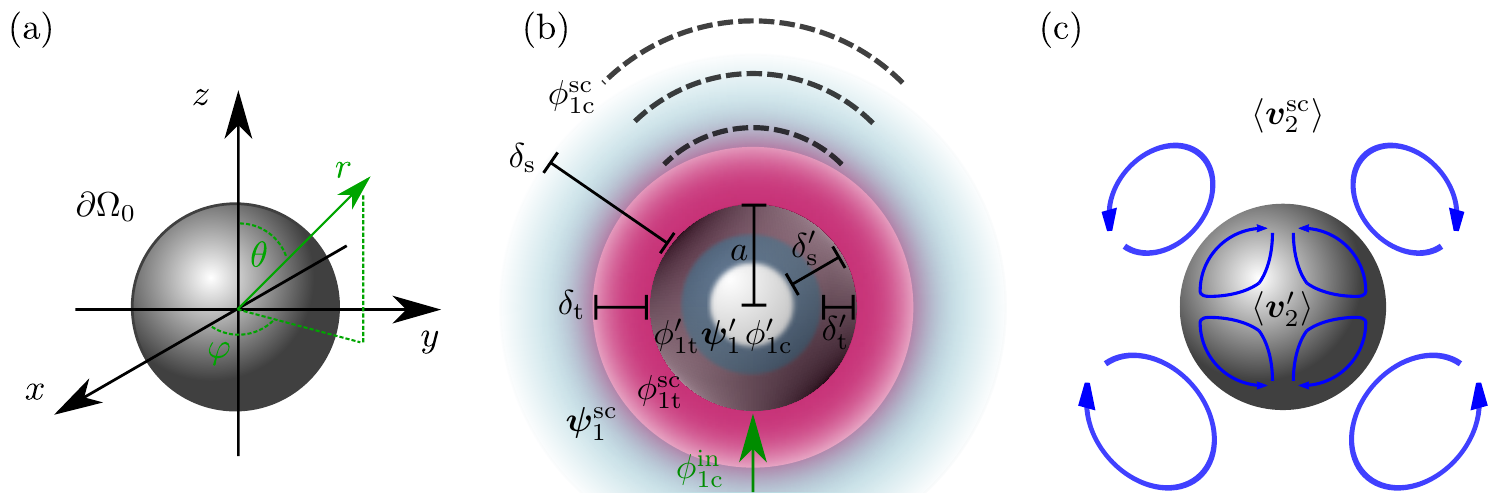}
\caption{\figlab{problem_sketch} (a) The spherical particle with radius $a$, and the referenced coordinate systems used in the calculation. (b) The incident pressure wave, described by the potential $\phiIcin$ and the corresponding scattered response, described by $\phiIc^\mr{sc}$, $\phiItsc$, $\psibfIsc$, $\phiIc'$, $\phiIt'$, and $\psibfIp$. (c) the second order time-averaged streaming rolls generated by the first-order wave scattering. }
\end{figure*}

\section{The first-order problem}
\seclab{FirstOrder}
We consider a spherical particle with radius $a$ centered in a spherical coordinate system $(r,\theta,\varphi)$ as shown in \figref{problem_sketch}(a). For a solid particle, we must solve \eqref{solid_potential_governing} for $r<a$ and \eqref{fluid_potential_governing} for $r>a$, whereas for a fluid particle, we must solve \eqref{fluid_potential_governing} in both regions. Clearly, these two systems of equations have the same structure, and following Karlsen \citep{Karlsen2015}, they can be treated in the same manner using unified potential theory.

\subsection{Unified potential theory for fluids and solids}
To determine $\psibfI$ in either a solid or a fluid, we must solve \eqsref{psi1_fl}{psi1_solid}, which both can be written as
 \beq{psi1_unified}
 \nabla^2 \psibfI+\ks^2 \psibfI =\zerovec, \qquad \text{(fluids and solids)}.
 \eeq
By defining the kinematic viscosity $\nuO$ of a solid as,
 \beq{solid_viscosity}
 \nuO = \ii\:\frac{\cTO^2}{\omega}, \qquad \text{(solids)},
 \eeq
the square of the shear wave number $\ks$ in both fluids and solids can be written as,
 \beq{ks}
 \ks^2 = \ii\: \frac{\omega}{\nuO}, \qquad \text{(fluids and solids)}.
 \eeq
To determine $\phi_1$ and $T_1$ we solve \eqsref{phi1_fl}{T1_fl} for fluids and \eqsref{phi1_solid}{T1_solid} for solids. In both cases, combining the pair of equations, a single equation for $\phi_1$ is obtained,
 \beq{Biharmonic}
 \alpha_\mr{xl} \nabla^2 \nabla^2 \phi_1 +\beta_\mr{xl}k_0^2\nabla^2\phi_1+k_0^4\phi_1
 = 0, \quad k_0=\frac{\omega}{c_0}.
 \eeq
Here, $\alpha_\mr{xl}$ and $\beta_\mr{xl}$ for fluids ($\xl = \fl$) and solids ($\xl = \sl$) are given by
 \bsubal{alpha_xl_and_beta_xl}
 \eqlab{xl_fluid}
 \alpha_\mr{fl} &= -\ii(1-\ii\gamO \Gams)\Gamt, \quad &\beta_\mr{fl} &= 1-\ii(\Gams+\gamO\Gamt),
 \\
 \eqlab{xl_solid}
 \alpha_\mr{sl} &= -\ii(1+X)\Gamt, \quad &\beta_\mr{sl} &=1-\ii\gamO\Gamt,
 \\
 \eqlab{defGamsGamt}
 \Gams&= \frac{(\etaB_0+\frac{4}{3}\eta_0)\omega}{\rho_0 \cO^2},
 & \Gamt&=	\frac{\DthO \omega}{\cO^2}	,
 \\
 \eqlab{defXchi}
 X &=	(\gamO-1)(1-\chi),
 & \chi &=\frac{1}{\rho_0\kapSO \cO^2},
 \esubal
where we have introduced the viscous and thermal damping factors, $\Gams$ and $\Gamt$, and the two parameters $X$ and $\chi$.
The biharmonic equation \eqnoref{Biharmonic} is solved by factorization,
 \bal\eqlab{factorized_Biharmonic}
 (\nabla^2+\kc^2)(\nabla^2+\kt^2)\phi_1=0,
 \eal
which combined with \eqnoref{Biharmonic} yields $\kc^2+\kt^2=\beta_\mr{xl}k_0^2/\alpha_\mr{xl}$ and $\kc^2\kt^2=k_0^4/\alpha_\mr{xl}$, and thus
 \bsubal{kc_kt_general}
 \eqlab{kc_general}
 \kc^2&=k_0^2 \frac{1}{2\alpha_\mr{xl}}\bigg(\beta_\mr{xl}-\sqrt{\beta_\mr{xl}^2-4\alpha_\mr{xl}}\bigg), \\
 \eqlab{kt_general}
 \kt^2&=k_0^2 \frac{1}{2\alpha_\mr{xl}}\bigg(\beta_\mr{xl}+\sqrt{\beta_\mr{xl}^2-4\alpha_\mr{xl}}\bigg).
 \esubal
The solution for $\phi_1$ is then split into two components
 \beq{phi_components}
 \phi_1=\phiIc+\phiIt,
 \eeq
where
 \bsubal{phic_phit_eqs}
 \eqlab{phic_eq}
 \nabla^2\phiIc+\kc^2\phiIc&=0,  \\
 \eqlab{phit_eq}
 \nabla^2\phiIt+\kt^2\phiIt&=0.
 \esubal
After solving for $\phi_1$, $T_1$  can be determined in fluids and solids from \eqsref{phi1_fl}{phi1_solid},
 \bsubal{T1_from_phi1}
 \eqlab{T1_fl_from_phi1}
 T_1&= \frac{\ii \omega \rho_0 \kapTO}{\alphapO}\bigg[\phi_1+\frac{\cO^2}{\omega^2}\frac{1-\ii\gamO\Gams}{\gamO}\nabla^2 \phi_1\bigg]  \text{ (fluids)},
 \\
 \eqlab{T1_sl_from_phi1}
 T_1&= 	\frac{\ii \omega \rhoO \kapTO}{\alphapO}\bigg[\phi_1+\frac{\cLO^2}{\omega^2}\nabla^2 \phi_1\bigg]  \text{\hspace*{13mm} (solids)}.
 \esubal
For the systems we are considering here, $\Gams,\Gamt\ll 1$, and expanding to first order in these parameters, we find the following wave numbers when using $\kO = \omega/\cO$,
 \bsubal{wavenumbers_fluids}
 \eqlab{kc_fluids}
 \kc &= \kO\:\Big[1+\frac{\ii}{2}[\Gams+(\gamO-1)\Gamt]\Big], \\
 \eqlab{kt_fluids}
 \kt &= \frac{1+\ii}{\delt}\Big[1+\frac{\ii}{2}(\gamO-1)(\Gams-\Gamt)\Big], \\
 \eqlab{ks_fluids}
 \ks &= \frac{1+\ii}{\dels},\quad \text{\underline{for fluids}} ,
 \esubal
and
 \bsubal{wavenumbers_solids}
 \eqlab{kc_solids}
 \kc &= \kO\:\Big[1+\frac{\ii}{2}(\gamO-1)\chi\Gamt\Big],
 \\
 \eqlab{kt_solids}
 \kt &= \frac{1+\ii}{\delt}\frac{1}{\sqrt{1+X}}\Big[1-\frac{\ii}{2}(\gamO-1)\chi\Gamt\Big],
 \\
 \eqlab{ks_solids}
 \ks &= \frac{\omega}{\cTO},\quad \text{\underline{for solids}}.
 \esubal
Here, we have introduced the thermal and viscous boundary layer thicknesses $\delt$ and $\dels$ as,
 \beq{boundary_layers}
 \delt = \sqrt{\frac{2\DthO}{\omega}},
 \qquad
 \dels = \sqrt{\frac{2\nu_0}{\omega}}.
 \eeq
To lowest order in $\Gams$ and $\Gamt$, we also find that
 \bsubal{T1_potential_form}
 \eqlab{T1_potential}
 T_1 &=\bc \phiIc + \bt \phiIt,\; \text{ (fluids and solids)},
 \\
 \eqlab{bc_and_bt}
 \bc &= \frac{\ii\omega(\gamO-1)}{\alphapO \cO^2}, \quad
 \bt = \frac{1}{\chi \alphapO \DthO},
 \esubal
where $\chi=1$ for fluids. Thus, the governing first-order equations are the same for fluids and solids, only differing by the form of their wave numbers. In fluids, both $\kt$ and $\ks$ describe strongly damped waves, whereas in solids $\ks$ describes a propagating transverse wave.

\subsection{Partial-wave expansions and scattering coefficients}
\seclab{FirstOrderPartialWave}
Following Karlsen \cite{Karlsen2015}, quantities inside the particle ($r<a$) are denoted with a prime,  whereas outside in the surrounding fluid medium ($r>a$) they remain unprimed, see \figref{problem_sketch}(b) and (c). Moreover, a tilde denotes the ratio of a parameter  $q'_0$ inside a particle and $q_0$ in the surrounding fluid (in contrast to Doinikov, who uses the tilde to denote particle parameters \cite{Doinikov1994, Doinikov1994a, Doinikov1997, Doinikov1997a, Doinikov1997b}),
 \beq{tilde_def}
 \tilde{q}_0 = \frac{q'_0}{q_0}.
 \eeq
To determine the field solutions of the first-order problem, we consider an axisymmetric incident wave in the surrounding fluid medium, which is far enough away from solid boundaries, such that thermal and viscous modes from there are vanishingly small. When the incident field interacts with the particle, scattered fields are generated. Thus, we write $\phiIc$ outside the particle as the sum of an incident field and a scattered field, while the remaining potentials only contain a scattered part,
 \bsubal{pot_I_all}
 \eqlab{phiIc}
 \phiIc &= \left\{\begin{array}{ll}
 \phiIcin + \phiIcsc,& r > a,\\
 \phiIcp, & r < a, \end{array} \right.
 \\
 \eqlab{phiIt}
 \phiIt &= \left\{\begin{array}{ll}
 \phiItsc,& r > a,\\
 \phiItp, & r < a, \end{array} \right.
 \\
 \eqlab{psibfI}
 \psibfI &= \left\{\begin{array}{ll}
 \psibfIsc,& r > a,\\
 \psibfIp, & r < a. \end{array} \right.
 \esubal
All potentials $\phiIcin$, $\phiIcsc$, $\phiItsc$, $\psibfIsc$, $\phiIc'$, $\phiIt'$, and $\psibfIp$ are solutions to Helmholtz equations. Inside the particle for $r<a$, the solutions are written in terms spherical Bessel functions $j_n(k'_ir)$ and Legendre polynomials $P_n(\cos\theta)$ to avoid singularities at $r=0$. Outside the particle for $r>a$, the scattered fields  are written in terms of the decaying outgoing spherical Hankel functions $h_n(k_ir)$.
\\[1mm]
\underline{In the surrounding fluid (unprimed), $r>a$}
 \bsuba{potential_solutions}
 \bal
 \eqlab{phicin_solution}
 \phiIcin &= \sum_{n=0}^\infty \phiIcinn(r) P_n(\cos\theta),
 \\ \nn 
 &\text{\; with }\; \phiIcinn(r) = A_n j_n(\kc r)							,
 \\ \eqlab{phicsc_solution}
 \phiIcsc &= \sum_{n=0}^\infty \phiIcscn(r) P_n(\cos\theta),
 \\ \nn 
 &\text{\; with }\; \phiIcscn(r) = A_n \acnsc h_n(\kc r)					,
 \\ \eqlab{phit_solution}
 \phiItsc &= \sum_{n=0}^\infty \phiItscn(r) P_n(\cos\theta)	,
 \\ \nn 
 &\text{\; with }\; \phiItscn(r) = A_n \atnsc h_n(\kt r)				,
 \\ \eqlab{psi_solution}
 \psibfIsc &= \eee_\varphi\sum_{n=1}^\infty \psiIscn(r) \pp_\theta P_n(\cos\theta) ,
 \\ \nn 
 &\text{\; with }\; \psiIscn(r) = A_n \asnsc h_n(\ks r).
 \eal
 \esuba
\underline{Inside the particle (primed), $r<a$}
 \bsuba{potential_solutions_solid}
 \bal
 \eqlab{phicm_solution}
 \phiIcp &= \sum_{n=0}^\infty \phiIcnp(r) P_n(\cos\theta), 	
 \\ \nn 
 &\text{\; with }\; \phiIcnp(r) = A_n \acnp j_n(\kc' r), 	
 \\
 \eqlab{phitm_solution}
 \phiItp &= \sum_{n=0}^\infty \phiItnp(r) P_n(\cos\theta), 	
 \\ \nn 
 &\text{\; with }\; \phiItnp(r) = A_n \atnp j_n(\kt' r),	
 \\
 \eqlab{psim_solution}
 \psibfIp &= \eee_\varphi\sum_{n=1}^\infty \psiInp(r) \pp_\theta P_n(\cos\theta) ,
 \\ \nn 
 &\text{\; with }\; \psiInp(r) = A_n \asnp j_n(\ks' r).
 \eal
 \esuba
Here, the wave coefficients $A_n$ defines the incident wave, and the scattering coefficients $\acnsc$, $\atnsc$, $\asnsc$, $\acnp$, $\atnp$, and $\asnp$ are determined from the boundary conditions (continuous stress, velocity, temperature, and heat flux) at the particle-fluid interface $r = a$. A compact notation $\ain$, with index  $i=0,1,2,3,4,5,6 = 0, c, t, s, c', t', s'$ and mode number $n=0,1,\ldots$, is introduced for the six scattering coefficients, and when we include unity for the incoming wave for each mode $n$, we obtain
 \bal
 \eqlab{alphaDef}
 &\big\{\ain\}_{i=0, 1,\ldots,6} =
 \big\{1, \acnsc, \atnsc, \asnsc, \acnp, \atnp, \asnp \big\},
 \nn \\
 &\quad \text{with }\; i = 0,1,2,3,4,5,6 = 0, c, t, s, c', t', s'.
 \eal
When imposing the boundary conditions at the particle surface, it is convenient to express the first-order pressure in terms of the potentials,
 \beq{p1_from_potentials}
 p_1 =  \ii\omega\rho_0(\phiIc+\phiIt) -(\etaB_0+\tfrac{4}{3}\eta_0)(\kc^2\phiIc+\kt^2\phiIt).
 \eeq
The velocity in a solid is given by $\vvv_1=-\ii\omega\uuu_1$, and using the definition~\eqnoref{solid_viscosity} of solid viscosity, we express the first-order stress $\sigmabf_1$ for both fluids and solids as,
 \bal\eqlab{stress_from_potentials}
 \sigmabf_1 &=\eta_0 [(2\kc^2-\ks^2)\phiIc+(2\kt^2-\ks^2)\phiIt] \,\Imat
 \nn \\
 &\quad + \eta_0\Big[\nablabf \vvv_1+(\nablabf\vvv_1)^\textsf{T}\Big],
 \;\; \text{(fluids and solids)}.
 \eal
Lastly, from \eqsref{potential_solutions}{potential_solutions_solid} and the definition $\vvv_1=\nablabf \phiIc+\nablabf \phiIt+\rot\psibfIsc$ for both the unprimed and primed first-order velocity, we have the following expressions:
 \bsuba{v1_solutions}
 \bal
 \eqlab{v1r_sol}
 \vIr &=\sum_{n=0}^\infty \vIrn(r)P_n(\cos\theta),
 \\ \nn
 &\text{with }\; \vIrn(r) = A_n\frac{1}{a}\bigg[\xc j'_n(\kc r)+\acnsc\xc h'_n(\kc r)
 \\ \nn
 &\qquad +\atnsc\xt h'_n(\kt r)-\asnsc a\frac{n(n+1)}{r}h_n(\ks r)\bigg],
 \\
 \eqlab{v1th_sol}
 \vIth &= \sum_{n=1}^\infty \vIthn(r) \pp_\theta P_n(\cos\theta),
 \\ \nn
 &\text{with }\; \vIthn(r) = A_n\frac{1}{a} \bigg[\frac{a}{r}j_n(\kc r)
 + \acnsc \frac{a}{r}h_n(\kc r)
    \nn\\ & \qquad +\atnsc \frac{a}{r}h_n(\kt r) -\asnsc \Big(\xs h_n'(\ks r)+\frac{a}{r}h_n(\ks r)\Big)  \bigg], \nn
 \\
 \eqlab{v1rm_sol}
 \vIrp &=\sum_{n=0}^\infty \vIrnm(r)P_n(\cos\theta),
 \\ \nn
 &\text{with }\; \vIrnm(r) = A_n\frac{1}{a}\bigg[\acnp\xcp j'_n(\kc' r)
 \\ \nn
 &\qquad +\atnp\xtp j'_n(\kt' r)-\asnp a\frac{n(n+1)}{r}j_n(\ks' r)\bigg],
 \\
 \eqlab{v1thm_sol}
 \vIthp &= \sum_{n=1}^\infty \vIthnm(r) \pp_\theta P_n(\cos\theta),
 \\ \nn
 &\text{with }\; \vIthnm(r) = A_n\frac{1}{a} \bigg[\acnp \frac{a}{r}j_n(\kc' r)+\atnp \frac{a}{r}j_n(\kt' r)
   \\ &\qquad  -\asnp \Big(\xsp j_n'(\ks' r)+\frac{a}{r}j_n(\ks' r)\Big)\bigg], \nn
 \eal
 \esuba
where primed spherical Bessel and Hankel functions refer to differentiation with respect to the argument, and where we have introduced the normalized wave numbers,
\bsubalat{normalized_wavenumbers}{4}
\eqlab{normalized_wavenumbers_fluid}
\xc &= \kc a,& \quad \xt &= \kt a,&  \quad \xs &= \ks a,& &\quad \text{for }\; r>a,\\
\eqlab{normalized_wavenumbers_particle}
\xcp &= \kc' a,& \quad  \xtp &= \kt' a,& \quad \xsp &= \ks' a,& &\quad  \text{for }\; r<a.
\esubalat

The boundary conditions on the particle surface are continuity of velocity, stress, temperature, and heat flux. For a fluid particle, the Young--Laplace pressure $\pST$ due to the surface tension $\gamST$ is explicitly added as a discontinuity in the normal-stress boundary condition,
 \bsubal{BCs_1st_order}
 \eqlab{v1_continuous}
 \vIr &=\vIrp, &\vIth &= \vIthp,
 \\
 \eqlab{T1_continuous}
 T_1 &= T_1' ,  &\kthO \pp_r T_1 &= \kthOp \pp_r T_1' ,
 \\
 \eqlab{stress_continuous}
 \sigma_{1\theta r} & = \sigma_{1\theta r}' ,   &\sigma_{1rr} &= \sigma_{1rr}'+\pST,
 \\
 \eqlab{pST_def}
 \pST &= -\frac{\ii\gamST}{a^2\omega} && \hspace*{-15mm}
 \big(\pp_\theta^2 v_{1r}' +\pp_\theta v_{1r}' \cot\theta+2 v_{1r}'\big)\big|^{}_{r=a}.
 \esubal
Here, \eqref{pST_def} for $\pST$ is valid for an axisymetrically perturbed spherical surface \cite{Doinikov1997b,Jackson1988}. Substituting \eqref{v1rm_sol} into \eqref{pST_def}, and using $\pp_\theta^2P_n(\cos\theta)+\pp_\theta P_n(\cos\theta)\cot \theta+2P_n(\cos\theta) = -(n-1)(n+2)P_n(\cos\theta)$, yields,
 \beq{ST_sol}
 \pST = \sum_{n=0}^\infty \ii \frac{\gamST}{a^2 \omega}(n-1)(n+2) \vIrnm(a) P_n(\cos\theta).
 \eeq
For each value of $n$, the final expressions of the boundary conditions from \eqnoref{BCs_1st_order} are,
 \bsuba{BCs_n_1st_order}
 \bal
 \eqlab{BC_n_v1r}
 &\text{\underline{$\vIr=\vIrp$}} \nn\\
 &\;\; \acnsc\xc h'_n(\xc)+\atnsc\xt h'_n(\xt)-\asnsc n(n+1)h_n(\xs) \nn\\
 &\;\; -\acnp\xcp j'_n(\xcp) -\atnp\xtp j'_n(\xtp)+\asnp n(n+1) j_n(\xsp) \nn\\
 &\;\; =-\xc j'_n(\xc),
 \\
 \eqlab{BC_n_v1t}
 &\text{\underline{$\vIth= \vIthp$}} \nn\\
 &\;\; \acnsc h_n(\xc)+\atnsc h_n(\xt)-\asnsc \big[\xs h_n'(\xs)+h_n(\xs)\big] \nn\\
 &\;\; -\acnp j_n(\xcp)-\atnp j_n(\xtp)+\asnp \big[\xsp j_n'(\xsp)+j_n(\xsp)\big] \nn\\
 &\;\; = -j_n(\xc),
 \\
 \eqlab{BC_n_T1}
 &\text{\underline{$T_1=T_1'$}} \nn\\
 &\;\; \acnsc \bc h_n(\xc)+\atnsc \bt h_n(\xt)\nn\\
 &\;\;  - \acnp \bc' j_n(\xcp) -  \atnp \bt' j_n(\xtp) = -\bc j_n(\xc),
 \\
 \eqlab{BC_n_T1_flux}
 &\text{\underline{$\kthO \pp_rT_1=\kthOp \pp_r T_1'$}} \nn\\
 &\;\; \acnsc\kthO \bc \xc h_n'(\xc)+\atnsc \kthO \bt \xt h_n'(\xt)  \nn\\
 &\;\; -\acnp\kthOp \bc' \xcp j_n'(\xcp)-\atnp \kthOp \bt' \xtp j_n'(\xtp) \nn \\
 &\;\;= -\kthO\bc\xc j_n'(\xc),
 \eal
 \bal
 \eqlab{BC_n_sigma1tr}
 &\text{\underline{$\sigma_{1\theta r}=\sigma_{1\theta r}'$}}
 \nn\\
 &\;\; \acnsc 2\eta_0 \big[\xc h_n'(\xc)-h_n(\xc)\big] \nn\\
 &\;\; +\atnsc 2\eta_0\big[\xt h_n'(\xt)-h_n(\xt)\big]
 \nn\\
 &\;\; -\asnsc \eta_0\big[\xs^2 h_n''(\xs)+(n^2+n-2)h_n(\xs)\big]
 \nn\\
 &\;\; -\acnp 2\eta_0' \big[\xcp j_n'(\xcp)\!-\!j_n(\xcp)\big] \nn\\
 &\;\;-\atnp 2\eta_0'\big[\xtp j_n'(\xtp)\!-\!j_n(\xtp)\big]
 \nn\\
 &\;\;  +\asnp \eta_0'\big[\xs^{\prime 2} j_n''(\xsp)+(n^2+n-2)j_n(\xsp)\big]
 \nn\\
 &\;\;   =-2\eta_0\big[\xc j_n'(\xc)-j_n(\xc)\big],
 \\
 \eqlab{BC_n_sigma1rr}
 &\text{\underline{$\sigma_{1rr}=\sigma_{1rr}'+\pST$}} \nn\\
 &\;\; \acnsc \eta_0 \big[(2\xc^2-\xs^2)h_n(\xc)+2\xc^2 h_n''(\xc)\big]
 \nn\\
 &\;\; +\atnsc \eta_0 \big[(2\xt^2-\xs^2)h_n(\xt)+2\xt^2 h_n''(\xt)\big]
 \nn\\
 &\;\; -\asnsc \eta_0 2n(n+1) \big[\xs h_n'(\xs)-h_n(\xs)\big]
 \nn\\
 &\;\; -\acnp \eta_0' \big[(2 x_\mr{c}^{\prime 2}-\xs^{\prime 2})j_n(\xcp)
 +2x_\mr{c}^{\prime 2} j_n''(\xcp)\nn\\
 &\;\; \qquad \qquad +(n-1)(n+2)\frac{\ii \gamST}{a \omega \eta_0'}\xcp j'_n(\xcp)\big]
 \nn\\
 &\;\; -\atnp \eta_0' \big[(2 x_\mr{t}^{\prime 2}-\xs^{\prime 2})j_n(\xtp)
 +2x_\mr{t}^{\prime 2} j_n''(\xtp)
 \nn\\
 &\;\; \qquad \qquad  +(n-1)(n+2)\frac{\ii \gamST}{a \omega \eta_0'}\xtp j'_n(\xtp)\big]
 \nn\\
 &\;\; +\asnp \eta_0' \big[ 2n(n+1) (\xsp j_n'(\xsp)-j_n(\xsp))
 \nn\\
 &\;\; \qquad \qquad  +(n-1)n(n+1)(n+2)\frac{\ii \gamST}{a\omega \eta_0'}j_n(\xsp) \big]
 \nn\\
 &\;\; = -\eta_0\big[(2\xc^2-\xs^2)j_n(\xc)+2\xc^2j_n''(\xc)\big].
 \eal
 \esuba
We notice that the terms with $n=1$, describing translation of the sphere, are unaffected by the surface tension terms, and that for a solid particle, we have $\gamST=0$. The expressions \eqnoref{BCs_n_1st_order} show that the scattering coefficients $\big\{\ain\big\}$, $i = 1,2,\ldots,6$, for each mode $n$ can be found by solving the following 6-by-6 matrix equation,
 \bsub
 \beq{matrix_system}
 \sum_{i=1}^6 M_{ki,n} \ain = N_{k,n},\; k=1,2,\ldots,6, \text{ mode } n,
 \eeq
where the 36 matrix elements $(\MMM_n)_{ik}$ for a given mode $n$ can be read off from \eqref{BCs_n_1st_order}, as can the 6 vector components $(\NNN_n)_k$ on the right-hand side,
 \beq{Nvec}
 \NNN_n = \begin{bmatrix}
           -\xc j'_n(\xc)  			\\
           -j_n(\xc) 					\\
           -\bc j_n(\xc) 				\\
           -\kthO\bc\xc j_n'(\xc)		\\
           -2\etaO[\xc j_n'(\xc)-j_n(\xc)] 	\\
            -\etaO[(2\xc^2-\xs^2)j_n(\xc)+2\xc^2j_n''(\xc)]
         \end{bmatrix}.
 \eeq
Note that \eqsref{BC_n_v1t}{BC_n_sigma1tr} are void for mode $n=0$, leading to a $4\times 4$ matrix system in this case of the monopole coefficients. For a given mode $n$, the scattering coefficients can be found by Cramer's rule,
 \beq{scatter_coef_compute}
 \ain = \big|\MMM_n\big|^{-1}\:\big|\MMM_n^{(i)}\big|, \quad i = 1, 2, \ldots, 6,
 \eeq
 \esub
where $|\dots|$ indicates the determinant, and $\MMM_n^{(i)}$ is the matrix $\MMM_n$ with its $i$'th column replaced by $\NNN_n$.

We consider the long-wavelength limit, where the incident wavelength $\lambda=\cO/f$ is much longer than the particle radius and the boundary layer thicknesses, $\lambda\gg a,\delt,\delt',\dels,\dels'$ for fluids and $\lambda, \dels' \gg a,\delt,\delt',\dels$ for solids. No restrictions are put on the particle radius relative to the boundary-layer thicknesses. The long-wavelength limit is characterized by the  small parameter $\xO$, in which we develop leading order expressions,
 \beq{xOdef}
 \xO = \kO a = 2\pi \frac{a}{\lambda} \ll 1.
 \eeq
It is assumed that all fluids and solids have speeds of sounds---and therefore wavelengths---of roughly the same order of magnitude, leading to
 \bal\eqlab{general_scaling_1}
 |\xc|^2, |\xcp|^2 \sim \xO^2.
 \eal
For the case of either a solid or a fluid particle, the following general scalings hold,
 \bal\eqlab{general_scaling_2}
 \Gams, \Gamt, \frac{|\bc|}{|\bt|}, \frac{|\bc'|}{|\bt|} \sim \xO^2.
 \eal
Lastly, due to the definition of solid viscosity \eqnoref{solid_viscosity} and the difference in solid and fluid shear wave numbers $\ks$, one has the following different scalings for the case of a solid sphere and a fluid sphere, respectively:
 \bsubal{specific_scalings}
 \eqlab{solid_scalings}
 &|\xsp|^2,|\etaOTi|^{-1} \sim \xO^2 \ll |\xs|^2,|\xt|^2,|\xtp|^2,  &&\text{ (solids)},\\
 \eqlab{fluid_scalings}
 &\xO^2 \ll |\xs|^2,|\xsp|^2,|\xt|^2,|\xtp|^2, &&\text{ (fluids)}.
 \esubal
We expand the determinants in \eqnoref{scatter_coef_compute} to lowest order in $\xO$, while respecting the scalings in \eqssref{general_scaling_1}{general_scaling_2}{specific_scalings}, in order to find explicit expressions for the scattering coefficients in the long-wavelength limit. We performed this laborious procedure with the algebraic tool Maple \cite{Maple}. For a solid, we only need the unprimed scattering coefficients to calculate the radiation force, whereas for a fluid, we also need the primed ones. Only scattering coefficients for $n=0,1,2$ turn out to give significant contributions to the final force expressions. The scattering coefficients for a solid and for a fluid particle are given in \appsref{ScatteringCoeffSolids}{ScatteringCoeffFluids}, respectively. We find that the expressions for $\acsc0$ and $\acsc1$ in solids are identical to those given by Karlsen \cite{Karlsen2015}, whereas $\acsc0$ for a fluid differ for a non-zero surface tension, $\gamST\neq 0$.

\section{The second-order problem}
\seclab{SecondOrder}

We now turn to the solution of the second-order equations~\eqnoref{2nd_order_fluid_governing} for the pressure $\avr{p_2}$ and streaming $\avr{\vvv_2}$. We largely follow the method used by Doinikov \cite{Doinikov1994a, Doinikov1994}.

\subsection{The general solution}
Firstly, we split the fields into incident and scattered fields (superscript "in" and "sc", respectively) outside the particle, and inside a fluid particle they are described by the transmitted (superscript "prime") fields,
 \bsubal{v2p2_tot}
 \eqlab{v2_tot}
 \avr{\vvv_2} &= \left\{
 \begin{array}{ll}
 \avr{\vvvIIin}+\avr{\vvvIIsc}, & \quad r > a,
 \\[2mm]
 \avr{\vvvIIp}& \quad r < a, \end{array}
  \right.
 \\[2mm]
 \eqlab{p2_tot}
 \avr{p_2} &= \left\{
 \begin{array}{ll}
 \avr{\pIIin}+\avr{\pIIsc}, & \quad r > a,
 \\[2mm]
 \avr{\pIIp}& \quad r < a. \end{array}
  \right.
 \esubal
Here, the second-order incident field $\avr{\vvvIIin}$ is calculated from the governing equations \eqnoref{2nd_order_fluid_governing} using the given first-order potential $\phi^\mr{in}_{1c}$ in \eqref{phicin_solution} describing the incident first-order wave. Solutions for $\avr{\vvvIIin}$ in the case of standing plane waves, traveling plane waves, and diverging spherical waves are given by Doinikov in Refs.~\cite{Doinikov1994, Doinikov1994b}. Below, we show how to solve for the scattered fields "sc" outside the particle, and note without showing it that the transmitted fields "prime" inside the particle are solved identically. We do, however, write explicitly when differences in the two solutions occur.

Secondly, we express $\avr{\vvvIIsc}$ and the right-hand side of the Navier--Stokes equation~\eqnoref{2nd_order_NS} by Helmholtz decompositions,
 \bsubal{Helmholtz_v2NS}
 \eqlab{v2_Helmholtz}
 &\avr{\vvvIIsc} =\nablabf \phiIIsc + \rot \psibfIIsc,
 \\[2mm]
 \eqlab{rhs_v2_Helmholtz}
 &-\div\avr{\nuI\Big[\nablabf \vvv_1 + (\nablabf \vvv_1)^\textsf{T}\Big]
 +[\nuIB-\tfrac{2}{3}\nuI](\div \vvv_1) \Imat }_\mr{nii}
 \nn\\
 &\qquad +\div\avr{(\vvv_1\vvv_1)}_\mr{nii}
 = \nuO\big(\nablabf Q + \rot \qqq\big).
 \esubal
The subscript "nii" (read as "no-incident-incident") indicates that all terms containing products of two incident first-order fields are discarded. The source terms $Q$ and $\qqq$ are found from the first-order fields $\vvvI$, $\nuI$, and $\nuIB$, and subsequently we determine the potential fields $\phiIIsc$ and $\psibfIIsc$.
We express the terms in \eqref{rhs_v2_Helmholtz} by partial-wave expansions,
 \bsuba{Q_q_RHSNS_expansion}
 \bal
 \eqlab{Q_expansion}
 Q &= \frac{1}{a^2} \sum_{n=0}^{\infty}  Q_n(\rhat) P_n(\cos\theta),
 \\
 \eqlab{q_expansion}
 \qqq &= \eee_\varphi \frac{1}{a^2}\sum_{n=1}^\infty q_n(\rhat) \pp_\theta P_n(\theta),
 \\
 \eqlab{expansion_Psi2_rhs}
 \nablabf\scap\Big\langle \vvv_1&\vvv_1 -\! \nuI\Big[\nablabf \vvv_1 \!+\! (\nablabf \vvv_1)^\textsf{T}\Big]
 \!-\! \Big[\nuIB \!-\! \tfrac{2}{3}\nuI\Big](\div \vvv_1) \Imat \Big\rangle_\mr{nii}
 \nn\\
 &=   \eee_r \frac{\nuO}{a^3}\sum_{n=0}^\infty (2n+1)\chi_{rn}(\rhat)P_n(\cos\theta)
 \nn\\
 &\quad -\eee_\theta \frac{\nuO}{a^3}
 \sum_{n=1}^\infty (2n+1)\chi_{\theta n}(\rhat)\pp_\theta P_n(\cos\theta),	
 \\
 \eqlab{chir}
 \chi_{rn}(\rhat) &=
 \frac{a^3}{2\nu_0} \int_0^\pi \dm\theta\: P_n(\cos\theta)\sin\theta\:
 \eee_r \cdot \bigg\{\div
 \nn\\
 \Big\langle \vvv_1\vvv_1
 &\!-\!\nuI\Big[\nablabf \vvv_1 \!+\! (\nablabf \vvv_1)^\textsf{T}\Big]
 \!-\! \Big[\nuIB \!-\! \tfrac{2}{3}\nuI\Big] (\div \vvv_1) \Imat \Big\rangle_\mr{nii}\bigg\},
 \eal
 \bal
 \eqlab{chit}
 &\chi_{\theta n}(\rhat) = \frac{-a^3}{2n(n+1)\nu_0}
 \int_0^\pi \! \dm\theta\:\big[\pp_\theta P_n(\cos\theta)\big] \sin\theta\:
 \eee_\theta\scap\bigg\{\div
 \nn\\
 &\Big\langle \vvv_1\vvv_1
 - \nuI\Big[\nablabf \vvv_1+(\nablabf \vvv_1)^\textsf{T}\Big]
 - \Big[\nuIB-\tfrac{2}{3}\nuI\Big](\div \vvv_1) \Imat \Big\rangle_\mr{nii}
 \bigg\},
 \eal
\esuba
where we have introduced the nondimensional radial coordinate $\rhat = r/a$. Combining \eqsref{rhs_v2_Helmholtz}{Q_q_RHSNS_expansion}, we find the following equations for $Q_n$ and $q_n$:
 \bsubal{Qn_and_qn_eqs}
 \eqlab{Qn_eq}
 \rhat \pp_\rhat Q_n(\rhat) -n(n+1)q_n(\rhat) &=(2n+1)\rhat\chi_{rn}(\rhat) ,
 \\
 \eqlab{qn_eq}
 -Q_n(\rhat)+\rhat\pp_\rhat q_n(\rhat)+q_n(\rhat) &= (2n+1)\rhat\chi_{\theta n}(\rhat) .
 \esubal
By differentiating \eqnoref{qn_eq} and using \eqnoref{Qn_eq}, we obtain an equation for $q_n$,
 \bal\eqlab{qn_only}
 \bigg[\pp_\rhat^2 +\frac{2}{\rhat}\pp_\rhat &-\frac{n(n+1)}{\rhat^2}\bigg]q_n(\rhat)
 \nn\\
 & = \frac{2n+1}{\rhat}
 \big[\chi_{\theta n}(\rhat) + \rhat\pp_\rhat\chi_{\theta n}(\rhat) + \chi_{rn}(\rhat)\big].
 \eal
The homogeneous solution to this equation is,
 \bal\eqlab{hom_solution_qn}
 q_n^\mr{hom}(\rhat)= c_{3n} \rhat^{-(n+1)}+c_{4n}\rhat^n.
 \eal
A particular solution is found from the Green's function $G_n(\xi,y)$ that solves the inhomogeneous equation,
 \beq{Greens_eq}
 \bigg[\pp_\rhat^2 +\frac{2}{\rhat}\pp_\rhat -\frac{n(n+1)}{\rhat^2}\bigg] G_n(\xi,\rhat) = \delta(\rhat-\xi).
 \eeq
From standard methods, we determine $G_n(\xi,\rhat)$ to be
 \bal\eqlab{Greens_function_done}
 G_n(\xi,\rhat)=
 \begin{cases}
      -\frac{1}{2n+1}\xi^{-(n-1)} \rhat^n, 		& \rhat<\xi, 	\\[2mm]
      -\frac{1}{2n+1}\xi^{n+2}\rhat^{-(n+1)}, 	& \rhat>\xi.
 \end{cases}
 \eal
By combining the expressions~\eqsnoref{hom_solution_qn}{Greens_function_done} for $q_n^\mr{hom}(\rhat)$ and $G_n(\xi,\rhat)$, we determine $q_n(\rhat)$, and subsequently, $Q_n(\rhat)$ is found directly from \eqref{qn_eq}. After using integration by parts on the term containing $\pp_\rhat \chi_{\theta n}(\rhat)$, the two solutions become,
 \bsubal{Qn_and_qn_done}
 \eqlab{qn_done}
 &q_n(\rhat) =
 \nn\\
 &-\rhat^{-(n+1)}\bigg\{\!
 \int_1^\rhat \! \xi^{n+1}\big[\chi_{rn}(\xi)-(n+1)\chi_{\theta n}(\xi)\big]\,\dm \xi
 -\frac{c_{3n}}{n}\bigg\}
 \nn\\
 & +\rhat^n \bigg\{\int_1^\rhat \xi^{-n} [\chi_{rn}(\xi)+n\chi_{\theta n}(\xi)] \,\dm\xi -\frac{c_{4n}}{n+1}\bigg\},
 \\[2mm]
 \eqlab{Qn_done}
  &Q_n(\rhat) =
 \nn\\
 &\rhat^{-(n+1)}\bigg\{n \int_1^\rhat \xi^{n+1}[\chi_{rn}(\xi)
 -(n+1)\chi_{\theta n}(\xi)]\,\dm \xi - c_{3n}\bigg\}
 \nn\\
 &+\rhat^n \bigg\{(n+1)\int_1^\rhat \xi^{-n}
 [\chi_{rn}(\xi)+n\chi_{\theta n}(\xi)] \,\dm\xi -c_{4n}\bigg\}.
 \esubal
Substituting unprimed fields and parameters outside the particle, by primed fields and parameters, the same equations are obeyed inside the particle for $q'_n$ and $Q'_n$. The integration constants $c_{3n}$, $c_{4n}$, $c'_{3n}$, and $c'_{4n}$ are determined later by the boundary conditions at $\rhat \rightarrow 0$, $\rhat = 1$, and $\rhat \rightarrow \infty$.

Substituting \eqsref{v2_Helmholtz}{rhs_v2_Helmholtz} into \eqnoref{2nd_order_fluid_governing} yields
 \bsubal{2nd_order_potential_governing}
 \eqlab{Phi2}
 \nabla^2 \phiIIsc &= -\frac{1}{\rho_0}\div\avr{\rho_1\vvv_1}_\mr{nii}, \\
 \eqlab{Psi2}
 \nabla^2 \psibfIIsc &= \qqq,
 \\
 \eqlab{p2}
 \frac{1}{\rho_0}\avr{p_2^\mr{sc}} &= -\nuO Q
 - (\nuOB+\tfrac{4}{3}\nuO)\:\frac{1}{\rho_0}\div\avr{\rho_1\vvv_1}_\mr{nii}.
 \esubal
To determine $\phiIIsc$ and $\psibfIIsc$, we insert the partial-wave expansions
 \bsubal{rhs_phi2_expansion}
 \eqlab{Phi2_ansatz}
 \phiIIsc &= \sum_{n=0}^\infty \phiIInsc(\rhat) P_n(\cos\theta),
 \\
 \eqlab{Psi2_ansatz}
 \psibfIIsc &= \eee_\varphi \sum_{n=1}^\infty \psiIInsc(\rhat) \pp_\theta P_n(\cos\theta),
 \\
 \eqlab{rhs_phi2_expanded}
 \frac{1}{\rho_0}\div\avr{\rho_1\vvv_1}_\mr{nii}
 &= \sum_{n=0}^\infty \frac{2n+1}{a^2}\mu_n(\rhat) P_n(\cos\theta),
 \\
 \eqlab{munr}
 \mu_n(\rhat) &= \frac{a^2}{2\rho_0} \int_0^\pi \div\avr{\rho_1\vvv_1}_\mr{nii} P_n(\cos\theta)\sin\theta \,\dm \theta,
 \esubal
into \eqref{2nd_order_potential_governing}, and obtain
 \bsubal{Phi2n_Psi2n_eq}
 \eqlab{Phi2n_eq}
 \bigg[\pp_\rhat^2 +\frac{2}{\rhat}\pp_\rhat -\frac{n(n+1)}{\rhat^2}\bigg]
 \phiIInsc(\rhat) &= -(2n+1)\mu_n(\rhat),
 \\
 \eqlab{Psi2n_eq}
 \bigg[\pp_\rhat^2 +\frac{2}{\rhat}\pp_\rhat -\frac{n(n+1)}{\rhat^2}\bigg]
 \psiIInsc(\rhat) &= q_n(\rhat).
 \esubal
The left-hand sides contain the same differential operator as in \eqref{qn_only}, so using the same Greens function $G_n$~\eqnoref{Greens_eq} as before, we obtain the solution for $\phiIInsc$ and $\psiIInsc$,
 \bsubal{Phi2scPsi2sc}
 \eqlab{Phi2n_general}
 & \phiIInsc(\rhat) = \rhat^{-(n+1)}\bigg(\int_1^\rhat \xi^{n+2}\mu_n(\xi)\,\dm\xi -c_{1n}\bigg)
 \nn\\&
 \hspace*{15mm}
 -\rhat^{n}\bigg(\int_1^\rhat \xi^{-(n-1)}\mu_n(\xi)\,\dm\xi -c_{2n}\bigg),
 \\
 \eqlab{Psin_done}
 &\psiIInsc(\rhat)=
 \nn\\
 &-\frac{\rhat^{-(n+1)}}{2(2n+3)}\bigg\{\int_1^\rhat
 \!\xi^{n+3}[\chi_{rn}(\xi)-(n\!+\!3)\chi_{\theta n}(\xi)] \,\dm \xi -c_{6n}  \bigg\}
 \nn \\
 &-\frac{\rhat^n}{2(2n-1)}\bigg\{\int_1^\rhat
 \!\xi^{-n+2}[\chi_{rn}(\xi)+(n\!-\!2)\chi_{\theta n}(\xi)]\,\dm \xi\!-\!c_{5n}\! \bigg\}
 \nn \\
 &+\frac{\rhat^{-(n-1)}}{2(2n-1)}\bigg\{\int_1^\rhat \xi^{n+1}[\chi_{rn}(\xi)-(n+1)\chi_{\theta n}]\,\dm\xi -\frac{c_{3n}}{n} \bigg\}
 \nn\\
 &+\frac{\rhat^{n+2}}{2(2n+3)} \bigg\{\int_1^\rhat \xi^{-n}[\chi_{rn}(\xi)+n\chi_{\theta n}(\xi)]\,\dm \xi -\frac{c_{4n}}{n+1} \bigg\}.
\esubal
The same equations are obeyed by the transmitted potentials $\phiIInp$ and $\psiIInp$  with primed integration constants $c'_{in}$. We determine the constants  $\big\{c'_{1n}, c'_{3n}, c'_{6n}\big\}$ and $\big\{c_{2n}, c_{4n}, c_{5n}\big\}$ by insisting that {$\avr{\vvvIIp}$ does not diverge for $\rhat \rightarrow 0$, and that $\avr{\vvvIIsc}\rightarrow 0$ for $\rhat \rightarrow \infty$,
 \bsuba{cin_cinp}
 \bal
 \eqlab{c1np}
 c_{1n}' & = -\int_0^1 \xi^{n+2}\mu_n'(\xi)\,\dm \xi,
 \\
 \eqlab{c2n}
 c_{2n} &= \int_1^\infty \xi^{-(n-1)}\mu_n(\xi)\,\dm \xi,
 \\
 \eqlab{c3nm}
 c_{3n}'&= -n\int_0^1 \xi^{n+1}[\chi'_{rn}(\xi)-(n+1)\chi'_{\theta n}(\xi)]\,\dm\xi,
 \\
 \eqlab{c4n}
 c_{4n} &= (n+1)\int_1^\infty \xi^{-n}[\chi_{rn}(\xi)+n\chi_{\theta n}(\xi)] \,\dm \xi, \\
 \eqlab{c5n}
 c_{5n}&=\int_1^\infty \xi^{-(n-2)}[\chi_{rn}(\xi)+(n-2)\chi_{\theta n}(\xi)] \,\dm \xi, \\
 \eqlab{c6nm}
 c_{6n}'&=-\int_0^1 \xi^{n+3}[\chi'_{rn}(\xi)-(n+3)\chi'_{\theta n}(\xi)] \,\dm \xi, \\
 \eqlab{c5nm_and_c6n}
 c_{5n}'&=c_{6n}=0.
 \eal
 \esuba
The two constants $c'_{5n}$ and $c_{6n}$ are set to zero because $c_{5n}'$ leads to the same kind of terms in $\avr{\vvv_2'}$ as those containing $c_{2n}'$, and similarly for $c_{6n}$ with respect to $c_{1n}$ in $\avr{\vvvIIsc}$. The remaining four coefficients $c_{1n}$, $c'_{2n}$, $c_{3n}$, and $c'_{4n}$ are discussed in \secsref{BC_fluid_solid}{BC_fluid_fluid} when applying the second-order boundary conditions at $\rhat = 1$.

\subsection{Fluid-solid boundary conditions}
\seclab{BC_fluid_solid}
For the case of a solid particle, we have no second-order time-averaged motion inside the sphere. Therefore, we only need to determine the two last unprimed constants $c_{1n}$ and $c_{3n}$. This is done through the no-slip boundary condition to second order at $\rhat=1$, including the Stokes-drift terms \cite{Bach2018},
 \beq{Second_order_no_slip_sl}
 \avr{\vvvIIsc}= -\avr{\vvvIIin}-\avr{(\sss_1\cdot \nablabf)\vvv_1},
 \quad \sss_1 = \frac{\ii}{\omega}\vvv_1.
 \eeq
This boundary condition is different from the one used by Doinikov, who replaced $\avr{(\sss_1\cdot \nablabf)\vvv_1}$ by $\avr{s_{1r}\pp_r\vvv_1}$ \cite{Doinikov1994a,Doinikov1994}. While this is equivalent for the  case of a rigid sphere, it is generally not true for the case of a compressible sphere. We make another expansion on the right-hand side of \eqref{Second_order_no_slip_sl}, where all terms are evaluated at $\rhat = 1$,
 \bsuba{Second_order_no_slip_sl_expansion}
 \bal
 \eqlab{expansion_Second_order_no_slip_sl}
 \big\langle\vvvIIin&\big\rangle + \avr{(\sss_1\cdot \nablabf)\vvv_1}
 \nn\\
 &=\eee_r \frac{1}{a}\sum_{n=0}^\infty (2n+1)\ansl P_n(\cos\theta)
 \nn\\
 &\quad -\eee_\theta \frac{1}{a}\sum_{n=1}^\infty (2n+1)\bnsl \pp_\theta P_n(\cos\theta),
 \\
 \eqlab{an_sl}
 \ansl &= \frac{a}{2}\int_0^\pi\! \eee_r \!\cdot\! \Big[\avr{\vvvIIin}
 +\avr{(\sss_1\cdot \nablabf)\vvv_1}\Big] P_n(\cos\theta)\sin\theta \,\dm\theta,
 \eal
 \bal
 \eqlab{bn_sl}
 \bnsl &= -\frac{a}{2n(n+1)}
 \int_0^\pi \! \eee_\theta \!\cdot\! \Big[\avr{\vvvIIin}\nn
 \\
 &\qquad+\avr{(\sss_1\cdot \nablabf)\vvv_1}\Big]\pp_\theta P_n(\cos\theta)\sin\theta \,\dm\theta.
 \eal
 \esuba
Using that
 \bal\eqlab{vvv2sc_explicit}
 &\avr{\vvvIIsc}=
 \eee_r \sum_{n=0}^\infty \frac{1}{a}\bigg[\pp_\rhat \phiIInsc(\rhat)-\frac{n(n+1)}{\rhat}\psiIInsc(\rhat) \bigg] P_n(\cos\theta)
 \nn\\
 & +\eee_\theta \sum_{n=1}^\infty \frac{1}{a}\bigg[\frac{1}{\rhat} \phiIInsc(\rhat)\!-\!\frac{1}{\rhat}\psiIInsc(\rhat) \!-\! \pp_\rhat \psiIInsc(\rhat) \bigg] \pp_\theta P_n(\cos\theta),
 \eal
and inserting the explicit forms~\eqnoref{Phi2scPsi2sc} for $\phiIInsc$ and $\psiIInsc$ at $\rhat=1$, we find two equations for the remaining constants,
 \bsubal{c1n_and_c3n_eqs}
 \eqlab{c1n_and_c3n_eq_1}
 &n(n+1)\bigg[\frac{c_{3n}}{2n(2n-1)} + \frac{c_{4n}}{2(n+1)(2n+3)} -\frac{c_{5n}}{2(2n-1)}\bigg]
 \nn\\
 &\qquad + (n+1) c_{1n} + nc_{2n} = -(2n+1)\ansl ,
 \\
 \eqlab{c1n_and_c3n_eq_2}
 &-c_{1n}+c_{2n} + \frac{(2-n)c_{3n}}{2n(2n-1)} + \frac{(n+3)c_{4n}}{2(n+1)(2n+3)}
 \nn\\
 &\qquad -\frac{(n+1)c_{5n}}{2(2n-1)} = (2n+1)\bnsl.
 \esubal
As we show in \secref{Evaluating_Frad}, only $c_{31}$ is needed to evaluate the acoustic radiation force $\FFFrad$, so we refrain from determining $c_{3n}$ for $n \neq 1$ and $c_{1n}$. From \eqnoref{c1n_and_c3n_eqs}, we obtain
\bal\eqlab{c31_sl}
c_{31}= -\frac{3}{2}a_1^\mr{sl} + 3b_1^\mr{sl} -\frac{3}{2}c_{21}-\frac{1}{4}c_{41}+\frac{3}{2}c_{51}.
\eal

\subsection{Fluid-fluid boundary conditions}
\seclab{BC_fluid_fluid}
For a fluid particle, the second-order time-averaged equations of motion are solved both inside and outside the particle. For the boundary conditions, we follow the method used in Ref.~\cite{Baasch2020}. In this case, we assume that the tangential motion and the tangential stress over the momentary boundary of the particle are continuous. Meanwhile, the structural integrity of the particle-fluid interface is assumed to be upheld by surface tension, such that no time-averaged flow can happen across the momentary boundary, which has no perpendicular motion to second order. The boundary conditions at $\rhat=1$ for a fluid particle can then be formulated as,
 \bsubal{Second_order_no_slip_fl}
 \eqlab{v2r_no_slip_fl}
 &\avr{\vvvIIrsc} = -\avr{\vvvIIrin}-\avr{(\sss_1\cdot \nablabf)\vvv_1} \cdot \eee_r ,
 \\
 \eqlab{v2rm_no_slip_fl}
 &\avr{\vvvIIrm} = -\avr{(\sss_1\cdot \nablabf)\vvv_1'} \cdot \eee_r ,
 \\
 \eqlab{tangential_velocity_continuous}
 &\avr{\vvvIItsc}-\avr{\vvvIItm} =
 \nn\\
 & \qquad -\avr{\vvvIItin}-\avr{(\sss_1\cdot \nablabf)(\vvv_1-\vvv_1')} \cdot \eee_\theta ,
 \\
 \eqlab{tangential_stress_continuous}
 &\avr{\sigma_{2\theta r}^\mr{sc}}-\avr{\sigma_{2\theta r}'} =
 \nn\\
 &\qquad
 -\avr{\sigma_{2\theta r}^\mr{in}} -
 \avr{(\sss_1\cdot \nablabf)\big[(\sigmabf_{1}-\sigmabf_{1}')\cdot \eee_r\big]}\cdot \eee_\theta.
 \esubal
The time-averaged stress component $\avr{\sigma_{2\theta r}}$ for the full wave outside the particle is given by
 \bal
 \eqlab{sigma2rth}
 \avr{\sigma_{2\theta r}} &= \frac{\etaO}{a}
 \bigg[\frac{1}{\rhat}\pp_\theta \avr{v_{2r}} - \frac{1}{\rhat} \avr{v_{2\theta}}
 + \pp_\rhat \avr{v_{2\theta}}\bigg]
 \nn \\
 & \quad +\frac{1}{a}\bigg\langle  \etaI \bigg[\frac{1}{\rhat}\pp_\theta v_{1r}
 - \frac{1}{\rhat} v_{1\theta} + \pp_\rhat v_{1\theta}\bigg] \bigg\rangle.
 \eal
The same expression is obeyed by the stress component $\avr{\sigma_{2\theta r}^\prime}$ with primed fields for the transmitted wave inside the particle. Similar to \secref{BC_fluid_solid}, we expand the second-order incident fields and the products of first-order fields into partial-wave expansions and insert into \eqsref{Second_order_no_slip_fl}{sigma2rth}  the explicit expressions \eqnoref{vvv2sc_explicit} for the second-order scattered and primed fields. At $\rhat = 1$ we then obtain the following four expansion coefficients,
 \bsuba{ananpbndnfl}
 \bal
 \eqlab{anfl}
 \anfl &=  \frac{a}{2}\int_0^\pi \eee_r \scap \Big[\avr{\vvvIIin}
 + \avr{(\sss_1\cdot \nablabf)\vvv_1}\Big]P_n(\cos\theta)\sin\theta \,\dm\theta,
 \\
 \eqlab{anpfl}
 \anflp &= \frac{a}{2}\int_0^\pi \eee_r \scap \avr{(\sss_1\cdot \nablabf)\vvv_1'} P_n(\cos\theta)\sin\theta \,\dm\theta ,
 \\
 \eqlab{bnfl}
 \bnfl &= \frac{-a}{2n(n+1)}\int_0^\pi \eee_\theta \scap \Big[\avr{\vvvIIin}
 +\avr{(\sss_1\cdot \nablabf)(\vvv_1-\vvv_1')}\Big]
 \nn\\
 &\hspace*{30mm}\times
 \pp_\theta P_n(\cos\theta)\sin\theta \,\dm\theta ,
 \\
 \eqlab{dnfl}
 \dnfl &= \frac{-a}{2n(n+1)\eta_0}\int_0^\pi \bigg\{\eta_0 \Big[\pp_\rhat \avr{\vvvIItin}+
 \pp_\theta\avr{\vvvIIrin} -\avr{\vvvIItin}\Big]
 \nn\\
 &\quad +a\avr{(\sss_1\cdot \nablabf)\big[(\sigmabf_{1}-\sigmabf_{1}')\cdot \eee_r\big]}\cdot \eee_\theta
 \nn\\
 &\quad +\avr{\eta_1(\pp_\theta \vIr+\pp_\rhat\vIth-\vIth)}
 \nn\\
 &\quad -\avr{\eta_1'(\pp_\theta \vIr'+\pp_\rhat \vIth'-\vIth')} \bigg\}
 \;\pp_\theta P_n(\cos\theta)\sin\theta \,\dm\theta.
 \eal
 \esuba
The resulting four equation for the determination of the unknown constants $c_{1n}$, $c_{2n}'$, $c_{3n}$, and $c_{4n}'$, become
 \bsuba{fl_constants}
 \bal
 \eqlab{fl_BC_1}
 &-(2n+1)\anfl = (n+1)c_{1n}+ nc_{2n} +n(n+1)
 \nn\\
 &\quad \times\bigg(\frac{c_{3n}}{2n(2n-1)}+\frac{c_{4n}}{2(n+1)(2n+3)}-\frac{c_{5n}}{2(2n-1)}\bigg),
 \\
 \eqlab{fl_BC_2}
 &-(2n+1)\anflp =
 (n+1)c_{1n}'+nc_{2n}' + n(n+1)
 \nn\\
 &\quad \times \bigg(\frac{c_{3n}'}{2n(2n-1)}
 +\frac{c_{4n}'}{2(n+1)(2n+3)}-\frac{c_{6n}'}{2(2n+3)}\bigg) ,
 \\
 \eqlab{fl_BC_3}
 &(2n+1)\bnfl =  -(c_{1n}-c_{1n}')+(c_{2n}-c_{2n}')
 \nn\\
 &\quad -\frac{(n-2)(c_{3n}-c_{3n}')}{2n(2n-1)} +\frac{(n+3)(c_{4n}-c_{4n}')}{2(n+1)(2n+3)}
 \nn\\
 &\quad -\frac{(n+1)c_{5n}}{2(2n-1)}-\frac{n c_{6n}'}{2(2n+3)},
 \eal
 \bal
 \eqlab{fl_BC_4}
 &(2n+1)\dnfl = 2(n\!+\!2)(c_{1n}\!-\!\etaOTi c_{1n}')+2(n\!-\!1)(c_{2n}\!-\!\etaOTi c_{2n}')
 \nn\\
 &\quad +\frac{n^2-1}{n(2n-1)}(c_{3n}-\etaOTi c_{3n}')+\frac{n(n+2)(c_{4n}-\etaOTi c_{4n}')}{(n+1)(2n+3)}
 \nn\\
 &\quad -\frac{(n^2-1)c_{5n}}{2n-1}+\frac{\etaOTi n(n+2)c_{6}'}{2n+3}.
 \eal
 \esuba
As for the solid particle in \secref{BC_fluid_solid}, we only need $c_{31}$ to evaluate the acoustic radiation force on a fluid particle,
 \bal\eqlab{c31_fl}
 c_{31} & =
 -\frac{3}{2} a_1^\mr{fl} +3 b_1^\mr{fl}  -\frac{3}{2} c_{21}
 -\frac{1}{4} c_{41}+\frac{3}{2} c_{51}
 \nn \\
 &\quad +\frac{1}{1+\etaOTi}\bigg[\bigg(-3a_1^\mr{fl\prime}-5c_{11}'-\frac{1}{2}c_{31}'
 + \frac{1}{2}c_{61}'\bigg)\etaOTi
 \nn\\
 &\quad -\frac{3}{2}a_1^\mr{fl} - 3b_1^\mr{fl} - d_1^\mr{fl}
 +\frac{1}{2}c_{21}+\frac{1}{4}c_{41}-\frac{1}{2}c_{51}\bigg].
\eal

\section{The acoustic radiation force}
\seclab{Evaluating_Frad}
With the axisymmetric solutions of the first-order acoustic fields in \secref{FirstOrder} and the second-order steady fields in \secref{SecondOrder} at hand, we can now evaluate the radiation force $\FFFrad$ from \eqref{Frad}. We note that only terms containing scattered fields contribute to $\FFFrad$, such that
 \bal\eqlab{Frad_sc}
 &\FFFrad = \oint_{\surfP}
 \Big[\avr{\sigmabf^\mr{sc}_2}-\rho_0\avr{\vvvI\vvvI}_\mr{nii}\Big] \cdot \nnn \,\dm S
 \nn \\
 & = \rhoO \oint_{\surfP} \dm S\: \bigg\{
 \nuO \Big[\pp_r \avr{v_{2\theta}^\mr{sc}}-\frac{1}{r}\avr{v_{2\theta}^\mr{sc}}
 + \frac{1}{r}\pp_\theta \avr{v_{2r}^\mr{sc}}\Big]\: \eee_\theta
 \nn \\
 &\quad + \Big[2\nuO \pp_r \avr{v_{2r}^\mr{sc}}
 +\big(\nuOB\!-\!\tfrac{2}{3}\nuO\big)\div \avr{\vvvIIsc}
 - \frac{1}{\rhoO}\avr{p_2^\mr{sc}}\Big]\:\eee_r
 \nn\\
 &\quad -\Big\langle \vvv_1 \vvv_1
 - \nuI \Big[\!\nablabf \vvv_1 \!+\! (\nablabf \vvv_1)^\textsf{T}\!\Big]
 \nn\\
 &\hspace*{20mm} -\Big[\nuIB \!-\! \tfrac{2}{3}\nuI\Big](\nablabf\scap\vvv_1) \Imat \Big\rangle_\mr{nii}
 \cdot\: \eee_r \bigg\}.
 \eal
Using $\dm S = a^2\sin\theta\:\dm\theta\:\dm\varphi$ and writing the spherical unit vectors in terms of their cartesian components, only a contribution along the direction of wave propagation $\eee_z$ will remain after integrating over the azimuthal angle $\varphi$. In \eqref{Frad_sc} the terms containing $\avr{\vvvIIsc}$ and $\avr{p^\mr{sc}_2}$ can be substantially simplified as follows. We first express $\div \avr{\vvvIIsc} = \Lapl \phiIIsc$ and $\avr{p^\mr{sc}_2}$ by the right-hand sides of \eqref{Phi2} and  \eqref{p2}, respectively. Then we insert the expressions \eqssnoref{vvv2sc_explicit}{Q_expansion}{rhs_phi2_expanded} for $\avr{\vvvIIsc}$, $Q$, and $\frac{1}{\rhoO}\avr{\rhoI\vvvI}_\mr{nii}$, respectively. Using the Legendre orthogonal relations~\eqref{legendre_23}, the $\theta$ integrals are carried out, leaving only partial-wave contributions with $n=1$. A further simplification is obtained by grouping the terms with $\phi^\mr{sc}_{2,1}$ and $\psi^\mr{sc}_{2,1}$, respectively, and noting that the left-hand-sides of \eqref{Phi2n_Psi2n_eq} appear with $\rhat = 1$, and thus they can be substituted by the much simpler right-hand sides. In the final expression, only $Q_1(1)$ and $q_1(1)$ appear, with values given by \eqref{Qn_and_qn_done} for $n=1$ and $\rhat=1$. The resulting simplified expression for $\FFFrad$ \eqref{Frad_sc} becomes,
 \bal\eqlab{Frad_simplified}
 &\FFFrad = -4\pi \rhoO \nuO c_{31}\:\eee_z
 -\rhoO \oint_{\partial\Omega_0} \dm S\: \Big\langle\vvv_1 \vvv_1
 \nn\\
 &
 -\! \nuI \Big[ \nablabf \vvv_1 \!+\! (\nablabf \vvv_1)^\textsf{T}\Big]
 \!-\!\Big[\nuIB \!-\! \tfrac{2}{3}\nuI\Big] (\div \vvv_1) \Imat \Big\rangle_\mr{nii}
 \!\!\cdot \eee_r.
 \eal
Inspection of the constant $c_{31}$, which is given by \eqsref{c31_sl}{c31_fl} for a solid and fluid particle, respectively, reveals two types of contributions: terms containing the steady $\avr{\vvvIIin}$ and terms containing time-averaged products of the first-order acoustic fields. Consequently, it is convenient to split $\FFFrad$ into two contributions: $\FFFradII$ containing the time-averaged first-order products, and $\FFFradin$ containing the streaming velocity $\avr{\vvvIIin}$ of the incident wave,
 \beq{Frad_form}
 \FFFrad = \FFFradII + \FFFradin.
 \eeq
The final step in the calculation is the evaluation of $\FFFradII$ and $\FFFradin$. This is a particularly tedious part of the calculation that is best treated for solid and fluid particles separately, in the ensuing two subsections.

\subsection{Solid particle}
\seclab{Ssolid}
For a solid particle, we evaluate expression~\eqnoref{Frad_simplified} for $\FFFrad$  by inserting $c_{31}$ from \eqref{c31_sl}, which can be expressed explicitly in terms of the time-averaged products of first-order acoustic fields known from \secref{FirstOrderPartialWave}, and terms including the known field $\avr{\vvvIIin}$ discussed after \eqref{v2p2_tot}. To do this, we insert the explicit expressions~\eqsnoref{an_sl}{bn_sl} for $a_1^\mr{sl}$ and $b_1^\mr{sl}$, as well as the relations~\eqnoref{cin_cinp} for $c_{21}$, $c_{41}$, and $c_{51}$ with $\mu_1(\xi)$, $\chi_{r1}(\xi)$, and $\chi_{\theta 1}(\xi)$ from \eqssref{munr}{chir}{chit}, respectively. We notice that $\avr{\vvvIIin}$ appears only in some terms of $a_1^\mr{sl}$ and $b_1^\mr{sl}$, and these are thus grouped and manipulated to obtain $\FFFradin$ for a solid particle as,
 \beq{F2_sl}
 \FFFradin = \eee_z\:\frac{3\eta_0}{2a}\oint_{\partial \Omega_0}
 \dm S\: \avr{v_{2z}^\mr{in}}.
 \eeq
This corresponds to the Stokes drag on the particle generated by the steady streaming of the incident wave.

Next, we compute $\FFFradII$, which contains the remainder of the terms, all involving time-averaged products of first-order acoustic fields. It is convenient to rewrite the two terms arising from $c_{41}$ and $c_{51}$, \eqsref{c4n}{c5n} with $n=1$, as detailed in \appref{c41c51},
\begin{widetext}
 \bal\eqlab{c41_and_c51_sl}
 &-\eee_z 4\pi\rhoO\nuO \bigg(\frac{3}{2}c_{51}-\frac{1}{4}c_{41}\bigg)
 \nn\\
 & = \rhoO \oint_{\partial\Omega_0} \dm S\: \Big\langle \vvv_1 \vvv_1
 - \nuI\Big[\nablabf \vvv_1+(\nablabf \vvv_1)^\textsf{T}\Big]
 -[\nuIB-\tfrac{2}{3}\nuI](\div \vvv_1) \Imat \Big\rangle_\mr{nii} \! \cdot \eee_r
 \nn\\
 &\quad -\eee_z 3\pi \rhoO a^2 \int_1^\infty \dm \xi
 \int_0^\pi \dm \theta \: \sin\theta \:
 \big[1-\xi^{-2}\big]\; \eee_r \cdot \Big\langle \vvv_1 \vvv_1
 -\nuI\Big[\nablabf \vvv_1+(\nablabf \vvv_1)^\textsf{T}\Big]
 -[\nuIB-\tfrac{2}{3}\nuI](\div \vvv_1) \Imat \Big\rangle_\mr{nii}\! \cdot \eee_z
 \nn\\
 &\quad -\eee_z 3\pi\rhoO a^3 \int_1^\infty \dm \xi
 \int_0^\pi \dm \theta\: \sin^2\theta
 \frac{1}{2} \big[\xi \!-\! \xi^{-1}\big]  \Big\{ \div \Big\langle \vvv_1 \vvv_1
 -\nuI \Big[\nablabf \vvv_1 \!+\! (\nablabf \vvv_1)^\textsf{T}\Big]
 -[\nuIB \!-\! \tfrac{2}{3}\nuI](\div \vvv_1 ) \Imat \Big\rangle_\mr{nii}\Big\} \cdot \eee_\theta.
 \eal
We note that the term with the surface integral "$\oint_{\partial\Omega_0}$" also appears in \eqref{Frad_simplified}, but with opposite sign, and it thus cancels out. After similarly manipulating  $a_1^\mr{sl}$, $b_1^\mr{sl}$, and $c_{21}$, $\FFFradII$ becomes,
 \bal\eqlab{F1_sl}
 \FFFradII &=  -\eee_z 3\pi\rhoO \bigg\{
 -\frac{a^2 \nu_0}{\rhoO} \int_1^\infty \dm\xi
 \int_0^\pi \dm\theta\: \cos\theta\sin\theta
 \div \frac{1}{2}\re[\rhoI \vvv_1^*]_\mr{nii}
 \\
 &+ a^2 \int_1^\infty \dm\xi
 \int_0^\pi \dm\theta \: \sin\theta
 \big[1-\xi^{-2}\big] \frac{1}{2}\re\bigg[\eee_r\cdot \Big(
 \vvv_1 \vvv_1^* -\nuI\Big[\nablabf \vvv_1+(\nablabf \vvv_1)^\textsf{T}\Big]^*
 -\Big[\nuIB-\tfrac{2}{3}\nuI\Big](\div \vvv_1^*) \Imat \Big)  \cdot \eee_z
 \nn\\
 &\hspace*{20mm} +\frac{1}{2}a\xi \Big\{ \div \Big( \vvv_1 \vvv_1^*
 -\nuI\Big[\nablabf \vvv_1+(\nablabf \vvv_1)^\textsf{T}\Big]^*
 -\Big[\nuIB-\tfrac{2}{3}\nuI\Big](\div \vvv_1^*) \Imat \Big)\Big\} \cdot \eee_\theta \sin\theta
 \bigg]_\mr{nii}
 \nn\\
 \nn
  &+\int_0^\pi  \!\dm\theta\:\frac{\sin\theta}{2}\re\bigg[\frac{a^3}{\xs^2}\Big(\vIr\pp_r \vIr^*+\frac{1}{r}\vIth \pp_\theta \vIr^* -\frac{1}{r}\vIth\vIth^*\Big)\cos\theta
 -\frac{a^3}{\xs^2}\Big(\vIr\pp_r \vIth^*+\frac{1}{r}\vIth\pp_\theta \vIth^* -\frac{1}{r}\vIr\vIth^* \Big)\sin\theta\bigg]_{\rhat=1}
 \bigg\}.
 \eal
\end{widetext}
At this point, we need explicit relations for $\rhoI$, $\nuI$, and $\nuIB$. Combining \eqssref{1st_order_continuity}{Helmholtz_decomp_v1}{phic_phit_eqs} leads to $\rhoI$,
 \bsuba{rho1nu1nuB1}
 \bal\eqlab{rho1_wavenumb}
 \rho_1 &=  -\frac{\rhoO}{\nuO \xs^2}\big(\xc^2 \phiIc + \xt^2 \phiIt\big).
 \eal
This is used alongside $T_1$ from \eqref{T1_potential_form} in the first-order expansions for $\eta_1$ and $\etaB_1$ (see \eqref{pertubation_params_1}) to reach,
 \bal
 \eqlab{eta1_pot}
 \nuI &= \frac{\xc^2}{\xs^2} \Bc \phiIc+ \frac{\xt^2}{\xs^2} \Bt \phiIt,
 \\
 \eqlab{etab1_pot}
 \nuIB &= \frac{\xc^2}{\xs^2} \Bbc \phiIc+ \frac{\xt^2}{\xs^2} \Bbt \phiIt.
 \eal
 \esuba
Here, we have introduced $\Bc$, $\Bt$, $\Bbc$, and $\Bbt$ that to lowest order in $\xO$ become,
 \bsubal{B_coeff}
 \eqlab{Bc}
 \Bc &=\bigg[\frac{1-\gamO}{\alphapO \etaO} \bigg(\frac{\pp \eta}{\pp T}\bigg)_{T_0}- \frac{\rhoO}{\etaO}\bigg(\frac{\pp \eta}{\pp  \rho}\bigg)_{\rhoO}\bigg] ,
 \\
 \eqlab{Bt}
 \Bt&=\bigg[\frac{1}{\alphapO \etaO} \bigg(\frac{\pp \eta}{\pp T}\bigg)_{T_0}- \frac{\rhoO}{\etaO}\bigg(\frac{\pp \eta}{\pp  \rho}\bigg)_{\rhoO}\bigg] ,
 \\
 \eqlab{Bbc}
 \Bbc&=\bigg[\frac{1-\gamO}{\alphapO \etaO} \bigg(\frac{\pp \etaB}{\pp T}\bigg)_{T_0}- \frac{\rhoO}{\etaO}\bigg(\frac{\pp \etaB}{\pp  \rho}\bigg)_{\rhoO}\bigg] ,
 \\
 \eqlab{Bbt}
 \Bbt & =\bigg[\frac{1}{\alphapO \etaO} \bigg(\frac{\pp \etaB}{\pp T}\bigg)_{T_0}-  \frac{\rhoO}{\etaO}\bigg(\frac{\pp \etaB}{\pp \rho}\bigg)_{\rhoO}\bigg] .
 \esubal
Now, we insert \eqref{rho1nu1nuB1} into \eqnoref{F1_sl} with $\phiIc$, $\phiIt$, and $\vvvI$ from \eqsref{potential_solutions}{v1_solutions} to reach an explicit relation for $\FFFradII$ that can be evaluated analytically. The result is a double sum over mode indices $n$ and $m$, where all terms contain quadratic combinations of scattering coefficients $\ain$, but only the unprimed ones
   $\big\{\ain\}_{i=0, 1, 2, 3} = \big\{1, \acnsc, \atnsc, \asnsc \big\}$. After evaluating the integrals over $\theta$, see \eqref{legendre_23}, and switching dummy indices (e.g.\ $[n-1] \rightarrow n$) as well as complex conjugation (e.g.\ $\re\big[v_{1r}^{n*}v_{1r}^{n+1}\big] = \re\big[v_{1r}^{n}v_{1r}^{n+1*}\big]$), one obtains a single sum where all terms contain $A_n A^*_{n+1} \ain \aknI^*$, as detailed in the Supplemental Material~\cite{Note1}. The resulting form of $\FFFradII$ is,
 \bsubal{F1_form_sl}
 \eqlab{F1_expr_sl}
 \FFFradII &= -\eee_z 3\pi \rho_0 \sum_{n=0}^\infty \frac{n+1}{(2n+1)(2n+3)}\re[A_nA_{n+1}^*D_n],
 \\
 \eqlab{Dn_expr_sl}
 D_n &= \sum_{i,k=0}^3 S_{ik,n} \ain \aknI^*,
 \esubal
where we have introduced the force coefficient $D_n$ and the 16 second-order coefficients $S_{ik,n}$ for $i,k = 0,1,2,3$ for each mode $n$. The explicit expressions for $S_{ik,n}$ are determined by grouping all terms with specific quadratic combinations $\ain \aknI^*$ of the first-order scattering coefficients $\ain$. We have made extensive use of the algebraic tool Maple \cite{Maple} to derive $S_{ik,n}$ analytically, as listed in Section~S3~A of the Supplemental Material~\cite{Note1}. This ends the analysis of $\FFFrad$ for a solid particle, and we now proceed to the case of a fluid particle.

\subsection{Fluid particle}
\seclab{Sfluid}

\vspace*{-3mm}

For a fluid particle, we use the same computational method as for a solid particle. We insert expression~\eqnoref{c31_fl} for $c_{31}$ into \eqref{Frad_simplified} for $\FFFrad$. We subsequently substitute the explicit expressions~\eqnoref{ananpbndnfl} for $\anfl$, $\anflp$, $\bnfl$ and $\dnfl$, as well as the relations~\eqnoref{cin_cinp} for $c'_{11}$, $c_{21}$, $c'_{31}$, $c_{41}$, $c_{51}$, and $c'_{61}$ with $\mu_1(\xi)$, $\chi_{r1}(\xi)$, and $\chi_{\theta 1}(\xi)$ from \eqssref{munr}{chir}{chit}, respectively, and with $\mu'_1(\xi)$, $\chi'_{r1}(\xi)$, and $\chi'_{\theta 1}(\xi)$ also from from \eqssref{munr}{chir}{chit} but with all unprimed fields and parameters substituted by primed quantities. Each term is then evaluated separately, and their contributions to the force are summed up. The resulting expression for $\FFFradII$ for a fluid particle has the same form as for a solid particle, however, now with 49 second-order coefficients $S_{ik,n}$ with $i,k = 0,1,2,\ldots,7$ for each mode $n$, because in contrast to a solid particle, the three transmitted scattering coefficients $\acnp$, $\atnp$, and $\asnp$ are also included,
 \bsubal{F1_form_fl}
 \eqlab{F1_expr_fl}
 \FFFradII &= -3\pi \rho_0 \sum_{n=0}^\infty \frac{n+1}{(2n+1)(2n+3)}\re[A_nA_{n+1}^*D_n]\:\eee_z,
 \\
 \eqlab{Dn_expr_fl}
 D_n &= \sum_{i,k=0}^6 S_{ik,n} \ain \aknI^* .
 \esubal
General expressions for the 49 second-order coefficients $S_{ik,n}$ for a fluid are given in Section~S3~B of the Supplemental Material~\cite{Note1}.

The contribution to $\FFFrad$ from the second-order incident field $\avr{\vvvIIin}$ in \eqref{c31_fl}, is found in analogy with the solid-particle case \eqref{F2_sl},
 \bal\eqlab{F2_fl}
 \FFFradin &= \eee_z\bigg\{
 \frac{3\eta_0}{2a}\oint_{\partial \Omega_0}\dm S\:\avr{v_{2z}^\mr{in}}
 + \frac{\pi\eta_0 a}{1+\etaOTi}
  \int_0^\pi  \dm\theta \: \sin\theta
 \nn\\
 &
 \times \bigg[
 3 \Big(\avr{\vvvIIrin}\cos\theta  +\avr{\vvvIItin} \sin\theta\Big)
 \nn\\
 &
 + \Big( \pp_\rhat \avr{\vvvIItin}+\pp_\theta\avr{\vvvIIrin}-\avr{\vvvIItin}\Big)
 \sin\theta \bigg]_{\rhat=1} \bigg\}.
 \eal

\subsection{\textit{\textbf{F}}$^{\textbf{rad}}$ in the long-wavelength limit
\textit{\textbf{x}}$_{\textbf{0}}$ $\ll$ 1}}
\seclab{FradLongWave}

\vspace*{-3mm}

To leading order in the dimensionless wave number~\eqnoref{xOdef} $\xO = \kO a$, only $D_0$ and $D_1$ contribute to $\FFFradII$ for both solid and fluid particles,
 \bsuba{F1_final_form}
 \bal
 \FFFradII & = - \pi \rho_0 \re\bigg[A_0A_1^*D_0+\frac{2}{5}A_1A_2^* D_1\bigg]\eee_z,
 \\
 \eqlab{D0_final_form}
 D_0 &= \sum_{i,k=0}^N S_{ik,0} \alpha_{i,0} \alpha^*_{k,1} \propto \xO^3,
 \\
 \eqlab{D1_final_form}
 D_1 &= \sum_{i,k=0}^N S_{ik,1} \alpha_{i,1} \alpha^*_{k,2} \propto \xO^3,
 \\ \nn
 &\quad \text{for }\; x_0 \ll 1\; \text{with}\;  N = 3\; (6) \text{ for solids (fluids)},
 \eal
 \esuba
where the leading order $\xO^3$ of $D_0$ and $D_1$  is shown. The terms contributing to this leading order are given in Appendices~\appnoref{ScatteringCoeffSolids} (solids) and \appnoref{ScatteringCoeffFluids} (fluids) for $\alpha_{i,n}$ and in Appendices~\appnoref{S_coeff_sl} (solids) and \appnoref{S_coeff_fl} (fluids) for $\Smn{ik}0$ and $\Smn{ik}1$.

The result in \eqref{F1_final_form} can be expressed in terms of $\pIin$ and $\vvvIin$ and brought to the familiar form used in the literature~\cite{Gorkov1962, Settnes2012, Karlsen2015}. In the long-wavelength limit, $\Gams$ and $\Gamt$ are negligible compared to unity, so $\kc\approx \kO$ in \eqref{kc_fluids}. From the properties of $j_n(\kc r)$ and $P_n(\cos\theta)$, we derive the following from \eqref{phicin_solution},
 \bsubal{phicin_r_zero_relations}
 \eqlab{phicin_grad_phicin}
 \phiIcin\nablabf \phiIcinC\big|_{\rhat=0} &= \frac{1}{3}\kcO A_0 A_1^*\: \eee_z ,
 \\
 \eqlab{vin_grad_vin}
 (\nablabf \phiIcin \cdot \nablabf)\nablabf \phiIcinC\big|_{\rhat=0}
 &=\bigg[\frac{2A_1A_2^* \kcO^3}{45}-\frac{A_1A_0^* \kcO^3}{9}\bigg] \eee_z .
 \esubal
We use $\vvv_1^\mr{in}=\nablabf \phiIcin$ and $p_1^\mr{in}\approx \ii\omega \rho_0 \phiIcin$ from \eqref{p1_from_potentials} and combine \eqref{phicin_r_zero_relations} with \eqref{F1_final_form}, to derive
 \bal\eqlab{F1_familiar_form}
 \FFFradII &=-\pi a^3 \bigg\{\frac{2\kapSO}{3} \re\bigg[
 \frac{9(D_0+D_1^*)}{2\xO^3}p_1^\mr{in}\nablabf p_1^\mr{in *}\bigg]
 \nn\\
 &\qquad-\rho_0 \re\bigg[-\frac{9D_1}{\xO^3}(\vvv_1^\mr{in}\cdot \nablabf)\vvv_1^\mr{in *}
 \bigg]\bigg\}^{{}}_{\rhat=0}.
 \eal
$\FFFradII$ is equal to $\FFFrad$ derived by Settnes and Bruus \citep{Settnes2012} and by Karlsen and Bruus \cite{Karlsen2015} when substituting their monopole and dipole coefficients $f_0$ and $f_1$ by,
 \beq{f0f1_D0D1}
 f_0 \rightarrow \frac{9(D_0+D_1^*)}{2\xO^3},
 \qquad
 f_1 \rightarrow -\frac{9D_1}{\xO^3}.
 \eeq
For a solid particle, the long-wavelength limit of $\FFFradin$ is found from \eqref{F2_sl} by noting that $\avr{v_{2z}^\mr{in}}$ does not vary much across the particle surface, thus
 \beq{F2_sl_longwave}
 \FFFradin \approx 6\pi \eta_0 a \avr{v_{2z}^\mr{in}}\big|_{r=0} \: \eee_z.
 \eeq

For a fluid particle, the long-wavelength limit of $\FFFradin$ is found from \eqref{F2_fl} by first noting that
 $\big|\pp_\rhat \avr{\vvvIItin}\big| \sim \big|\frac{a}{\lambda}\avr{\vvvIItin}\big|$, which is neglected compared to the remaining terms in \eqref{F2_fl}, and then by making the approximation
 \bal\eqlab{v2z_approx}
 \avr{\vvvIIin}_ {\rhat=1} \approx \avr{v_{2z}^\mr{in}}_{\rhat=0} \big[\cos\theta\:\eee_r
 - \sin\theta\:\eee_\theta\big].
 \eal
Finally, the components $\avr{\vvvIIin}_ {\rhat=1}$ are inserted in \eqref{F2_fl}, and we arrive at
 \beq{F2_fl_longwave}
 \FFFradin \approx 2\pi \frac{2+3\etaOTi}{1+\etaOTi}\:\eta_0 \avr{v_{2z}^\mr{in}}\Big|_{r=0}\:\eee_z,
 \eeq
which is the well known result for the drag force on a droplet in a constant Stokes flow \cite{Hetsroni1970}.

We conclude that in the long-wavelength limit, $\FFFrad$ exerted on a suspended particle by an arbitrary axisymmetric incident wave defined by $\phiIcin$ is given by,
 \bal
 \eqlab{Frad_longwave}
 &\FFFrad = \FFFradII + \FFFradin,
 \\[2mm]
 \nn
 &\;\textit{\underline{Solid particle}}\!:\; \FFFradII  \text{ from \eqref{F1_familiar_form}, }
 \FFFradin \text{ from \eqref{F2_sl_longwave}},
 \\ \nn
 &\quad D_0  \text{ from \eqref{D0_final_form}, }
 D_1 \text{ from \eqref{D1_final_form}},
 \\ \nn
 &\; \ain \text{ from \appref{ScatteringCoeffSolids}, }
 \text{ and } \Smn{ik}n \text{ from \appref{S_coeff_sl},}
 \\[2mm]
 \nn
 &\; \textit{\underline{Fluid particle}}\!:\; \FFFradII  \text{ from \eqref{F1_familiar_form}, }
 \FFFradin \text{ from \eqref{F2_fl_longwave}},
 \\ \nn
 &\; D_0  \text{ from \eqref{D0_final_form}, }
 D_1 \text{ from \eqref{D1_final_form}},
 \\ \nn
 &\; \ain \text{ from \appref{ScatteringCoeffFluids}, }
 \text{ and } \Smn{ik}n \text{ from \appref{S_coeff_fl}.}
 \eal
These closed-form analytical expressions for $\FFFrad$ acting on a spherical solid particle and on a spherical fluid particle in the long-wavelength limit constitute the primary result of this work. \textsc{Matlab} scripts for computing $D_0$ and $D_1$ in the long-wavelength limit are included in Section~S1 in the Supplemental Material~\cite{Note1}.
\\[0mm]

\section{\textit{D}\textbf{$_0$} and \textit{D}\textbf{$_1$}
in the limit of very thin and very thick boundary layers}
\seclab{D0_D1_limits}

\vspace*{-3mm}

The force coefficients $D_0$ and $D_1$ are complicated functions of the physical parameters of the system. To facilitate direct comparison of the expression~\eqnoref{Frad_longwave} for $\FFFrad$ with the results in the literature by Gor'kov~\cite{Gorkov1962}, Doinikov~\cite{Doinikov1997, Doinikov1997a, Doinikov1997b}, and Karlsen and Bruus~\cite{Karlsen2015}, we derive some limiting cases for $D_0$ and $D_1$. Specifically, we consider the weakly dissipative limit where thermal and viscous boundary-layer thicknesses $\delt$  and $\dels$ are much smaller than the particle radius $a$, that is $\delt, \dels \ll a$, and the opposite strongly dissipative limit, where $\delt, \dels \gg a$.

\vspace*{-4mm}

\subsection{Limiting cases for solid particles}
\seclab{D0D1_limit_sl}

\vspace*{-3mm}

The weakly dissipative limit for a solid particle in a thermoviscous fluid in the long-wavelength limit is characterized by $\dels,\delt,\delt^\prime \ll a \ll \lambda$. By keeping terms up to first order in $\delta/a$ in $D_0$ and $D_1$ from~\eqref{Frad_longwave}, we derive
 \bsuba{small_BL_solid}
 \bal
 \eqlab{D0_small_BL_solid}
 D_0 &= \frac{2\ii}{3}\big(\alpha_{c,0}^\mr{sc,wd}+\alpha_{c,1}^\mr{sc,wd *}\big),
 \\
 \eqlab{D1_small_BL_solid}
 D_1 &= \frac{2\ii}{3} \alpha_{c,1}^\mr{sc,wd},
 \\
 \eqlab{acsc0_small_BL_solid}
 \alpha_{c,0}^\mr{sc,wd} &= -\frac{\ii \xc^3}{3} \Bigg[1-\kapSOTi
 \nn\\
 &\hfill - \frac{3}{2}\frac{(1+\ii)(\gamO-1)\big(1-\frac{\alphapOTi}{\rhoOTi \cpOTi}\big)^2}{1+(1+X')^{1/2}(\DthOTi)^{1/2}(\kthOTi)^{-1}} \frac{\delt}{a} \Bigg],
 \\
 \eqlab{acsc1_small_BL_solid}
 \alpha_{c,1}^\mr{sc,wd} &= \frac{\ii \xc^3}{3}\frac{\rhoOTi-1}{2\rhoOTi+1}\Bigg[1+3(1+\ii)\frac{\rhoOTi-1}{2\rhoOTi+1}\frac{\dels}{a}\Bigg].
 \eal
 \esuba
When inserting \eqref{small_BL_solid} into \eqref{f0f1_D0D1} and noting that  $\xc\approx \xO$, we obtain the same monopole and dipole coefficients $f_0$ and $f_1$ as derived by Karlsen and Bruus~\cite{Karlsen2015} in their Eqs.~(66) and (71). The only minor discrepancy is that Karlsen and Bruus have $(1-X')^{1/2}$ instead of $(1+X')^{1/2}$ in the denominator of the thermal correction to $\alpha_{c,0}^\mr{sc,wd}$. This sign difference, which is insignificant since $X'\ll 1$ for most solids, is due to a simple sign error by Karlsen and Bruus in their expressions~(48b) for the thermal wave number.

In Ref. \cite{Doinikov1997a}, Doinikov computed $D_0$ and $D_1$ for a rigid solid particle in his Eqs.~(21) and (22). Our result~\eqnoref{small_BL_solid} reduces exactly to his, when we model a rigid sphere without thermal and mechanical expansion by setting $\alpha'_{p0}=0$, $\cTO'/c_0 \rightarrow \infty$, and $\cLO'/c_0 \rightarrow \infty$,
 \bsubal{small_BL_solid_rigid}
 \eqlab{D0_small_BL_solid_rigid}
 D_0 &= \frac{2\rhoOTi \xO^3}{3(2\rhoOTi+1)}\Bigg[1+ \frac{(1-\ii) (\rhoOTi-1)^2 }{\rhoOTi(2\rhoOTi+1)}\frac{\dels}{a}
 \nn\\
  &\hspace*{20mm} -\frac{(1+\ii)(\gamO-1)(2\rhoOTi+1)}{2\rhoO \big[1+(\delt \kthOTi)^{-1}\delt' \big]}\frac{\delt}{a} \Bigg]  ,
  \\
 \eqlab{D1_small_BL_solid_rigid}
 D_1 &= -\frac{2(\rhoOTi-1)\xO^3}{9(2\rhoOTi+1)} \bigg[1+3(1+\ii)\frac{\rhoOTi-1}{2\rhoOTi+1}\frac{\dels}{a}\bigg].
 \esubal
It should be noted that Doinikov in his 1997 work \cite{Doinikov1997} defined the coefficients $Z_n$, which have an opposite sign to the corresponding $D_n$ coefficients defined in his own 1994 work \cite{Doinikov1994a, Doinikov1994} and in our present work. We also note that Doinikov in his 1997 work \cite{Doinikov1997} moved the contributions to the $\FFFradII$ from what would correspond to his $\Smn{00}n$ coefficients into $\FFFradin$, because they only result from the incident wave. However, these coefficients do not contribute to the weakly dissipative limit.

Lastly, we note that expressions~\eqnoref{small_BL_solid} reduce to the results for $\FFFrad$ obtained by Settness and Bruus for non-thermal viscous fluids \cite{Settnes2012}  by letting $\delt/a\rightarrow 0$, and by Gor'kov for ideal fluids \cite{Gorkov1962} by letting $\dels/a\rightarrow 0$ and $\delt/a\rightarrow 0$.

We now pass to the strongly dissipative limit of large boundary layers characterized by $a \ll  \dels,\delt,\delt^\prime \ll \lambda$. The dominant terms in $D_0$ and $D_1$ for $\dels/a, \delt/a, \delt'/a \rightarrow \infty$ are simple and purely imaginary,
 \beq{D0D1_large_delta}
 D_0 = D_1 =
 -\frac{\ii}{6}\:\xO^3\:\frac{\dels^2}{a^2}.
 \eeq
However, we see from expression~\eqnoref{F1_familiar_form} for $\FFFradII$, that using \eqref{D0D1_large_delta}  leads to $D_0+D_1^* = 0$ in the first term, and if furthermore $(\vvvIin\cdot\nablabf)\vvvIinC$ is real to leading order (e.g.\ for standing waves), also the second term is zero. Consequently, we need to include the dominant real terms in $D_0$ and $D_1$,
 \bsuba{large_BL_solid}
 \bal
 \eqlab{D0_large_BL_solid}
 D_0 &= \xO^3 \Bigg[-\frac{\ii}{6}\frac{\dels^2}{a^2}
 -\frac{2(\rhoOTi-1)}{27} (3B_c+2) \Bigg]
 \nn \\
 & +\xO^3\:\frac{2(\rhoOTi+2)}{27\Big[1-\frac43 \frac{\cTO^{\prime 2}}
 {c_0^2}\frac{\gamO-1}{1+X'}\frac{\alphapOTi^2 \chi'}{\cpOTi}\Big]}
 \nn \\ \nn
 &\quad\times \Bigg\{1-\kapSOTi + \frac{(\gamO-1)\rhoOTi}{1+X'}\:
 \alphapOTi \chi'
 \eal
 \bal
 &\quad\times
 \bigg[\frac{1}{\rhoOTi}\Big(1-\frac{\alphapOTi}{\rhoOTi \cpOTi}\Big)
 + \frac{4\cTO^{\prime 2}\kapSOTi}{3c_0^2} \Big(1-\frac{\alphapOTi}{\rhoOTi \cpOTi \kapSOTi}\Big)
 \bigg]\Bigg\},
 \\
 \eqlab{D1_large_BL_solid}
 D_1 &= \xO^3 \bigg[-\frac{\ii}{6}\frac{\dels^2}{a^2} + \frac{2(\rhoOTi-1)}{27} \bigg],
 \eal
 \esuba
In the strongly dissipative limit, the included temperature and density dependency of the fluid viscosity enters through $B_c$ defined in \eqref{Bc}. We note that $|B_c| \gtrsim 1$ causes large changes to $\FFFradII$, if simultaneously the density contrast $\rhoOTi$ between the particle and the surrounding fluid deviates sufficiently from unity.

In general, the expressions~\eqnoref{f0f1_D0D1} for $f_0$ and $f_1$ in the strongly dissipative limit differ from the results derived by Karlsen and Bruus, Eqs.~(65) and (73) in Ref.~\cite{Karlsen2015}, which signals the importance of the inclusion of acoustic microstreaming in the calculation of $\FFFrad$ on small particles. We note that the difference is most dramatic when $\rhoOTi$ differs significantly from unity.

For a rigid particle, $D_0$ and $D_1$ in \eqref{large_BL_solid}, are reduced by applying the same assumptions as used in connection with \eqref{small_BL_solid_rigid},
 \bsubal{large_BL_solid_rigid}
 \eqlab{D0_large_BL_solid_rigid}
 D_0 &= -\frac{2\xO^3}{9}\bigg[\frac{3\ii}{4}\frac{\dels^2}{a^2} + B_c(\rhoOTi-1) +\frac{\rhoOTi-4}{3} \bigg] ,  \\
 \eqlab{D1_large_BL_solid_rigid}
 D_1 &= -\frac{2\xO^3}{9}\bigg[\frac{3\ii}{4}\frac{\dels^2}{a^2} - \frac{\rhoOTi-1}{3} \bigg] .
 \esubal
We compare these expressions to the result by Doinikov, Eqs.~(25) and (26) in Ref.~\cite{Doinikov1997a} taken to leading order in $\delta/a$. Firstly, we set $B_c = 0$, because Doinikov does not treat the temperature and density dependency of the viscosity $\eta$, and then we obtain identical real parts. Secondly, as discussed after \eqref{small_BL_solid_rigid}, Doinikov has defined $\FFFradII$ and $\FFFradin$ by a different grouping of terms in \cite{Doinikov1997} than we have. When moving the terms in Doinikov's work ($S_{90}=S_{91}=\tfrac19\xO^3\xs^{-2}$ defined in the appendix of Ref.~\cite{Doinikov1994a}) corresponding to our second-order coefficients $\Smn{00}0$ and $\Smn{00}1$ from $\FFFradin$ back to $\FFFradII$, we obtain the same imaginary parts as well. We note that Doinikov's coefficients $S_{90}$ and $S_{91}$ differ from our $\Smn{00}0$ and $\Smn{00}1$ given in \appref{S_coeff_sl}, because in the second-order no-slip boundary conditions~\eqnoref{Second_order_no_slip_sl} used in our analysis, Doinikov replaced the full Stokes term $\avr{(\sss_1 \cdot \nablabf)\vvv_1}$ by the radial part $\avr{s_{1r}\pp_r\vvv_1}$. However, for the special case of a rigid sphere, these two versions of the boundary condition lead to the same final expression for $\FFFrad$.

\subsection{Limiting cases for fluid particles}
For a thermoviscous fluid particle of radius $a$ in a thermoviscous fluid, the weakly dissipative limit is characterized by $\dels,\delt,\dels',\delt'\ll a \ll \lambda$. The general expressions for $D_0$ and $D_1$ in \eqref{Frad_longwave} then reduce to,
 \bsubal{small_BL_fluid}
 \eqlab{D0_small_BL_fluid}
 D_0 &= \frac{2\xO^3}{9}\Bigg[1-\kapSOTi - \frac{3}{2}\frac{(1+\ii)(\gamO-1)}{1+(\DthOTi)^{1/2}(\kthOTi)^{-1}} \nn\\
 &\quad \times  \bigg(1-\frac{\alphapOTi}{\rhoOTi \cpOTi}\bigg)^2 \bigg(1-\frac{\rhoOTi}{(1+\etaOTi)(2\rhoOTi+1)}\bigg) \frac{\delt}{a}
 \nn\\
 & \quad + \frac{\rhoOTi-1}{2\rhoOTi+1} + \frac{3(1-\ii)(\rhoOTi-1)^2}{(2\rhoOTi+1)^2\big(1+\tilde{\nu}_0^{1/2}\etaOTi^{-1}\big)}\frac{\dels}{a} \Bigg] ,  \\
 \eqlab{D1_small_BL_fluid}
 D_1 &= -\frac{2\xO^3}{9}\Bigg[\frac{\rhoOTi-1}{2\rhoOTi+1}+\frac{3(1+\ii)(\rhoOTi-1)^2}{(2\rhoOTi+1)^2 \big(1+\tilde{\nu}_0^{1/2}\etaOTi^{-1}\big)} \nn \\
 & \quad \times\bigg(1-\frac{(3+\ii)(2\rhoOTi+1)}{4(3\rhoOTi+2)(1+\etaOTi)}\bigg) \frac{\dels}{a}\Bigg] .
 \esubal
Using \eqref{f0f1_D0D1}, we compare the result~\eqnoref{small_BL_fluid} to Eqs.~(33) and (34) in Ref.~\cite{Doinikov1997b} by Doinikov and to Eqs.~(60) and (69) in Ref.~\cite{Karlsen2015} by Karlsen and Bruus. We find that all three results agree, except that the factors $\big(1-\frac{\rhoOTi}{(1+\etaOTi)(2\rhoOTi+1)}\big)$ in $D_0$ and $\big(1-\frac{(3+\ii)(2\rhoOTi+1)}{4(3\rhoOTi+2)(1+\etaOTi)}\big)$ in $D_1$ in~\eqref{small_BL_fluid} are replaced by unity in the two other theories. This difference between the present theory and the previous work arises due to our inclusion of acoustic microstreaming inside the fluid particle, and we notice that the difference vanishes for very viscous droplets where $\etaOTi\gg 1$. To obtain the comparison with Doinikov's result, we keep only the linear terms in Eqs.~(35) and (36) in Ref.~\cite{Doinikov1997b}, and then we apply $\delt'/\delt=(\DthOTi)^{1/2}$ and $\dels'/\dels=(\tilde{\nu}_0)^{1/2}$ alongside the relation from \eqref{thermodynamic_relation_gamma}. Again, we note that the expressions converge to the ideal fluid result for vanishing boundary layers.

Passing on to the large boundary layer limit characterized by $a\ll\dels,\delt,\dels',\delt'\ll \lambda$, we reduce the force coefficients $D_0$ and $D_1$ in \eqref{Frad_longwave} to
%
 \bsuba{large_BL_fluid}
 \bal
 \eqlab{D0_large_BL_fluid}
 D_0 &=\frac{2\xO^3}{9(1+\etaOTi)}
 \Bigg\{-\ii\frac{\dels^2}{a^2}\frac{2+3\etaOTi}{4} -\frac{2}{3}(\rhoOTi-1)(1+\etaOTi)
 \nn\\		
 &
 +\bigg[1-\kapSOTi + (\gamO-1)\alphapOTi \tilde{K}\bigg]
 \nn \\
 &\quad
 \times\bigg[-\ii\frac{\dels^2}{a^2}+\frac{(1+\etaOTi)(3\etaOTi\rhoOTi+6\etaOTi-2\rhoOTi+8)}{3(3\etaOTi+2)}\bigg] \nn \\
 & +\frac{(\rhoOTi-1)\etaOTi}{3\etaOTi+2}
 \bigg[-\kapSOTi B_c' - \frac{B_c}{\etaOTi}(3\etaOTi^2+4\etaOTi+2)
 \nn\\
 &\qquad \qquad
 + B_t' (\gamO-1)\alphapOTi \tilde{K} \bigg]
 \nn \\ \nn
 &
 - \frac{2(\gamO-1)\tilde{K}}{15 D_0^\mr{th \prime} \nuO^{-1}}
\Big[\alphapOTi(\etaOTi-1)
 +5\kthOTi (1-\alphapOTi) \Big] \Bigg\},
 \eal
 \bal
 & \text{ for }\;
 \frac{2\gamST}{3a}\kapsO' \ll 1
 \text{ and }\; \tilde{K} = 1-\frac{\alphapOTi}{\rhoOTi \cpOTi},
 \\
 \eqlab{D1_large_BL_fluid}
 D_1 &= -\frac{\ii }{18}\:\xO^3\:\frac{\dels^2}{a^2}
 \nn \\
 &\quad \times
 \frac{\etaOTi \big[114\etaOTi^2+457\etaOTi+305-\ii\frac{40\gamST}{a\omega \etaO}(7+3\etaOTi)\big]}
 {(1+\etaOTi)\big[38\etaOTi^2+89\etaOTi +48 -\ii\frac{40\gamST}{a\omega \etaO}(1+\etaOTi)\big]} ,
 \eal
 \esuba
%
As in \eqref{large_BL_solid} for a solid particle, expressions~\eqnoref{large_BL_fluid} for $D_0$ and $D_1$ for a fluid particle differ significantly from what was found by Karlsen and Bruus \cite{Karlsen2015} Eqs.~(62) and (73), who neglected microstreaming, which tends to dominate the thermoviscous corrections for large boundary layers.
Moreover, result~\eqnoref{large_BL_fluid} also differs from what was found by Doinikov in 1997~\cite{Doinikov1997b} Eqs.~(45) and (46), where \textit{external} microstreaming was included. This discrepancy is caused by our inclusion of microstreaming \textit{inside} the droplet \big($\avr{\vvvIIp} \neq \zerovec$\big), the temperature and density dependency of the viscosities ($B_c, B_c',B_t' \neq 0$ in \eqref{B_coeff}), and the tangential component of the displacement $\sss_1$ \big($\avr{s_{1\theta}\frac1r \pp_\theta\vvv_1} \neq \zerovec$\big) in the Stokes terms in \eqsref{Second_order_no_slip_sl}{Second_order_no_slip_fl}. Agreement is obtained when turning off inner microstreaming \big($\avr{\vvvIIp} = \zerovec$\big) and making the viscosities constant ($B_c, B_c',B_t' = 0$) in our model~\eqref{Frad_longwave}, and by adding the tangential component $\avr{s_{1\theta}\frac1r \pp_\theta\vvv_1}$ to the Stokes term in Doinikov's boundary conditions. We note that the effects of inner microstreaming on $\FFFrad$ can almost be removed by simply taking the limit $\etaOTi \rightarrow \infty$, however in expression~\eqnoref{D0_large_BL_fluid} for $D_0$, a term $-\frac{4\xO^3(\gamO-1)}{135 D_0^\mr{th \prime} \nuO^{-1}}\alphapOTi\big(1- \frac{\alphapOTi}{\rhoOTi \cpOTi}\big)$ remains, which is independent of $\etaOTi$, but which is nevertheless induced by the inner microstreaming through  the second-order Stokes terms of the inner streaming in \eqref{Second_order_no_slip_fl} and the compressional term $c_{11}'$ in \eqref{c1np}.

\section{Results for a standing plane wave}
\seclab{ResultsPlaneWave}

To illustrate some of the implications of our theory, we consider the important case of a weakly damped, one-dimensional, standing incident pressure wave $p_1^\mr{in}(z)$ along the $z$-axis having the complex wave number $\kc$, amplitude $p_a$, and phase shift $\kc d$,
 \bsubal{p1_v1_in_standing}
 \eqlab{p1_in_standing}
 \pIin(z) &= p_a \cos[\kc (z+d)],
 \\
 \vvvIin(z) &= -\ii\frac{1}{\omega\rhoO}\:\nablabf \pIin
 = \ii\frac{\kc p_a}{\omega\rhoO}\:\sin[\kc (z+d)]\:\een_z.
 \esubal
Inserting $\pIin$ and $\vvvIin$ in \eqref{Frad_longwave}, we obtain $\FFFrad$ on a spherical particle of radius $a$ to leading order $\xO^3 = \kO^3a^3$,
 \bsuba{Frad_standing}
 \bal
 \eqlab{Frad_standing_form}
 \FFFrad &= 4\pi \Phiac a^3 k_0 \Eac \sin(2k_0d)\:\een_z,
 \\
 \eqlab{Eac}
 \Eac & = \tfrac{1}{4}\kapSO p_a^2,
 \\
 \eqlab{Phiac}
 \Phiac &= \tfrac{3}{2}\:x_0^{-3}\re[D_0-2D_1],
 \eal
 \esuba
where we have introduced the usual acoustic energy density $\Eac$ and acoustic contrast factor $\Phiac$ \cite{King1934, Yosioka1955, Settnes2012, Karlsen2015}. A main feature is that $\Phiac>0$ signifies particle migration towards pressure nodes, whereas $\Phiac<0$ directs particles towards antinodes. We note that the incident field \eqnoref{p1_v1_in_standing} leads to the following steady incident streaming $\avr{\vvvIIin}$ \cite{Doinikov1994a},
 \bal\eqlab{v2_in}
 \avr{\vvvIIin} &= \frac{\ii \big|p_a \kc\big|^2}{8\rhoO^2\omega^3} \Big\{(\kc-\kc^*)\sin[(\kc+\kc^*)(z+d)]
 \nn\\
 &\qquad -(\kc +\kc^*)\sin[(\kc - \kc^*)(z+d)] \Big\}\:\eee_z.
 \eal
Using this to compute $\FFFradin$ by \eqref{F2_sl_longwave} or \eqnoref{F2_fl_longwave} leads to contributions of order $\xO^5$ or higher, which can be neglected compared to $\FFFradII$. In the following we study the contrast factor $\Phiac$ defined in \eqref{Phiac} as a function of particle radius $a$ from 0.05 to $30~\SImum$ at a frequency of either $1~\SIkHz$ or $1~\SIMHz$ for selected fluid and solid particles in liquids or gasses.

\subsection{The contrast factor for solid particles}
\seclab{Phiac_sl}
We study the following four examples of solid microspheres in fluids:
(a) Polystyrene in water,
(b) po\-ly\-styrene in oil,
(c) polystyrene in air,  and
(d) copper in oil,
with the material parameters listed in \tabref{Param_table}.
Case (a) is chosen as it is studied extensively both experimentally and theoretically in the acoustofluidic literature \cite{Lenshof2012, Hammarstrom2012, Muller2013, Laurell2015}.
Case (b) is chosen for the increased density ratio $\rhoOTi$ and thermal factor $\gamO -1$ of an organic liquid (oil) compared to water \cite{Liu2021}.
Case (c) is chosen for its relevance to aerosol studies \cite{Ran2015}.
Case (d) is the one selected and studied numerically by Baasch, Pavlic, and Dual for its pronounced thermoviscous response \cite{Baasch2019}.
In \figref{Phiac_solid} we plot $\Phiac$ versus $\dels/a$  computed from \eqref{Phiac}
(blue full curve \qmarks{Winckelmann}) for these four cases and compare with the results for $\Phiac$ obtained by
Doinikov \cite{Doinikov1997a}  (red full curve \qmarks{Doinikov}),
Karlsen and Bruus \cite{Karlsen2015}  (purple full curve \qmarks{Karlsen}),
Settnes and Bruus \cite{Settnes2012}  (green dashed curve \qmarks{Settnes}), and
Gor'kov \cite{Gorkov1962}  (brown full curve \qmarks{Gor'kov}). For the following discussion of the role of the thermoviscous effects in this figure, we refer to the selected parameter values listed in \tabref{Solids_abcd_table}. The elastic solid-particle result can be compared directly with Doinikov's rigid solid-particle result~\cite{Doinikov1997a} for the low compressibility cases (c) of a polystyrene sphere in air, $\kapSOTi=3\times 10^{-5}$, and (d) of a copper sphere in oil, $\kapSOTi=10^{-2}$. However, for the cases  with a polystyrene sphere in (a) water, $\kapSOTi = 0.53$, and (b) oil,  $\kapSOTi = 0.46$, it is a poor approximation. Thus, to obtain a more illuminating comparison in (a) and (b), we add the compressibilty term $-\xO^3 \tfrac29  \kapSOTi$ to the $D_0$ coefficient (named $-Z_0$ in Ref.~\cite{Doinikov1997a}), which makes Doinikov's results converge to the ideal-fluid result by Gor'kov in the limit of vanishing boundary layers.

For \hfill large \hfill particles, \hfill $a \gg \dels$, \hfill we \hfill see \hfill that \hfill the \hfill four

\begin{widetext}
\mbox{}\\[-15mm]
\begin{table}[!h]
\caption{\tablab{Param_table} Parameters at $T_0=300$~K for all fluids and solids used in the examples in \figsref{Phiac_solid}{Phiac_fluid}. Parameters are given for water \cite{Muller2014,Wagner2002,Huber2009,Huber2012}, oil \cite{Coupland1997,Noureddini1992}, air \cite{CRC2016, EngineersEdge_air_thermal}, copper \cite{Selfridge1985,CRC2016}, and polystyrene \citep{Karlsen2015}. We have given the parameters necessary to compute the scattering coefficients $\ain$  and the second-order coefficients $S_{ik,n}$ found in the appendices. Note that $\kapSO$, $\kapTO$, and $\gamO$ can be found from Eqs. \eqnoref{thermodynamic_relations_general} and \eqnoref{isentropic_sound_speed_fluids} for a fluid, and that $\cLO$, $\kapSO$, $\kapTO$, and $\gamO$ can be found from Eqs. \eqnoref{thermodynamic_relations_general}, \eqnoref{kapT_solid}, and \eqnoref{c_sl} for a solid. We could only find data for $\big(\frac{\pp\eta}{\pp \rho}\big)_{\rho_0}$ for water, so the quantity is set to zero for all other fluids.}
\begin{ruledtabular}
\begin{tabular}{lcccccl}
Parameter             & Water                   & Oil                    & Air
& Copper & Polystyrene            & Unit
\\ \hline
$\cO$                                  & $1502$                  & $1445$                 & $347.4$                & 5010 &    2400           & $\mr{m\,s^{-1}}$
\\
$\cTO$ & -- & -- & -- & 2270 & 1150 & $\mr{m\,s^{-1}}$
\\
$\rhoO$                             & $996.6$                 & $922.6$                & $1.161$                & 8930 &  1050             & $\SIkgm$
\\
$\alphapO$                      & $2.75\times 10^{-4}$   & $7.05\times 10^{-4}$ & $3.35 \times 10^{-3}$ &  $1.65\times 10^{-5}$ & $2.09 \times 10^{-4}$ & $\mr{K^{-1}}$
\\
$\cpO$                           & $4181$                  & $2058.4$               & $1007$                 & 385 &  1220             & $\mr{J\,kg^{-1}\,K^{-1}}$
\\
$\kthO$                               & $6.10\times 10^{-1}$   & $0.166$                & $2.64 \times 10^{-2}$ & 401 & 0.154             & $\mr{W\,m^{-1}\,K^{-1}}$
\\
$\gamO-1$, \eqref{thermodynamic_relation_gamma}
& 0.012 & 0.15 & 0.40  & 0.004 & 0.04 & --
\\
$\etaO$                              & $8.54\times 10^{-4}$   & $5.74\times 10^{-2}$  & $1.85\times 10^{-5}$  & -- & --  & $\mr{Pa\,s}$
\\
$\etaO^\mr{b}$                       & $2.4\times 10^{-3}$     & $8.513\times 10^{-2}$  & $1.1 \times 10^{-5}$   & -- & -- & $\mr{Pa\,s}$
\\
$\frac{1}{\etaO}\big(\frac{\pp \eta}{\pp T}\big)_{T_0}$   & $-0.022$  & $-0.044$ & $0.0025$   & -- & --  & $\SIK^{-1}$
\\
$\frac{1}{\etaO}\big(\frac{\pp\eta}{\pp \rho}\big)_{\rho_0}$ & $-2.3\times 10^{-4}$ & --                      & --                      & -- & --                     & $\SIm^3\:\SIkg^{-1}$
\\
\end{tabular}
\end{ruledtabular}
\end{table}

\vspace*{-7mm}
\noindent
\begin{figure}[!h]
\centering
\includegraphics[width=0.95\linewidth]{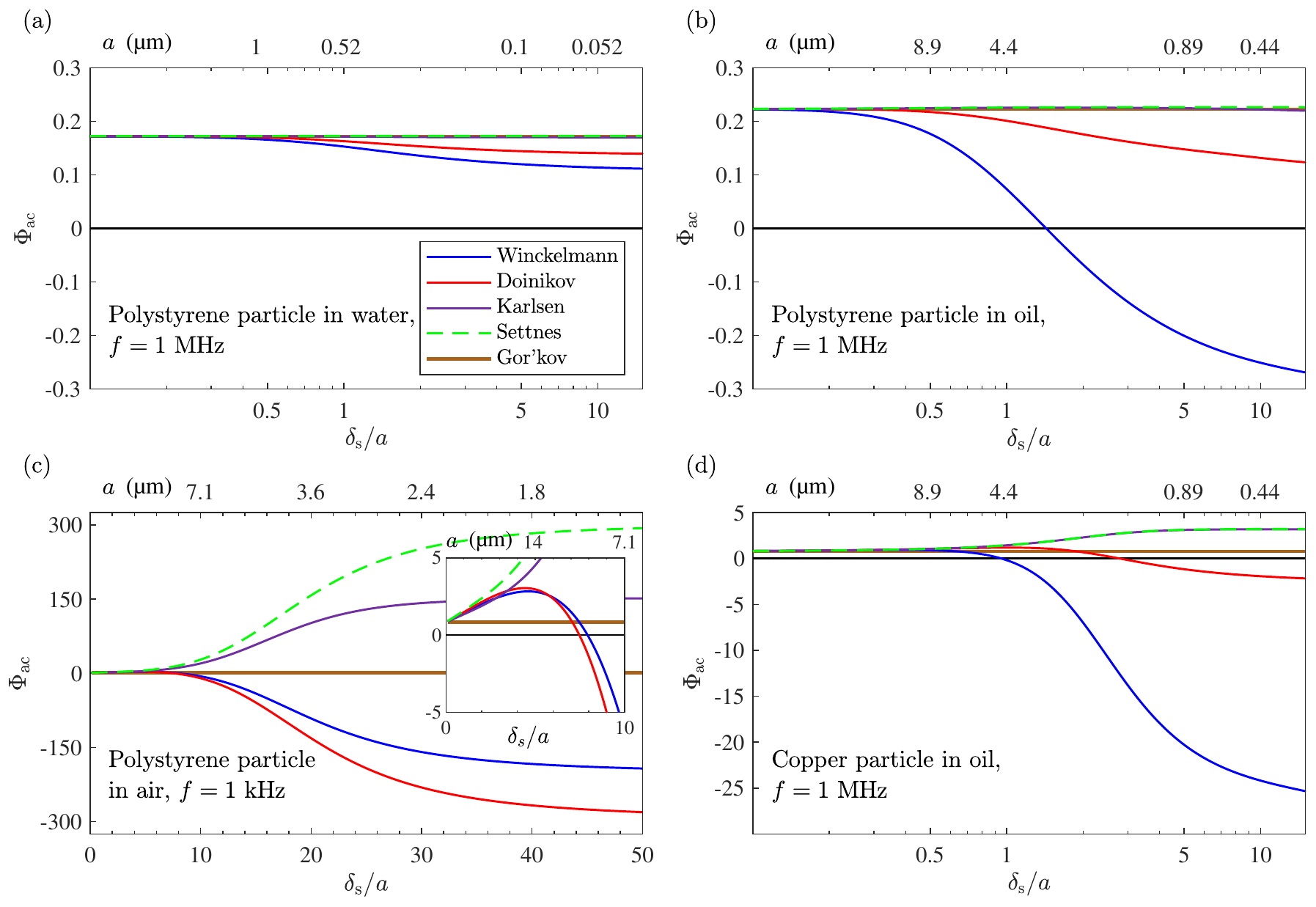}
\caption{\figlab{Phiac_solid} The acoustic contrast factor $\Phiac$ plotted versus normalized boundary-layer width $\dels/a$. Brown lines are the ideal-fluid theory following Gor'kov \cite{Gorkov1962}. Green dashed lines are from Settnes and Bruus \cite{Settnes2012}, magenta lines are from Karlsen and Bruus \cite{Karlsen2015}, red lines are from Doinikov 1997 \cite{Doinikov1997a}, and blue lines are the present theory by Winckelmann and Bruus.
(a) A polystyrene sphere in water at frequency $f=1$ MHz.
(b) A polystyrene sphere in oil at $f=1$ MHz.
(c) A polystyrene sphere in air at $f=1$ kHz. The inset shows the large-particle behavior.
(d) A copper sphere in oil at $f=1$ MHz.\\[-8mm]}
\end{figure}

\end{widetext}
\begin{table}[t]
\caption{\tablab{Solids_abcd_table} Parameters for the four combinations of solid particles in fluids discussed in \secref{Phiac_sl} and shown in \figref{Phiac_solid} for the cases (a) polystyrene in water, (b) polystyrene in oil, (c) polystyrene in air, and (d) copper in oil.}
\begin{ruledtabular}
\begin{tabular}{llcccl}
Case & Ref. & (a)	& (b)	& (c)	& (d)	
\\ \hline
$\rhoOTi-1$ & \eqref{tilde_def} 	& 0.05 	& 0.14 	& 903 				& 8.7
\\
$\kapSOTi$ & \eqref{tilde_def} 		& 0.53 	& 0.46 	& $3\times 10^{-5}$ & 0.01
\\
$B_c$ & \eqref{Bc} 			& 1.2 	& 9.5 	& $-$0.30 			& 9.5
\\
$\gamO-1$ & \eqref{thermodynamic_relation_gamma}	& 0.012 & 0.15 	& 0.40				& 0.15
\end{tabular}
\end{ruledtabular}
\vspace*{-5mm}

\end{table}

\noindent
boundary-layer theories converge towards the ideal-fluid result by Gor'kov for all cases. For a small particle, $a \lesssim \dels$, significant thermoviscous corrections to the ideal-fluid theory arise. We observe that the corrections found by each theory are most pronounced for a large relative density contrast $\rhoOTi-1$. Case (a) of polystyrene in water with $\rhoOTi - 1 = 0.05$ shown in \figref{Phiac_solid}(a) illustrates this point, as we see less dramatic corrections from the ideal-fluid theory compared to the three other solid-fluid combinations, only a modest 30-\% decrease for the smallest 50-nm-radius particles. It is clearly seen from cases (b)-(d) with $\rhoOTi - 1 = 0.14$, 903 and 8.7 that the microstreaming included by Doinikov and Winckelmann becomes dominant for small particles, in agreement with the recent numerical study by Pavlic \etal \cite{Pavlic2022}. The results by Doinikov and Winckelmann  exhibit the same qualitative behavior, but significant quantitative discrepancies between the two microstreaming models develop, when the temperature dependency of the viscosity, included by Winckelmann through the fluid parameter $B_c$, is large. This discrepancy is particularly large in cases (b) and (d) both having $B_c = 9.5$, but less prominent for cases (a) and (c) with  $B_c = 1.2$ and $-$0.30, respectively.

\begin{figure*}[t]
\centering
\includegraphics[width=0.95\linewidth]{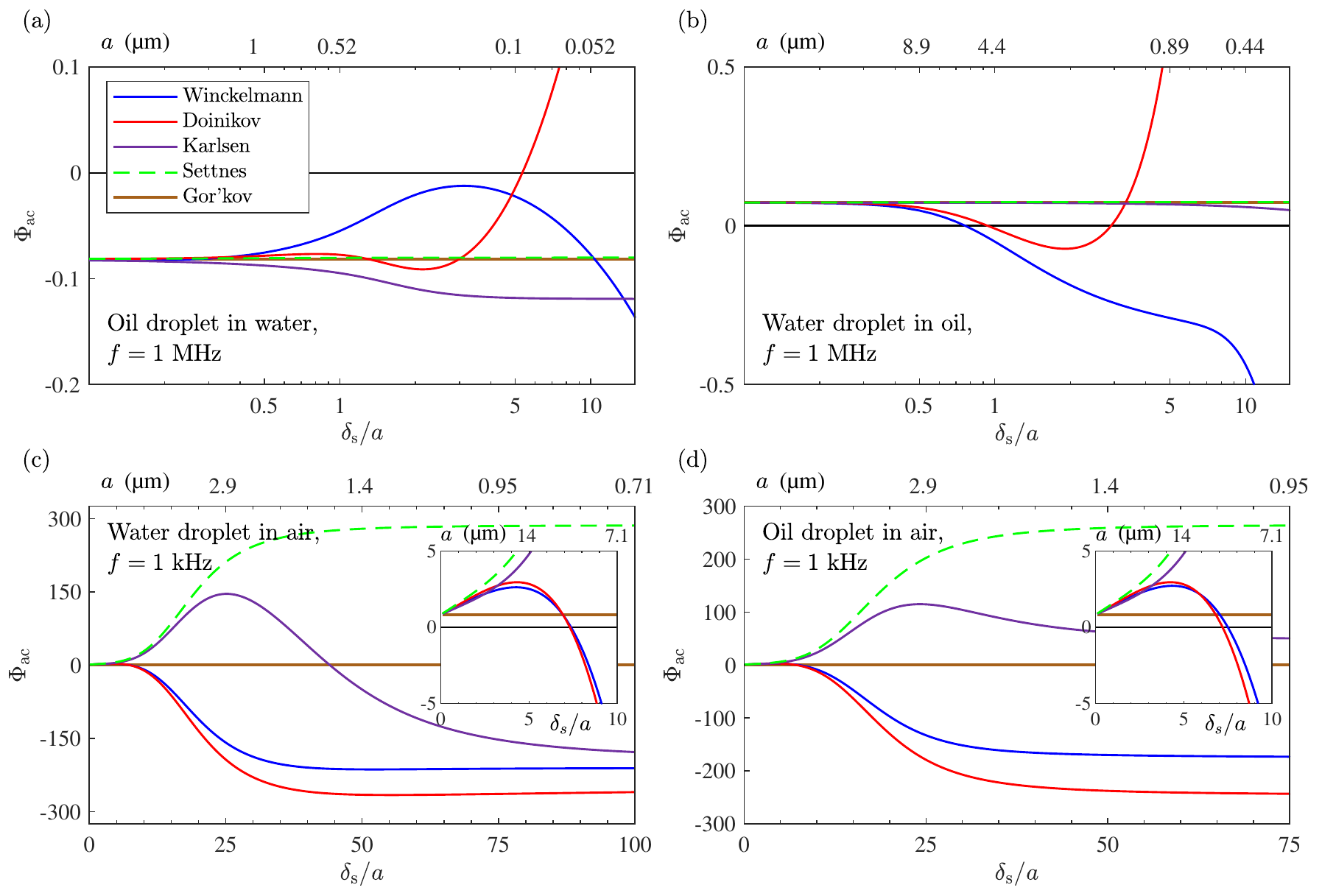}
\caption{\figlab{Phiac_fluid} The acoustic contrast factor $\Phiac$ plotted versus normalized boundary-layer thickness $\dels/a$. Brown lines are the ideal-fluid theory following Gor'kov \cite{Gorkov1962}. Green dashed lines are from Settnes and Bruus \cite{Settnes2012}, magenta lines are from Karlsen and Bruus \cite{Karlsen2015}, red lines are from Doinikov 1997 \cite{Doinikov1997b}, and blue lines are the present theory by Winckelmann and Bruus. (a) An oil droplet in water at frequency $f=1$~MHz. (b) A water droplet in oil at $f=1$~MHz. (c) A water droplet in air at $f=1$ kHz. (d) An oil droplet in air at $f=1$~kHz.  The insets in (c) and (d) show the large-particle behavior.}
\end{figure*}

We add the following specific comments to cases (b)-(d) in \figref{Phiac_solid}: In case (b) it is noteworthy that the full thermoviscous model by Winckelmann for a polystyrene particle in oil at $1~\SIMHz$, in contrast to Doinikov's model, predicts a sign reversal of the acoustic contrast factor $\Phiac$, and that this happens  at a relatively large particle radius $a = 3.1~\SImum$. In case (c), a polystyrene particle in air at $1~\SIkHz$, both Doinikov's and Winckelman's model predict a strong thermoviscous response, including a sign reversal in $\Phiac$ for relatively large particles at nearly the same radius $a=9.5~\SImum$ and $a=9.0~\SImum$, respectively. Finally, case (d) with a copper sphere in oil at $1~\SIMHz$ was treated more in depth in a numerical study by Baasch, Pavlic, and Dual \cite{Baasch2019}. They found a remarkable sign reversal in $\Phiac$ at a relatively large particle radius $a$, and they furthermore showed that this was correctly predicted by the viscous rigid-solid theory by Doinikov \cite{Doinikov1994a}. Qualitatively, we confirm this sign reversal, but quantitatively we find it to happen for $a = 4.7~\SImum$, which is nearly three times the value $1.6~\SImum$ predicted by the Doinikov model. Moreover, by including the temperature dependency of the viscosity, we find that at the smallest particle radius $a = 300$~nm, $\Phiac$  is an order of magnitude larger than predicted by the Doinikov model.

\vspace*{-3mm}

\subsection{The contrast factor for fluid particles}
\seclab{Phiac_fl}

\vspace*{-3mm}

We now replace the solid particle by a fluid particle, and study the following four examples of spherical microdroplets in fluids:
(a) An oil droplet in water,
(b) a water droplet in oil,
(c) a water droplet in air,  and
(d) an oil droplet in air,
with material parameters listed in \tabsref{Param_table}{Fluids_abcd_table}.
Cases (a) and (b) are chosen for their relevance to studies of droplets and emulsions \cite{Laurell2015, Liu2021, Pannacci2008}. Cases (c) and (d) are chosen for their relevance for aerosol studies \cite{Ran2015}.
In \figref{Phiac_fluid} we plot $\Phiac$ versus $\dels/a$  computed from \eqref{Phiac}
(blue full curve \qmarks{Winckelmann}) for these four cases and compare with the results for $\Phiac$ obtained by
Doinikov \cite{Doinikov1997b}  (red full curve \qmarks{Doinikov}),
Karlsen and Bruus \cite{Karlsen2015}  (purple full curve \qmarks{Karlsen}),
Settnes and Bruus \cite{Settnes2012}  (green dashed curve \qmarks{Settnes}), and
Gor'kov \cite{Gorkov1962}  (brown full curve \qmarks{Gor'kov}).
As for solid particles, we see that for large fluid particles, $a \gg \dels$, the four boundary-layer theories converge towards the ideal-fluid result by Gor'kov for all cases.

\begin{table}[t]
\caption{\tablab{Fluids_abcd_table} Parameters for the four combinations of fluid particles in fluids discussed in \secref{Phiac_fl} and shown in \figref{Phiac_fluid} for the cases (a) oil in water, (b) water in oil, (c) water in air, and (d) oil in air.}
\begin{ruledtabular}
\begin{tabular}{lcccc}
Case & (a)	& (b)	& (c)	& (d)	
\\ \hline
$\rhoOTi-1$ 	& $-$0.07 &  0.08 	&857 				& 794
\\
$\kapSOTi$ 		&1.17 	&  0.86 	& $6\times 10^{-5}$ & $7\times 10^{-5}$
\\
$\etaOTi$ 		& 67 	& 0.015 & 46				& $3.1\times 10^3$
\\
$B_c$ 			& 1.2 	& 9.5 	& $-$0.30 			& $-$0.30
\\
$B_t$ 			& $-$81 	& $-$63 	& 0.75 				& 0.75
\\
$B'_t$ 			&$-$63 	&  $-$81 	& $-$81 				& $-$63
\\
$\gamO-1$ 	&0.012 & 0.15 	&  0.40				& 0.40
\\
$\gamST$ 		& $38\:$mN$\:\SIm^{-1}$ 	& $38\:$mN$\:\SIm^{-1}$ 	
				& $72\:$mN$\:\SIm^{-1}$		& $22\:$mN$\:\SIm^{-1}$
\\
\end{tabular}
\end{ruledtabular}
\end{table}

Case (a), an oil droplet in water, is shown in \figref{Phiac_fluid}(a). Here, Karlsen predicted an approximate doubling of $\Phiac$ compared to the ideal-fluid theory. By adding microstreaming, Doinikov and Winckelmann also predict large variations from the ideal theory, but in contrast to the solid-particle cases, the two microstreaming models behave qualitatively different for small particles. We find that the sharp increase of $\Phiac$ for $\dels/a \gtrsim 2$ given by the Doinikov model is an artifact, which is due to the previously mentioned neglect of the tangential component $s_{1\theta}$ of the displacement $\sss_1$  in the Stokes terms in \eqsref{Second_order_no_slip_sl}{Second_order_no_slip_fl} combined with the effects of surface tension in the scattering coefficients. The downwards slope in the curve by Winckelmann for $\dels/a \gtrsim 2$ stems from a combination of surface tension and the inclusion of streaming inside the fluid droplet. The temperature dependency of the viscosities only make slight quantitative contributions to the behavior of the curve by Winckelmann.

In case (b), a water droplet suspended in oil shown in \figref{Phiac_fluid}(b), the Doinikov and the Winckelmann model have a similar behavior as in case (a), namely an increase and a decrease, respectively, of $\Phiac$ for $\dels/a \gtrsim 2$. The stronger decrease of $\Phiac$ in the Winckelmann model setting in at $\dels/a \gtrsim 5$ is due to inner streaming not included in the Doinikov model. In contrast to case (a), the microstreaming models in case (b) exhibit a sign change of $\Phiac$, and for both models this happens at nearly the same particle size $a \approx\dels$. The sharp increase in $\Phiac$ predicted by Doinikov is again due to the neglect of some of the Stokes terms as discussed before, while surface tension plays a minor role for the curves by Doinikov and Winckelmann in case (b). Like in case (a), temperature dependencies in the viscosities lead to quantitative changes to the curve by Winckelmann. Comparing \figref{Phiac_solid}(b), \figref{Phiac_solid}(d), and \figref{Phiac_fluid}(b), we note that the Winckelmann model predicts the sign change in $\Phiac$ to happen at $a \approx\dels$ for polystyrene, copper, and water particles in oil. Finally, we note that also the Karlsen-model predicts a sign change of $\Phiac$ for a water droplet in oil, but this happens for much smaller particles of size $a \approx 0.04\dels$, not shown. Clearly, microstreaming is dominating the behavior of $\FFFrad$ for microparticles with $a \lesssim \dels$ and a density ratio $\rhoOTi$ deviating sufficiently from unity.

Case (c) of a water droplet in air shown in \figref{Phiac_fluid}(c) was studied by Karlsen \cite{Karlsen2015}, who found a remarkable sign reversal in $\Phiac$. For an acoustic frequency of $1~\SIkHz$, this sign reversal happens for relatively large droplets of radius $a \approx 1.6~\SImum$ surrounded by a thick boundary layer with $\dels \approx 45 a$. Adding microstreaming, a corresponding sign reversal is also found by both Doinikov and Winckelmann, albeit for significantly larger particles, $a \approx 9.7~\SImum$. In this case, the discrepancy between Doinikov and Winckelmann is minute, due to a small value of $B_c = -0.3$ and a large value of $\etaOTi = 46$, making this case similar to the solid-particle-in-air case in \figref{Phiac_solid}(c).

Case (d) with an oil droplet in air shown in \figref{Phiac_fluid}(d) exhibits similar qualitative behavior as for the water droplet in case (c) for four of the five models. The exception is the Karlsen model, where no sign reversal occurs in  $\Phiac$ for the oil droplet, in contrast to the sign reversal in $\Phiac$ at $\delta \approx 45 a$ for the water droplet. The microstreaming models by Doinikov and Winckelmann predict a sign reversal in $\Phiac$ for relatively large oil droplets with $a\approx 9.9~\SImum$ and $a\approx 9.5~\SImum$, respectively, close to the value $a\approx 9.7~\SImum$ found above for water droplets.

We do not cover the case of gas bubbles in a surrounding fluids for several reasons. Bubble dynamics often involves nonlinearities and cavitation effects \cite{Eames2008, Wu2008a}, as well as gas diffusion across the interface \cite{Epstein1950}, effects that are not included in our work. Also, in microfluidic experiments, bubbles are typically stabilized by surface coatings, which alters the bubble dynamics \cite{Eames2008} and requires the addition of extra terms in the boundary conditions to describe the surface coating. But we do expect our model to be accurate for large bubbles in weak acoustic fields, where surface effects and nonlinearities are negligible.

\vspace*{-3mm}

\subsection{Rigidified droplets}
\seclab{rigidified_droplet}

\begin{figure}[b]
\centering
\includegraphics[width=0.95\columnwidth]{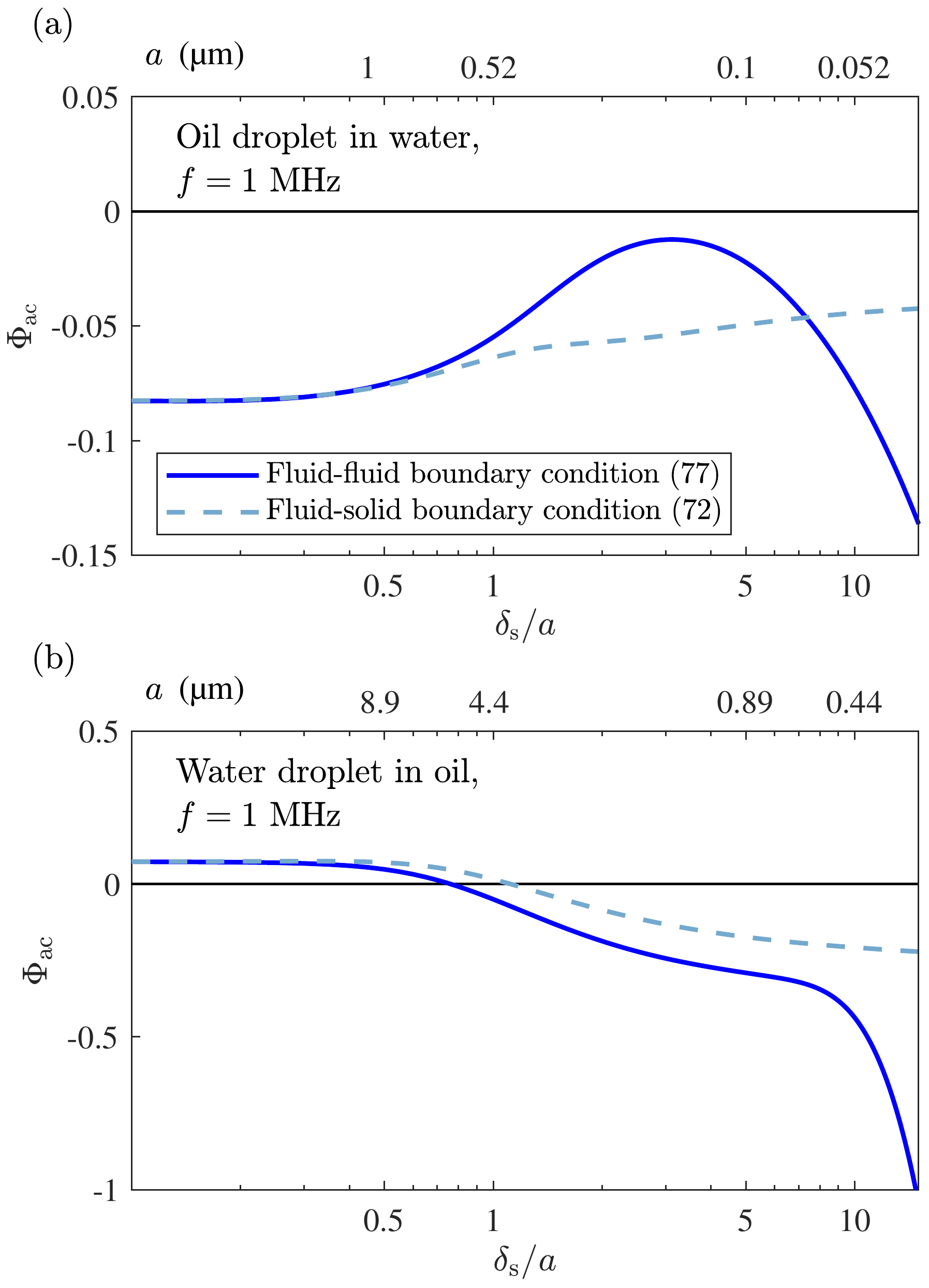}
\caption{\figlab{rigified_interface} The acoustic contrast factor $\Phiac$ from \eqref{Phiac} plotted versus $\dels/a$ computed for a fluid particle in a fluid using the fluid-fluid boundary condition~\eqnoref{Second_order_no_slip_fl} (full curve) and the fluid-solid boundary condition~\eqnoref{Second_order_no_slip_sl} (dashed curve). (a) An oil droplet in water at frequency $f=1$ MHz. (b) A water droplet in oil at $f=1$ MHz.}
\end{figure}

\vspace*{-3mm}

Impurities tend to collect at fluid-fluid interfaces, which can make the interface resemble that of a rigid boundary \cite{Maali2017}. It is also often desired to use surfactants to stabilize suspended droplets \cite{Yoon2018, Zembyla2019}. Our result for $\FFFrad$ in \eqref{Frad_longwave} for fluid particles in fluids, is derived for pure fluid-fluid interfaces under ideal conditions. A detailed description of impure interfaces and their complex dynamics is beyond the scope of this work, but in the following we discuss how to compute $\FFFrad$ on a fluid droplet with a rigidified interface in a simplified model. For a rigidified interface, which is strong enough to impose a no-slip boundary condition on the acoustic streaming, the fluid-fluid boundary condition~\eqnoref{Second_order_no_slip_fl} is replaced by the fluid-solid boundary condition~\eqnoref{Second_order_no_slip_sl}. This is equivalent to using the theory~\eqnoref{Frad_longwave} for solid particles, but with the scattering coefficients $\ain$ from \appref{ScatteringCoeffSolids} for solids particles replaced by $\ain$ from \appref{ScatteringCoeffFluids} for fluid particles with $i = 0,1,2,3$. In \figref{rigified_interface}, we show the result of computing $\Phiac$ from \eqref{Phiac} in this manner (dashed curves) compared to $\Phiac$ computed for a pure fluid-fluid interface (full curves) for the two cases \figref{rigified_interface}(a) an oil droplet in water and \figref{rigified_interface}(b) a water droplet in oil, corresponding to the pure fluid-fluid cases of \figref{Phiac_fluid}(a) and (b). It is seen that when assuming a rigidified interface, $\Phiac$ undergoes qualitative and quantitative changes in both cases. We note that the dashed curves in \figref{rigified_interface} for $\Phiac$ in rigidified droplets  have a quantitatively similar behavior to that found for $\Phiac$ in the Winckelmann model for solids shown in \figref{Phiac_solid}.

Furthermore, we remark that $\Phiac$ for rigidified oil in water (dashed curve) \figref{rigified_interface}(a) resembles $-\Phiac$ for polystyrene in water \figref{Phiac_solid}(a), which can be explained by $\kapSOTi-1 > 1$ for oil in water, whereas $\kapSOTi-1 < 1$ for polystyrene in water. In contrast,  we remark that $\Phiac$ for rigidified water in oil (dashed curve) \figref{rigified_interface}(b) resembles $+\Phiac$ for polystyrene in oil \figref{Phiac_solid}(b), because here $\kapSOTi-1 > 1$ in both cases. Finally, we note that the two remaining fluid-fluid cases studied in \figref{Phiac_fluid}(c) and (d) are only affected to a negligible degree by rigidified interfaces, and therefore they are not shown in this discussion.

\section{Conclusion}
\seclab{conclusion}
We have developed an extension of the model of the acoustic radiation force $\FFFrad$ presented by  Doinikov in 1997 \cite{Doinikov1997a, Doinikov1997, Doinikov1997b} for a rigid and for a fluid spherical particle suspended in a thermoviscous fluid.  Our extension comprises the inclusion of (1) elastic instead of rigid solid particles, (2) temperature- and density-dependent material parameters, in particular the viscosity, (3) the tangential part of the Stokes drift in the boundary condition of the acoustic streaming on the particle surface, and (4) inner streaming in a fluid particle. Using the method of Karlsen and Bruus~\cite{Karlsen2015} in \secref{FirstOrder} to compute the first-order fields and that of Doinikov \cite{Doinikov1994, Doinikov1994a} in \secref{SecondOrder} to compute the second-order fields, we arrive in \eqref{Frad_form} at $\FFFrad = \FFFradII + \FFFradin$, where $\FFFradin$ is the trivial Stokes drag force due to the streaming velocity of the incoming field, and $\FFFradII$ is the force in terms of the force coefficients $D_n$ due to time-averaged products of the first-order acoustic fields given by \eqsref{F1_form_sl}{F1_form_fl} for solid and fluid particles, respectively. The main result of our work is the long-wavelength limit of $\FFFrad$ presented in  \eqref{Frad_longwave}.

In \secref{D0_D1_limits} we compute the force coefficients $D_0$ and $D_1$ analytically in the limit of very thin and very thick boundary layers, and compare our results with results in the literature. We verify that we obtain the same result as Doinikov~\cite{Doinikov1997, Doinikov1997a, Doinikov1997b} when the differences in the above-mentioned model assumptions are negligible. Using \eqsref{F1_familiar_form}{f0f1_D0D1} allows for direct comparison to the models without microstreaming, namely the thermoviscous model by Karlsen and Bruus~\cite{Karlsen2015}, the viscous model by Settnes and Bruus~\cite{Settnes2012}, and the ideal-fluid model by Gor'kov~\cite{Gorkov1962}. We recover the results of these models for particles much larger than the acoustic boundary-layer width, but find large deviations for certain parameter values (in particular for $\rhoOTi \gg 1$) in the opposite limit. We thus conclude, in agreement with Doinikov for thermoviscous fluids and with recent numerical studies by Baasch, Pavlic, and Dual~\cite{Baasch2019} for viscous fluids, that microstreaming can dominate the acoustic radiation force $\FFFrad$ for small particles.

Further comparisons between the above five models are carried out in \secsref{Phiac_sl}{Phiac_fl} for the important special case of a standing incident wave by plotting the acoustic contrast factor $\Phiac$ of \eqref{Phiac} versus the normalized viscous boundary layer thickness $\dels/a$ for selected choices of particles and fluids. For the four solid-in-fluid examples (a) polystyrene in water, (b) polystyrene in oil, (c) polystyrene in air, and (d) copper in oil, shown in \figref{Phiac_solid}, the two main observations are: (1) The large values in (c) and (d) of the relative density contrast $\rhoOTi-1$ result in large thermoviscous deviations from the ideal-fluid Gor'kov model, and even in a sign reversal if microstreaming is included (the Doinikov and the Winckelmann models). (2) The two models including microstreaming exhibit the same qualitative trends, but quantitatively they differ significantly as shown for solid particles in oil in (b) and (d), where the Winckelmann model predicts a sign reversal for relatively large particle radii due to the inclusion of the temperature-dependency of the viscosity. For the four fluid-in-fluid examples (a) oil in water, (b) water in oil, (c) water in air, and (d) oil in air, shown in \figref{Phiac_fluid}, the two main observations are: (3) Each model predicts largely the same $\Phiac$ for droplets in air (c) and (d) as for solid particles in air, except for the Karlsen model. (4) The Winckelmann model deviates qualitatively from the Doinikov model for droplets in liquids (a) and (b), due to our inclusion of the tangential Stokes drift boundary condition and the inner streaming.

Finally, we have in \secref{rigidified_droplet} briefly addressed how to analyze the experimentally important case of droplets with rigidified surfaces due to impurities or surfactants, by combining fluid-fluid scattering coefficients with a second-order no-slip boundary condition. In \figref{rigified_interface} is shown for oil-in-water and water-in-oil systems, how the response of a rigidified droplet resembles that of an elastic solid particle.

We have extended the basic analytical theory of the acoustic radiation force on a single suspended particle by including temperature- and density dependent material parameters and by taking inner streaming in droplets into account. We have shown specific examples where these effects are important. We hope that our analysis will inspire related experimental efforts in the fields of microscale acoustofluidics, acoustic levitation, and aerosol dynamics.

\vspace*{-5mm}

\appendix\begin{widetext}

\section{Integrals of Legendre polynomials}
\seclab{Legendre_integrals}

\vspace*{-1mm}

Below, we list a series of useful integrals containing products of Legendre polynomials and their derivatives \cite{Gradshteyn1994},

 \bsuba{legendre_23}
 \bal
 \eqlab{legendre_double_1}
 \int_0^\pi P_n(\cos\theta)P_m(\cos\theta)\sin\theta\, \dm\theta &= \frac{2}{2n+1}\delta_{mn}
 \\
 \eqlab{legendre_double_2}
 \int_0^\pi \pp_\theta P_n(\cos\theta)\pp_\theta P_m(\cos\theta)\sin\theta\, \dm\theta &= \frac{2n(n+1)}{2n+1}\delta_{mn}
 \\
 \eqlab{legendre_triple_1}
 \int_0^\pi P_n(\cos\theta)P_m(\cos\theta)\cos\theta\sin\theta\, \dm\theta &= \begin{cases}
      \frac{2(n+1)}{(2n+1)(2n+3)} & m=n+1\\
      \frac{2n}{(2n-1)(2n+1)} & m=n-1\\
      0 & \text{otherwise}
 \end{cases}
 \\
 \eqlab{legendre_triple_2}
 \int_0^\pi P_n(\cos\theta)\pp_\theta P_m(\cos\theta)\sin^2\theta\, \dm\theta &=
 \begin{cases}
      -\frac{2(n+1)(n+2)}{(2n+1)(2n+3)} & m=n+1\\
      \frac{2(n-1)n}{(2n-1)(2n+1)} & m=n-1\\
      0 & \text{otherwise}
 \end{cases}
 \\
 \eqlab{legendre_triple_3}
 \int_0^\pi \pp_\theta P_n(\cos\theta)\pp_\theta P_m(\cos\theta)\cos\theta\sin\theta\, \dm\theta &= \begin{cases}
      \frac{2n(n+1)(n+2)}{(2n+1)(2n+3)} & m=n+1\\
      \frac{2(n-1)n(n+1)}{(2n-1)(2n+1)} & m=n-1\\
      0 & \text{otherwise}
 \end{cases}
 \\
 \eqlab{legendre_triple_4}
 \int_0^\pi \pp_\theta P_n(\cos\theta)\pp_\theta^2 P_m(\cos\theta)\sin^2\theta\, \dm\theta &=
 \begin{cases}
      -\frac{2n(n+1)(n+2)^2}{(2n+1)(2n+3)} & m=n+1\\
      \frac{2(n-1)^2n(n+1)}{(2n-1)(2n+1)} & m=n-1\\
      0 & \text{otherwise}
 \end{cases}
 \eal
 \esuba

\section{Scattering coefficients for solids}
\seclab{ScatteringCoeffSolids}

Here, we present the analytical expressions of the acoustic scattering coefficients $\ain$, $i=0,1,2,3$ and $n=0,1,2$ for solid particles. Only the coefficients that contribute to $\FFFrad$ to leading order in $\xO$ are given explicitly, whereas only the order in $\xO$ is stated for the remaining coefficients, which are $\atsc1$ and $\atsc2$,
 \beq{alpha0n_sl}
 \alpha_{0,n} = 1 ,  		
 \eeq
 \bal\eqlab{alpha0_sl}
 \acsc0 = -\frac{\ii \xc^3}{3}\frac{(1-\kapSOTi)+3(\gamO-1)\Big[\Big(1-\frac{\alphapOTi}{\rhoTi_0 \cpOTi}\Big)\Big(1-\frac{\chi'\alphapOTi}{\rhoTi_0 \cpOTi}\Big)-\frac{4}{3}\frac{\chi' \alphapOTi \kapSOTi}{\cpOTi}\frac{\cTO^{\prime 2}}{\cO^2}\Big(1-\frac{\alphapOTi}{\rhoTi_0 \cpOTi \kapSOTi}\Big)\Big]H(\xt,\xtp)}{1+4(\gamO-1)\frac{\chi' \alphapOTi^2\cTO^{\prime 2}}{\rhoTi_0 \cpOTi^2\cO^2}H(\xt,\xtp)} ,
 \eal
 \bal\eqlab{alphat0_sl}
 \atsc0 =-\frac{\xc^2 \xt}{\ii+\xt}\frac{(\gamO-1)\Big(1-\frac{\alphapOTi}{\rhoTi_0 \cpOTi}\Big)H(\xt,\xtp)\ee^{-\ii\xt}}{1+4(\gamO-1)\frac{\chi' \alphapOTi^2\cTO^{\prime 2}}{\rhoTi_0\cpOTi^2 \cO^2}H(\xt,\xtp)} ,
 \eal
 \bal\eqlab{scatter_coeff_1st_order_solid_n1}
 \acsc1 = -\frac{\ii\xc^3}{3}\frac{(\rhoTi_0-1)(3-3\ii \xs-\xs^2)}{\xs^2(2\rhoTi_0+1)-9(1-\ii \xs)}, \quad
 \atsc1 \sim \calO(x_0^3), \quad
 \assc1 =\frac{\ii\xc(\rhoTi_0-1)\xs^2 \ee^{-\ii\xs}}{\xs^2(2\rhoTi_0+1)-9(1-\ii \xs)},
 \eal
 \bal\eqlab{scatter_coeff_1st_order_solid_n2}
 \acsc2 = \frac{2\ii\xc^5}{135 \xs^2}\frac{15(1-\ii\xs)-6\xs^2+\ii \xs^3}{\ii\xs-1},\quad
 \atsc2 \sim \calO(x_0^4), \quad
 \assc2 = -\frac{\ii\xc^2}{9}\frac{\xs \ee^{-\ii\xs}}{\ii\xs-1},
 \eal
 \bal\eqlab{function_H}
 \text{with the definition }\;
 H(\xt,\xtp) = \frac{1}{\xt^2}\bigg[\frac{1}{1-\ii\xt}
 -\frac{1}{\kthOTi}\frac{\tan\xtp}{\tan\xtp-\xtp}\bigg]^{-1}.
 \eal

\section{Scattering coefficients for fluids}
\seclab{ScatteringCoeffFluids}

\vspace*{-3mm}

Here, we present the analytical expressions of the acoustic scattering coefficients $\ain$, $i=0,1,2,3,4,5,6$ and $n=0,1,2$ for fluid particles. As above, only the coefficients that contribute to $\FFFrad$ to leading order in $\xO$ are given explicitly, while only the order in $\xO$ is stated for the remaining coefficients, which are $\atsc1$, $\atp1$, $\atsc2$, and $\atp2$,
 \beq{alpha0n_fl}
 \alpha_{0,n} = 1 	,	
 \eeq
 \bal\eqlab{alpha0_fl}
 \acsc0 = -\frac{\ii\xc^3}{3}\frac{1-\kapSOTi+3(\gamO-1)\Big(1-\frac{\alphapOTi}{\rhoOTi \cpOTi}\Big)^2 H(\xt,\xtp)-\frac{2\gamST \kapsO'}{3a}\Big[1+3(\gamO-1)\Big(1- \frac{\alphapOTi^2}{\rhoOTi^2 \cpOTi^2 \kapSOTi}\Big)H(\xt,\xtp) \Big]}{1-\frac{2\gamST \kapsO'}{3a}\Big[1-3(\gamO-1)\frac{\alphapOTi^2}{\rhoOTi^2 \cpOTi^2 \kapSOTi}H(\xt,\xtp) \Big]} ,
 \eal
 \bal\eqlab{alphat0_fl_and_alpha0m_fl}
 \atsc0 =-\frac{\xc^2 \xt}{\ii+\xt}\frac{(\gamO-1)\Big[\Big(1-\frac{\alphapOTi}{\rhoOTi\cpOTi}\Big)-\frac{2\gamST \kapsO'}{3a}\Big]H(\xt,\xtp)\ee^{-\ii\xt}}{1-\frac{2\gamST \kapsO'}{3a}
 \Big[1-3(\gamO-1)\frac{\alphapOTi^2}{\rhoOTi^2 \cpOTi^2 \kapSOTi}H(\xt,\xtp) \Big]},
 \quad
 \acp0 = \frac{\rhoOTi^{-1}+(\gamO-1)\frac{2\gamST \kapsO'}{a}\frac{\alphapOTi}{\rhoOTi^2 \cpOTi \kapSOTi}H(\xt,\xtp)}{1-\frac{2\gamST \kapsO'}{3a}\Big[1-3(\gamO-1) \frac{\alphapOTi^2}{\rhoOTi^2 \cpOTi^2 \kapSOTi}H(\xt,\xtp) \Big]} ,
 \eal
 \bal\eqlab{alphat0m_fl}
 \atp0 = \frac{\xc^2\xtp}{\sin\xtp-\xtp\cos\xtp}\frac{(\gamO-1)\frac{\alphapOTi}{\rhoOTi\cpOTi}\Big[\Big(1-\frac{\alphapOTi}{\rhoOTi\cpOTi}\Big)-\frac{2\gamST \kapsO'}{3a}\Big]H(\xt,\xtp)}{1-\frac{2\gamST \kapsO'}{3a}\Big[1-3(\gamO-1)\frac{\alphapOTi^2}{\rhoOTi^2 \cpOTi^2 \kapSOTi}H(\xt,\xtp) \Big]} ,
 \eal
 \bal\eqlab{alpha1_alphat1_and_beta1_fl}
 \acsc1 = \frac{\ii\xc^3}{3}\frac{(\rhoOTi-1)[1+F(\xs,\xsp)-G(\xs)]}{(2\rhoOTi+1)[1+F(\xs,\xsp)]-3G(\xs)},
 \qquad
 \atsc1 \sim \calO(x_0^3),
 \qquad
 \assc1 = \frac{\ii\xc(\rhoOTi-1)\ee^{-\ii\xs}}{(2\rhoOTi+1)[1+F(\xs,\xsp)]-3G(\xs)} ,
 \eal
 \bal\eqlab{alpha1m_and_beta1m_fl}
 \acp1 = \frac{3\xc}{\xcp}\frac{1-\frac{2+\rhoTi_0}{3\rhoTi_0}G(\xs)+F(\xs,\xsp)\Big[1+\frac{4}{\rhoTi_0\xs^2}(\rhoTi_0-1)(\etaOTi-1)\Big]}{(2\rhoTi_0+1)[1+F(\xs,\xsp)]-3G(\xs)}, \quad
  \atp1 \sim \calO(x_0^3) ,
 \eal
 \bal\eqlab{beta1_fl}
  \asp1 = \frac{\xc}{\xsp j_2(\xsp)}\frac{(\rhoTi_0-1)F(\xs,\xsp)}{(2\rhoTi_0+1)[1+F(\xs,\xsp)]-3G(\xs)} ,
 \eal
 \bal\eqlab{alpha2_fl}
 \acsc2= -\frac{\ii\xc^5\xs}{45}\frac{32(\etaOTi-1)^2-2\xs^2(\etaOTi-1)(\rhoTi_0-1)+[10(\etaOTi-1)-(\rhoTi_0-1)\xs^2]\kappa_1(\xs,\xsp)+\frac{2\ii\gamST}{\etaO a \omega}[8(\etaOTi-1)+4\kappa_1(\xs,\xsp)]}{X_1(\xs)+\etaOTi X_2(\xs,\xsp)+\etaOTi^2 X_3(\xs,\xsp)+\frac{2\ii\gamST}{\etaO a \omega}X_4(\xs,\xsp)} ,
 \eal
 \bal\eqlab{beta2_fl}
 \assc2 = \frac{\xc^2}{6h_3(\xs)}\frac{(\etaOTi-1)\Big[32(\etaOTi-1)-2(\rhoTi_0-2)\xs^2\Big]-\etaOTi \Big[10(\etaOTi-1)-(\rhoTi_0-1)\xs^2 \Big]\kappa_2(\xsp)+\frac{2\ii\gamST}{\etaO a \omega}
 [8(\etaOTi-1)-4\etaOTi\kappa_2(\xsp)]}{X_1(\xs)+\etaOTi X_2(\xs,\xsp)+\etaOTi^2 X_3(\xs,\xsp)+
 \frac{2\ii\gamST}{\etaO a \omega}X_4(\xs,\xsp)} ,
 \eal
 \bal\eqlab{alpha2m_fl}
 \acp2 = \frac{5\xc^2\xs^2}{2\xc^{\prime 2}\xsp}\frac{\xs\xsp \kappa_1(\xs,\xsp)+2(\etaOTi-1)\kappa_3(\xs,\xsp)+\Big[3\xs^2(\rhoTi_0-1)-12\frac{2\ii\gamST}{\etaO a \omega}\Big]\kappa_4(\xs,\xsp)}{X_1(\xs)+\etaOTi X_2(\xs,\xsp)+\etaOTi^2 X_3(\xs,\xsp)+\frac{2\ii\gamST}{\etaO a \omega}
 X_4(\xs,\xsp)} ,
 \eal
 \bal\eqlab{beta2m_fl}
 \asp2 = \frac{\xc^2\xs^2}{6\xsp j_3(\xsp)}\frac{2\xs(\etaOTi-1)+\Big[(\rhoTi_0-1)\xs^2-10(\etaOTi-1)-4\frac{2\ii\gamST}{\etaO a \omega}\Big]\frac{h_2(\xs)}{h_3(\xs)}}{X_1(\xs)+\etaOTi X_2(\xs,\xsp)+\etaOTi^2 X_3(\xs,\xsp)+\frac{2\ii\gamST}{\etaO a \omega} X_4(\xs,\xsp)} ,
 \eal
 \bal\eqlab{thermal_coeff_scaling_n2_fl}
 \atsc2 \sim x_0^4,
 \qquad
 \atp2 \sim x_0^4 .
 \eal

Alongside $H(\xt,\xtp)$ from \appref{ScatteringCoeffSolids}, we have here defined the functions
\bsubal{functions_for_scatter_fl}
\eqlab{G_function_and_F_function}
G(\xs)&=\frac{3}{\xs}\bigg(\frac{1}{\xs}-\ii\bigg), \qquad
F(\xs,\xsp)=\frac{1-\ii\xs}{2(1-\etaOTi)+\frac{\etaOTi \xs^{\prime 2}(\tan\xsp-\xsp)}{(3-\xs^{\prime 2})\tan\xsp-3\xsp}}, \\
\eqlab{X1}
X_1(\xs)&=(240-30\xs^2+\xs^4)\frac{h_2(\xs)}{h_3(\xs)}-2\xs(\xs^2+24) , \\
\eqlab{X2}
X_2(\xs,\xsp)&=\frac{\xsp j_2(\xsp)}{j_3(\xsp)}\Big[\Big(75-\frac{15}{2}\xs^2\Big)\frac{h_2(\xs)}{h_3(\xs)}-\xs(\xs^2+15)\Big]\nn\\
&\quad +\bigg[\bigg(\frac{3}{2}\xs^{\prime 2}+30\bigg)\xs^2+15\xs^{\prime 2}-480\bigg]\frac{h_2(\xs)}{h_3(\xs)}+2\xs\Big(\xs^2-\frac{3}{2}\xs^{\prime 2}+48\Big), \\
\eqlab{X3}
X_3(\xs,\xsp)&=\frac{3[5h_2(\xs)-\xs h_3(\xs)]}{2h_3(\xs)}\frac{(\xs^{\prime 3}-10\xsp)j_2(\xsp)+(32-2\xs^{\prime 2})j_3(\xsp)}{j_3(\xsp)}, \\
\eqlab{X4}
X_4(\xs,\xsp)&=6\bigg\{\Big[10(\etaOTi-1)-\xs^2\Big]\frac{h_2(\xs)}{h_3(\xs)}-2\xs(\etaOTi-1)+\etaOTi  \frac{\xs h_3(\xs)-5h_2(\xs)}{h_3(\xs)}\frac{\xsp j_2(\xsp)}{j_3(\xsp)}\bigg\},  \\
\eqlab{kappa1_kappa2_kappa3}
\kappa_1(\xs,\xsp)&=\frac{h_2(\xs)\xs}{h_3(\xs)}-\etaOTi \frac{\xsp j_2(\xsp)}{j_3(\xsp)}, \quad
\quad\kappa_2(\xsp) = \frac{j_2(\xsp)\xsp}{j_3(\xsp)},  \\
\eqlab{kappa4_kappa5}
\kappa_3(\xs,\xsp)&=\frac{[3\xs h_3(\xs)-15 h_2(\xs)]j_2(\xsp)+[\xs h_3(\xs)+3 h_2(\xs)]\xsp j_3(\xsp)}{h_3(\xs)j_3(\xsp)}, \quad \kappa_4(\xs,\xsp)= \frac{h_2(\xs)j_2(\xsp)}{h_3(\xs)j_3(\xsp)}.
\esubal

\section{Derivation of \eqref{c41_and_c51_sl} involving \textit{\textbf{c}}$_{\textbf{41}}$ and \textit{\textbf{c}}$_{\textbf{51}}$}
\seclab{c41c51}
The details of the computation leading to \eqref{c41_and_c51_sl} involving $c_{41}$ and $c_{51}$ are given here. First, the expressions for $c_{41}$ and $c_{51}$ from \eqsref{c4n}{c5n} are used to obtain,
 \bal\eqlab{c41_and_c51_sl_app_main}
 &-\eee_z 4\pi\rhoO\nuO \bigg(\frac{3}{2}c_{51}-\frac{1}{4}c_{41}\bigg)
 =
 -\eee_z 2\pi\rhoO\nuO \int_1^\infty \Big[3\xi-\xi^{-1}\Big]
 \Big[\chi_{r1}(\xi)-2\chi_{\theta 1}(\xi)\Big]
 +3\chi_{\theta 1}(\xi)\Big[\xi-\xi^{-1}\Big]\:\dm\xi.
 \eal
Then $\chi_{r1}(\xi)$ and $\chi_{\theta n}(\xi)$ are inserted from \eqsref{chir}{chit} for $n=1$, and the second term in the integrand yields
 \bal\eqlab{c41_and_c51_sl_app_result_1}
 &-\eee_z\: 2\pi\rhoO\nuO \int_1^\infty 3\chi_{\theta 1}(\xi)(\xi-\xi^{-1})\dm\xi =
 \\ \nn
 &\hspace*{10mm} -\eee_z 3\pi\rhoO a^3 \int_1^\infty \int_0^\pi
 \frac{1}{2} \big[\xi - \xi^{-1}\big]  \bigg\{ \div \Big\langle \vvv_1 \vvv_1
 -\nuI\Big[\nablabf \vvv_1+(\nablabf \vvv_1)^\textsf{T}\Big]
 -\Big[\nuIB-\tfrac{2}{3}\nuI\Big] (\div \vvv_1) \Imat
 \Big\rangle_\mr{nii}\bigg\} \cdot \eee_\theta \sin^2\theta \, \dm \theta \dm \xi .
 \eal
To ease the notation in the following, we introduce
 \bsuba{c41_and_c51_sl_app_1st_term}
 \bal
 \XXX = \Big\langle \vvv_1 \vvv_1
 -\nuI\Big[\nablabf \vvv_1+(\nablabf \vvv_1)^\textsf{T}\Big]
 -\Big[\nuIB-\tfrac{2}{3}\nuI\Big] (\div \vvv_1) \Imat \Big\rangle_\mr{nii}.
 \eal
Using this expression for $\XXX$, the first term of the integrand in \eqref{c41_and_c51_sl_app_main} becomes
 \bal
 \eqlab{c41_and_c51_sl_app_last_term}
 &-\eee_z 2\pi\rhoO\nuO \int_1^\infty (3\xi-\xi^{-1})[\chi_{r1}(\xi)-2\chi_{\theta 1}(\xi)]\dm\xi
 \nn\\
 & \hspace*{7mm} =  -\eee_z 2\pi\rhoO\nuO \frac{a^3}{2\nuO}\int_1^\infty \int_0^\pi (3\xi-\xi^{-1}) \,
  \big(\div \XXX\big)  \! \cdot (\eee_r \cos\theta - \eee_\theta \sin\theta) \sin\theta \, \dm\theta \dm\xi
 \nn\\
 & \hspace*{7mm} =  -\eee_z \pi\rhoO a^3\int_1^\infty \int_0^\pi
 (3\xi^{-1}-\xi^{-3}) \, \big(\div \XXX\big)  \! \cdot \eee_z \, \xi^2 \sin\theta \, \dm\theta \dm\xi
 \nn\\
 & \hspace*{7mm} =  -\eee_z \pi\rhoO a^3 \int_1^\infty \int_0^\pi
 \Big\{\div\big[(3\xi^{-1}-\xi^{-3}) \XXX\big] + \frac{3}{a} \big( \xi^{-2} - \xi^{-4} \big)\,\eee_r
 \cdot \XXX \Big\}  \! \cdot \eee_z \, \xi^2 \sin\theta \: \dm\theta\: \dm\xi,
 \eal
and by inserting the definition of $\XXX$ in the second term of the integrand in \eqref{c41_and_c51_sl_app_last_term}, we obtain
 \bal\eqlab{c41_and_c51_sl_app_result_2}
 &  -\eee_z \pi\rhoO a^3 \int_1^\infty \int_0^\pi
 \Big\{\frac{3}{a} \big( \xi^{-2} - \xi^{-4} \big)\,\eee_r
 \cdot \XXX \Big\}  \! \cdot \eee_z \, \xi^2 \sin\theta \, \dm\theta \dm\xi
 \\ \nn
 & \hspace*{7mm} = -\eee_z 3\pi \rhoO a^2 \int_1^\infty \int_0^\pi \big[1-\xi^{-2}\big]\; \eee_r \cdot \Big\langle(\vvv_1 \vvv_1)
 -\nuI\Big[\nablabf \vvv_1+(\nablabf \vvv_1)^\textsf{T}\Big]
 -[\nuIB-\tfrac{2}{3}\nuI](\div \vvv_1) \Imat \Big\rangle_\mr{nii}\! \cdot \eee_z \sin\theta \, \dm \theta \dm \xi.
 \eal
Lastly, using Gauss's law, we evaluate the divergence term of the integrand in \eqref{c41_and_c51_sl_app_last_term},
 \bal\eqlab{c41_and_c51_sl_app_result_3}
 & -\eee_z \pi\rhoO a^3\int_1^\infty \int_0^\pi
 \Big\{\div\big[(3\xi^{-1}-\xi^{-3}) \XXX\big] \Big\}  \! \cdot \eee_z \, \xi^2 \sin\theta \, \dm\theta \dm\xi
 \nn\\
 & \hspace*{7mm} = -\eee_z \frac{\rhoO}{2}\int_1^\infty \int_0^\pi \int_0^{2\pi}
 \Big\{\div\big[(3\xi^{-1}-\xi^{-3}) \XXX\big] \Big\}  \! \cdot \eee_z \, (a\xi)^2 \sin\theta \, \dm\varphi \dm\theta \, a\dm\xi ,
 \nn\\
 & \hspace*{7mm} = - \frac{\rhoO}{2} \int_\Omega
 \div\big[(3\xi^{-1}-\xi^{-3}) \XXX\big]  \dm V ,
 \nn\\
 & \hspace*{7mm} = - \frac{\rhoO}{2} \oint_{\pp\Omega_0}
 (3\xi^{-1}-\xi^{-3}) \XXX \cdot \big(-\eee_r\big) \, \dm S
 - \frac{\rhoO}{2} \oint_{\pp\Omega_\infty}
 (3\xi^{-1}-\xi^{-3}) \XXX \cdot \eee_r \, \dm S,
 \nn\\
 & \hspace*{7mm} = \rhoO\oint_{\partial\Omega_0}\Big\langle \vvv_1 \vvv_1
 - \nuI\Big[\nablabf \vvv_1+(\nablabf \vvv_1)^\textsf{T}\Big]
 -\Big[\nuIB-\tfrac{2}{3}\nuI\Big](\div \vvv_1) \Imat \Big\rangle_\mr{nii} \! \cdot \eee_r \,\dm S .
 \eal
 \esuba
Here, we have introduced the volume $\Omega$ between the particle surface $\pp \Omega_0$ and a spherical surface $\pp \Omega_\infty$ centered at $r=0$ with a radius going to infinity. In the last line of \eqref{c41_and_c51_sl_app_result_3}, we have used that the integrand goes to zero on $\pp \Omega_\infty$, and the definition of $\XXX$ has been reinserted. With the three result from \eqssref{c41_and_c51_sl_app_result_1}{c41_and_c51_sl_app_result_2}{c41_and_c51_sl_app_result_3} inserted in \eqref{c41_and_c51_sl_app_main}, we obtain the result stated in \eqref{c41_and_c51_sl}.

\vspace*{-5mm}

\section{The second-order coefficients \textit{\textbf{S}}$_{\textit{\!\textbf{ik,n}}}$ for a solid particle in a fluid}
\seclab{S_coeff_sl}

\vspace*{-1mm}

The 16 $\Smn{ik}n$ coefficients that contribute to $D_n$ to leading order in $\xO$ are stated: $9$ coefficients for $n=0$ and $7$ for $n=1$. For the remaining $16$ coefficients in modes $n=0$ and $n=1$, we only state their order in $\xO$ here.

 \bal\eqlab{S_coeff_sl}
 \Smn{00}0 &= \frac{x_0^3}{3\xs^2}, \qquad
 \Smn{00}1 = \frac{x_0^3}{3\xs^2}, \qquad
 \Smn{0c}0 = \frac{2\ii}{3},  \qquad
 \Smn{0c}1 \sim \calO (1) , \qquad
 \Smn{0t}0 \sim \calO(\xO^2), \qquad
 \Smn{0t}1 \sim \calO(x_0), \nn\\
 \Smn{0s}0 &= -x_0^2\frac{2\ii (1+B_c)}{\xs^2}\ee^{-\xs},\qquad
 \Smn{0s}1 \sim \calO(x_0^3)  , \qquad
 \Smn{c0}0 = \frac{2\ii}{3} , \qquad
 \Smn{c0}1 = \frac{2\ii}{3} ,\qquad
 \Smn{cc}0 = \frac{6}{\xs^2 x_0^3}, \qquad
 \Smn{cc}1 = \frac{135}{\xs^2 x_0^5}, \nn\\
 \Smn{ct}0 &\sim \calO(\xO^{-1}) , \qquad
 \Smn{ct}1 \sim \calO(\xO^{-2}) , \nn
 \eal
 \bal
 \Smn{cs}0 &= \frac{1}{24\xs^4 x_0} \Big[(-\xs^7 + \xs^6 - 14\xs^5 + 18\xs^4 - 48\xs^3 - 96\xs^2 - 144\xs - 144)\ee^{-\xs} + E_1(\xs)\xs^6(\xs^2 + 12) \Big] ,
 \nn\\
 \Smn{cs}1 &= -\frac{3\ii}{4\xs^5 x_0^2} \Big[(-\xs^7 + \xs^6 - 2\xs^5 + 6\xs^4 + 48\xs^3 + 168\xs^2 + 360\xs + 360)\ee^{-\xs} + E_1(\xs)\xs^8 \Big], \nn\\
 \Smn{t0}0 &= \frac{2\ii x_0}{3\xtO} \ee^{\ii\xtO}, \qquad
 \Smn{t0}1 \sim \calO(\xO^2), \nn\\
 \Smn{tc}0 &= \frac{1}{24 \xtO \xs^2 x_0^2} \Big\{-\big[\xtO^4\Bt
 + (\xs^2 + 18\Bt + 6)\xtO^2
 + 12\xs^2\big]\xtO^4 E_1(-\ii\xtO)+ \Big[\xtO^7\Bt\ii + \xtO^6\Bt
 + (\xs^2\ii + 16\ii\Bt + 6\ii)\xtO^5
 \nn\\
 &\quad + (\xs^2 + 12\Bt + 6)\xtO^4 + (10\ii\xs^2 - 12\ii\Bt - 12\ii)\xtO^3 + (6\xs^2 + 12\Bt - 36)\xtO^2 - 144\ii\xtO + 144 \Big] \ee^{\ii\xtO} \Big\} ,
 \nn\\
 \Smn{tc}1 &\sim \calO(\xO^{-3}), \qquad
 \Smn{tt}0 \sim \calO(1) , \qquad
 \Smn{tt}1 \sim \calO(1) , \nn\\
 \Smn{ts}0 &= \frac{1}{24\xs^4 \xtO} \Big\{ \big[\xtO^4\Bt + \big((\Bt + 1)\xs^2 + 18\Bt + 6\big)\xtO^2 + \xs^4 + 12\xs^2\big](\xs^2 + \xtO^2)^2 E_1(\xs-\ii\xtO)
 \nn\\
 & \quad + \Big[ -\xtO^7\Bt\ii - (\xs + 1)\Bt\xtO^6 + \big((-2\ii\Bt - \ii)\xs^2 + 2\ii\Bt\xs - 16\ii\Bt - 6\ii\big)\xtO^5
 \nn\\
 &\quad + \big((-2\Bt - 1)\xs^3 - \xs^2 + (-12\Bt - 6)\xs - 12\Bt - 6\big)\xtO^4
 \nn\\
 &\quad + \big((-\Bt\ii - 2\ii)\xs^4 + (2\ii\Bt + 2\ii)\xs^3 + (-24\ii\Bt - 16\ii)\xs^2 + (12\ii\Bt + 12\ii)\xs + 12\ii\Bt + 12\ii\big)\xtO^3
 \nn\\
 &\quad + \big((-\Bt - 2)\xs^5 + \xs^4\Bt + (-20\Bt - 12)\xs^3 - 24\xs^2\Bt + (-12\Bt + 36)\xs - 12\Bt + 36\big)\xtO^2
 \nn\\
 &\quad + \big(-\xs^6\ii + 2\ii\xs^5 - 18\ii\xs^4 + 48\ii\xs^3 + 96\ii\xs^2 + 144\ii\xs + 144\ii\big)\xtO
 \nn\\
 &\quad - \xs^7 + \xs^6 - 14\xs^5 + 18\xs^4 - 48\xs^3 - 96\xs^2 - 144\xs - 144 \Big] \ee^{-\xs+\ii\xtO} \Big\},
 \nn\\
 \Smn{ts}1 &\sim \calO(1) , \qquad
 \Smn{s0}0 = 0 , \qquad
 \Smn{s0}1 = x_0^2\frac{6\ii }{5\xs^2}\ee^{\ii \xs} , \qquad
 \Smn{sc}0 = 0 ,
 \nn\\
 \Smn{sc}1 &= \frac{1}{32\xs^4 x_0^3} \Big[ (\xs^{10} + 18\xs^8)E_1( -\ii\xs) - \Big(\xs^9\ii + 16\ii\xs^7 + \xs^8 - 12\ii\xs^5 + 12\xs^6 - 288\ii\xs^3 + 12\xs^4 + 4320\ii\xs \nn\\
 &\quad + 1728\xs^2 - 4320 \Big)\ee^{\ii\xs}\Big] ,
 \nn\\
 \Smn{st}0 &= 0, \qquad
 \Smn{st}1 \sim \calO(1) , \qquad
 \Smn{ss}0 = 0 , \nn\\
 \Smn{ss}1 &= -\frac{\ii}{2}\xs E_1(\xs-\ii\xs) (\xs^2 + 9)+ \frac{1}{\xs^7}\ee^{(-1+\ii)\xs}\bigg[\frac{1}{4}(-1 + \ii)\xs^9 + \frac{1}{4}\xs^8 + \frac{1}{2}(-5 + 4\ii)\xs^7 + \frac{1}{4}(9 + 3\ii)\xs^6 + \frac{1}{4}(9 + 57\ii)\xs^5 \nn\\
 & \quad + \frac{1}{4}(-72 + 177\ii)\xs^4 + (-108 + 72\ii)\xs^3 - (270 + 18\ii)\xs^2 - 270(1 + \ii)\xs -270\ii \bigg].
 \eal

Here, we have used the exponential integral function defined as
$\dpst E_1(x)= \int_1^\infty \xi^{-1} \ee^{-x \xi} \: \dm \xi$.

\vspace*{-3mm}

\section{The second-order coefficients \textit{\textbf{S}}$_{\textit{\!\textbf{ik,n}}}$ for a fluid particle in a fluid}
\seclab{S_coeff_fl}

\vspace*{-3mm}

The 45 $\Smn{ik}n$ coefficients that contribute to $D_n$ to leading order in $\xO$ are stated: $21$ coefficients for $n=0$ and $24$ for $n=1$. For the remaining $53$ coefficients in modes $n=0$ and $n=1$, we only state their order in $\xO$.
 \bal\eqlab{S_coeff_fl}
 \Smn{00}0 &= \frac{2 + \etaOTi}{9(1 + \etaOTi)\xs^2}x_0^3 , \qquad
 \Smn{00}1 =  \frac{2(5 + 4\etaOTi)}{45(1 + \etaOTi)\xs^2}x_0^3 , \qquad
 \Smn{0c}0 = -\frac{2\ii(-\etaOTi\xs^2 - \xs^2 + 6\Bc + \etaOTi)}{3(1 + \etaOTi)\xs^2},
 \nn\\
 \Smn{0c}1 &= -\frac{4\ii(3\etaOTi + 10)}{(1 + \etaOTi)\xs^2x_0^2}, \qquad
 \Smn{0t}0 \sim \calO (\xO^2), \qquad
 \Smn{0t}1 \sim \calO (\xO),
 \nn\\
 \Smn{0s}0 &= \frac{-2\ii \ee^{-\xs}(3\Bc\etaOTi\xs^2 + 2\etaOTi\xs^2 + \xs^3 - 6\Bc\xs - \etaOTi\xs + 3\xs^2 - 6\Bc - \etaOTi)}{3\xs^4(1+\etaOTi)}x_0^2 ,
 \nn\\
 \Smn{0s}1 &= -\frac{3\xs^4 + (9\etaOTi + 25)\xs^3 + (33\etaOTi + 105)\xs^2 + (72\etaOTi + 240)\xs + 72\etaOTi + 240}{3\xs^5(1 + \etaOTi)}\ee^{-\xs}x_0,
 \nn\\
 \Smn{0c'}0 &\sim \calO (\xO^5) , \qquad
 \Smn{0c'}1 = \frac{13x_0^{\prime 2}x_0\etaOTi}{45\xs^2(1 + \etaOTi)} , \qquad
 \Smn{0t'}0 \sim \calO(\xO^2), \qquad
 \Smn{0t'}1 \sim \calO(\xO), \nn
 \eal
 \bal
 \Smn{0s'}0 &= -\frac{2\ii x_0^2\etaOTi}{9}\frac{\big[(\xs^{\prime 4} + 24\xs^{\prime 2} + 51)\sinh(\xsp) - (7\xs^{\prime 3} + 51\xsp)\cosh(\xsp)\big]}{(1 + \etaOTi)\xs^2\xs^{\prime 2}} ,
 \nn \\
 \Smn{0s'}1 &= \frac{\big[(\xs^{\prime 5} + 55\xs^{\prime 3} + 348\xsp)\cosh(\xsp) - 3(3\xs^{\prime 4} + 57 \xs^{\prime 2} + 116)\sinh(\xsp)\big] x_0\etaOTi}{3(1 + \etaOTi)\xs^2\xs^{\prime 3}} ,
 \nn \\
 \Smn{c0}0 &= \frac{2\ii(\etaOTi\xs^2 +  \xs^2 + \etaOTi + 2)}{3(1 + \etaOTi)\xs^2} , \qquad
 \Smn{c0}1 = \frac{2\ii(5\etaOTi\xs^2 + 5\xs^2 + 8\etaOTi + 12)}{15(1 + \etaOTi)\xs^2},  \qquad
 \Smn{cc}0 = \frac{4 \etaOTi + 12}{(1 + \etaOTi)\xs^2} x_0^{-3},
 \nn\\
 \Smn{cc}1 &= \frac{72\etaOTi + 258}{(1 + \etaOTi)\xs^2}x_0^{-5},\qquad
 \Smn{ct}0 \sim \calO (\xO^{-1}), \qquad
 \Smn{ct}1 \sim \calO (\xO^{-2}),
 \nn\\
 \Smn{cs}0 &= \frac{1}{24\xs^4x_0(1 + \etaOTi)}\Big\{\big[-\xs^7\etaOTi + \xs^6\etaOTi + (-14\etaOTi - 8)\xs^5 + (18\etaOTi + 8)\xs^4 + (-48\etaOTi - 64)\xs^3 + (-48\etaOTi - 96)\xs^2 \nn\\
 &\quad + (-96\etaOTi - 288)\xs - 96\etaOTi - 288\big]\ee^{-\xs} +\xs^6(\etaOTi\xs^2 + 12\etaOTi + 8) E_1(\xs)\Big\},
 \nn\\
 \Smn{cs}1 &= \frac{-\ii}{4(1 + \etaOTi)\xs^5x_0^2}\Big[\Big(-3\etaOTi\xs^7 + 3\etaOTi\xs^6 - 2\xs^7 - 6\etaOTi\xs^5 + 2\xs^6 + 18\etaOTi\xs^4 - 4\xs^5 + 72\etaOTi\xs^3   \nn\\
 &\quad + 36\xs^4 + 264\etaOTi\xs^2 + 200\xs^3 + 576\etaOTi\xs + 888\xs^2 + 576\etaOTi + 2064\xs + 2064\Big)\ee^{-\xs} + 3\xs^8\Big(\etaOTi + \frac{2}{3}\Big) E_1(\xs) \Big],
 \nn\\
 \Smn{cc'}0 &\sim \calO (\xO^2), \qquad
 \Smn{cc'}1 = -\frac{2\ii \etaOTi x_0^{\prime 2}}{3(1 + \etaOTi)\xs^2 x_0^2},\qquad
 \Smn{ct'}0 \sim \calO(\xO^{-1}), \qquad
 \Smn{ct'}1 \sim \calO(\xO^{-2}) ,
 \nn\\
 \Smn{cs'}0 &= -\frac{2\big[(\xs^{\prime 4} + 24\xs^{\prime 2} + 51)\sinh(\xsp) - (7\xs^{\prime 3} + 51\xsp)\cosh(\xsp)\big]\etaOTi}{3(1 + \etaOTi)\xs^2 x_0\xs^{\prime 2}},
 \nn\\
 \Smn{cs'}1 &= \frac{2\ii\big[(\xs^{\prime 5} + 55\xs^{\prime 3} + 510\xsp)\cosh(\xsp) - 3(3\xs^{\prime 4} + 75\xs^{\prime 2} + 170)\sinh(\xsp)\big]\etaOTi}{(1 + \etaOTi)\xs^2x_0^2\xs^{\prime 3}} , \nn\\
 \Smn{t0}0  &= \frac{x_0 \ee^{\ii\xtO}}{9\xtO \xs^2(1+\etaOTi)}\big[-4\ii\xtO^2 + (2\xs^2 + 6\etaOTi + 12)\xtO + 6\ii(1 + \xs^2)\etaOTi + 12\ii + 6\ii\xs^2 \big], \qquad
 \Smn{t0}1 \sim \calO (\xO^2),
 \nn\\
 \Smn{tc}0 &= \frac{1}{24\xtO \xs^2(1+\etaOTi)x_0^2}\Big\{-\Big[\etaOTi\Bt\xtO^4
 + \big((\xs^2 + 18\Bt + 6)\etaOTi + 12\Bt + 4\big)\xtO^2 + 4(3\etaOTi + 2)\xs^2\Big]\xtO^4 E_1(-\ii\xtO)
 \nn\\
 &\quad + \Big[\Bt\etaOTi\xtO^7\ii + \Bt\etaOTi\xtO^6 + \big((\xs^2\ii + 16\ii\Bt + 6\ii)\etaOTi
 + 12\ii\Bt + 4\ii\big)\xtO^5 + \big((\xs^2 + 12\Bt + 6)\etaOTi + 12\Bt + 4\big)\xtO^4
 \nn\\
 &\quad + \big((10\ii\xs^2 - 12\ii\Bt - 12\ii)\etaOTi + 8\ii\xs^2 - 24\ii\Bt - 8\ii\big)\xtO^3
 + \big((6\xs^2 + 12\Bt - 36)\etaOTi + 8\xs^2 - 72\Bt - 88\big)\xtO^2
 \nn\\
 &\quad - (96\ii\etaOTi + 288\ii)\xtO + 96\etaOTi + 288\Big]\ee^{\ii\xtO} \Big\} , \qquad
 \Smn{tc}1 \sim \calO(\xO^{-3}), \qquad
 \Smn{tt}0 \sim \calO (1), \qquad
 \Smn{tt}1 \sim \calO (1), \nn\\
 \Smn{ts}0 &= \frac{1}{24\xtO \xs^4 (1+\etaOTi)}\Big\{ \Big[\!-\! \Bt\etaOTi\xtO^7\ii - \etaOTi\Bt(\xs + 1)\xtO^6 \!+\! \big(-\ii (2\Bt + 1)\etaOTi\xs^2 + 2\ii\Bt\etaOTi\xs - (16\ii\Bt + 6\ii)\etaOTi - 12\ii\Bt - 4\ii\big)\xtO^5
 \nn\\
 &\quad  - \big((2\Bt + 1)\etaOTi\xs^3 + \xs^2\etaOTi + ((12\Bt + 6)\etaOTi + 12\Bt + 4)\xs + (12\Bt + 6)\etaOTi + 12\Bt + 4\big)\xtO^4
 \nn\\
 &\quad  + \big(-\etaOTi (\Bt + 2)\xs^4\ii + 2\ii\etaOTi (\Bt + 1)\xs^3 + (-(24\ii\Bt + 16\ii)\etaOTi - 12\ii\Bt - 12\ii)\xs^2 + ((12\ii\Bt + 12\ii)\etaOTi + 24\ii\Bt + 8\ii)\xs
 \nn\\
 &\quad + (12\ii\Bt + 12\ii)\etaOTi + 24\ii\Bt + 8\ii \big)\xtO^3 + \big(-\etaOTi (\Bt + 2)\xs^5 + \etaOTi\xs^4\Bt + (-( 20\Bt + 12)\etaOTi - 12\Bt - 12)\xs^3
 \nn\\
 &\quad + (-24\Bt\etaOTi + 12\Bt - 4)\xs^2 + ((-12\Bt + 36)\eta + 72\Bt + 88)\xs + (-12\Bt + 36)\etaOTi + 72\Bt + 88 \big)\xtO^2
 \nn\\
 &\quad + \big(-\etaOTi\xs^6\ii + 2\ii\etaOTi\xs^5 + (-18\ii\etaOTi - 8\ii)\xs^4
 + (48\ii \etaOTi + 64\ii)\xs^3 + (48\ii\etaOTi + 96\ii)\xs^2 + (96\ii\etaOTi + 288\ii)\xs
 + 96\ii\etaOTi + 288\ii \big)\xtO
 \nn\\
 &\quad - \xs^7\etaOTi \!+\! \etaOTi\xs^6 \!-\! (14\etaOTi + 8)\xs^5 \!+\! (18\etaOTi + 8)\xs^4 \!-\! (48\etaOTi + 64)\xs^3 \!-\! (48\etaOTi + 96)\xs^2 \!-\! (96\etaOTi + 288)\xs \!-\! 96\etaOTi \!-\! 288\Big] \ee^{-\xs +\ii \xtO}
 \nn\\
 &\quad +  \Big[\etaOTi\Bt\xtO^4 + \big((\Bt + 1)\etaOTi\xs^2 + (18\Bt + 6)\etaOTi + 12\Bt
 + 4\big)\xtO^2 + \xs^2(\etaOTi\xs^2 + 12\etaOTi + 8)\Big]
 \Big(\xs^2 + \xtO^2\Big)^2 E_1(-\ii\xtO + \xs) \Big\},
 \nn\\
 \Smn{ts}1 & \sim \calO(1), \qquad
 \Smn{tc'}0 \sim \calO(\xO^3), \qquad
 \Smn{tc'}1 \sim \calO(\xO^2), \qquad
 \Smn{tt'}0 \sim \calO(1), \qquad
 \Smn{tt'}1 \sim \calO(1), \nn\\
 \Smn{ts'}0 &= \frac{2\ii \ee^{\ii \xtO}(\ii + \xtO)\big[(\xs^{\prime 4} + 24\xs^{\prime 2} + 51)\sinh(\xsp) -(7\xs^{\prime 3} +51\xsp)\cosh(\xsp)\big]\etaOTi}{3(1+\etaOTi)\xs^2\xtO\xs^{\prime 2}}, \qquad
 \Smn{ts'}1 \sim \calO(1) , \qquad
 \Smn{s0}0 = 0 ,
 \nn\\
 \Smn{s0}1 &= \frac{2\ee^{\ii \xs}(4\ii\etaOTi\xs^2 + 3\ii\xs^2 + \xs^3 + 8\ii\etaOTi + 8\xs\etaOTi + 12\ii + 12\xs)}{15\xs^4(1 + \etaOTi)}x_0^2, \qquad
 \Smn{sc}0 = 0, \nn
 \eal
 \bal
 \Smn{sc}1 &= \frac{1}{32\xs^4x_0^3(1 + \etaOTi)}\Big\{\xs^8(\etaOTi\xs^2 + 18\etaOTi + 12)E_1( -\ii\xs) - \Big[\xs^9\etaOTi\ii + \xs^8\etaOTi + (16\ii\etaOTi + 12\ii)\xs^7 + (12\etaOTi + 12)\xs^6  \nn\\
 &\quad - (12\ii\etaOTi + 24\ii)\xs^5 + (12\etaOTi - 72)\xs^4 + (-288\ii\etaOTi - 384\ii)\xs^3 + (288\etaOTi + 576)\xs^2 + (2304\ii\etaOTi + 8256\ii)\xs
 \nn\\
 &\quad - 2304\etaOTi - 8256\Big]\ee^{\ii\xs}\Big\}, \qquad
 \Smn{st}0 = 0, \qquad
 \Smn{st}1 \sim \calO (1), \qquad
 \Smn{ss}0 = 0,
 \nn\\
 \Smn{ss}1 &= \frac{1}{4\xs^7(1 + \etaOTi)}\Big\{-2\ii\xs^8(\etaOTi\xs^2 + 9\etaOTi + 6)E_1(\xs - \ii \xs) + \Big[(-1 + \ii)\etaOTi\xs^9 + \xs^8\etaOTi + \big(-6 + 6\ii + (-10 + 8\ii)\etaOTi\big)\xs^7    \nn\\
 &\quad+ \big(10 - 4\ii + (9 + 3\ii)\etaOTi\big)\xs^6 + \big(2 - 14\ii + (9 + 9\ii)\etaOTi\big)\xs^5 + (9\ii\etaOTi - 104 - 6\ii)\xs^4 + (-792 - 56\ii - 192\etaOTi)\xs^3   \nn\\
 &\quad + \big(-2064 - 744\ii - (576 + 192\ii)\etaOTi\big)\xs^2 + \big(-2064 - 2064\ii - (576 + 576\ii)\etaOTi \big)\xs - 576\ii\etaOTi - 2064\ii\Big]\ee^{(-1 + \ii)\xs}\Big\} ,
 \nn\\
 \Smn{sc'}0 &= 0, \qquad
 \Smn{sc'}1 = \frac{2\etaOTi \ee^{\ii\xs} x_0^{\prime 2}(-5\xs + 6\ii\xs^2 - 5\ii)}{15\xs^4(1 + \etaOTi)}, \qquad
 \Smn{st'}0 = 0 , \qquad
 \Smn{st'}1 \sim \calO(1), \qquad
 \Smn{ss'}0 = 0,
 \nn\\
 \Smn{ss'}1 &= \frac{\ee^{\ii\xs}\etaOTi}{(1+\etaOTi)\xs^4\xs^{\prime 3}}\bigg[\bigg(-18(\ii + \xs)\xs^{\prime 4} + (36\ii\xs^2 - 450\ii - 450\xs)\xs^{\prime 2} + 108\ii\xs^2 - 1020(\ii +\xs)\bigg)\sinh(\xsp)
 \nn\\
 & \quad +\cosh(\xsp)\xsp\bigg(2(\ii + \xs)\xs^{\prime 4}
 + 110(\ii + \xs)\xs^{\prime 2} - 108 \xs^2\ii + 1020(\ii +\xs)\bigg) \bigg] ,
 \nn\\
 \Smn{c'0}0 &= \frac{2x_0^{\prime 2}x_0\etaOTi}{9\xs^2(1 + \etaOTi)} , \qquad
 \Smn{c'0}1 = -\frac{2\etaOTi x_0' x_0^2}{15(1 + \etaOTi)\xs^2} , \qquad
 \Smn{c'c}0 = -\frac{4\ii x_0^{\prime 2}\etaOTi}{3x_0^2\xs^2(1 + \etaOTi)}, \qquad
 \Smn{c'c}1 = -\frac{6\ii x_0'\etaOTi}{x_0^3\xs^2(1 + \etaOTi)},
 \nn\\
 \Smn{c't}0 &\sim \calO(\xO^2), \qquad
 \Smn{c't}1 \sim \calO(\xO), \qquad
 \Smn{c's}0 = \frac{4\ii x_0^{\prime 2}\ee^{-\xs}(\xs + 1)\etaOTi}{3\xs^4(1 + \etaOTi)} , \qquad
 \Smn{c's}1 = -\frac{2x_0' \ee^{-\xs} (\xs^3 + 3\xs^2 + 6\xs + 6)\etaOTi}{\xs^5(1 + \etaOTi)}, \nn\\
 \Smn{c'c'}0 &\sim \calO (\xO^5) , \qquad
 \Smn{c'c'}1 \sim \calO (\xO^5) , \qquad
 \Smn{c't'}0 \sim \calO(\xO^2), \qquad
 \Smn{c't'}1 \sim \calO(\xO) ,
 \nn\\
 \Smn{c's'}0 &= -\frac{2\ii x_0^{\prime 2}\etaOTi}{\xs^{\prime 2}(1 + \etaOTi)\xs^2} \bigg[\bigg(-\frac{\xs^{\prime 4}}{9} + (\Bc' - 1)\xs^{\prime 2} + 3\Bc' - \frac{5}{3}\Bigg)\sinh(\xsp) - 3\cosh(\xsp)\bigg(-\frac{4\xs^{\prime 2}}{27} + \Bc' - \frac{5}{9}\bigg)\xsp\bigg],
 \nn\\
 \Smn{c's'}1 &= x_0'\frac{\big[(6\xs^{\prime 2} + 15)\sinh(\xsp)
 - (\xs^{\prime 3} + 15\xsp)\cosh(\xsp)\big]\etaOTi}{3(1 + \etaOTi)\xs^2\xsp},
 \nn\\
 \Smn{t'0}0 &= \frac{2\etaOTi x_0 \big(\sin(\xtO')\xtO^{\prime 2} + 2\xtO'\cos(\xtO') - 2\sin(\xtO')\big)}{3\xtO' (1 + \etaOTi)\xs^2} , \qquad
 \Smn{t'0}1 = \calO(\xO^2),
 \nn\\
 \Smn{t'c}0 &= -\frac{4\ii\etaOTi \big(\sin(\xtO')\xtO^{\prime 2} + 2\xtO'\cos(\xtO') - 2\sin(\xtO')\big)}{\xtO' (1 + \etaOTi)\xs^2 x_0^2} , \qquad
 \Smn{t'c}1 \sim \calO(\xO^{-3}), \qquad
 \Smn{t't}0 \sim \calO(1) , \qquad
 \Smn{t't}1 \sim \calO(1) ,
 \nn\\
 \Smn{t's}0 &= \frac{4\ii\etaOTi (\xs + 1)\ee^{-\xs}\big(\sin(\xtO')\xtO^{\prime 2} + 2 \xtO'\cos(\xtO') - 2\sin(\xtO')\big)}{\xs^4\xtO'(1 + \etaOTi)}, \qquad
 \Smn{t's}1 \sim \calO(1),
 \nn\\
 \Smn{t'c'}0 &= -\frac{2\big[\big(5\xtO^{\prime 4} - (6\xs^{\prime 2} + 15)\xtO^{\prime 2}
 + 15\xs^{\prime 2}\big)\sin(\xtO') + \cos(\xtO')\xtO'\big((\xs^{\prime 2} + 15)\xtO^{\prime 2}
 - 15\xs^{\prime 2}\big)\big]x_0'\etaOTi}{9\xtO^{\prime 3}\xs^2(1 + \etaOTi)}, \qquad
 \Smn{t'c'}1 \sim \calO(\xO^2),
 \nn\\
 \Smn{t't'}0 &\sim \calO (1), \qquad
 \Smn{t't'}1 \sim \calO (1),
 \nn\\
 \Smn{t's'}0 &= -\frac{2\ii\etaOTi}{9\xs^{\prime 2}\xs^2\xtO'(\xs^{\prime 2} + \xtO^{\prime 2})^2
 (1 + \etaOTi)} \Big\{\Big[\big(-3\xs^{\prime 6} + 3\xtO^{\prime 2}(3\Bt' - 4)\xs^{\prime 4}
 - 30\xtO^{\prime 2}\xs^{\prime 2} - 30\xtO^{\prime 4}\big)(\xs^{\prime 2} + \xtO^{\prime 2})\sin(\xtO')
 \nn\\
 &\quad + 3\cos(\xtO')\big(\xs^{\prime 6} + 2\xtO^{\prime 2}\xs^{\prime 4}
 + \xtO^{\prime 2}(\xtO^{\prime 2} + 9\Bt' + 5)\xs^{\prime 2}
 + \xtO^{\prime 4}(3\Bt' + 11)\big)\xs^{\prime 2}\xtO'\Big]\sinh(\xsp)  \nn\\
 &\quad - \Big[\big(-12\xs^{\prime 6} + 6(3\Bt' - 5)\xtO^{\prime 2}\xs^{\prime 4}
 - 30\xtO^{\prime 4}\xs^{\prime 2} - 30\xtO^{\prime 6}\big)\sin(\xtO')   \nn\\
 &\quad  + \big(12\xs^{\prime 2} + 3\xtO^{\prime 2}(3\Bt' + 1)\big)\cos(\xtO')
 \xs^{\prime 2} \xtO'(\xs^{\prime 2} + \xtO^{\prime 2})\Big]\xsp\cosh(\xsp) \Big\}, \qquad
 \Smn{t's'}1 \sim \calO(1),
 \nn\\
 \Smn{s'0}0 &= 0 , \qquad
 \Smn{s'0}1 = \frac{2 x_0^2\etaOTi\big[(\xs^{\prime 4} - 6\xs^{\prime 2} + 15)\sin(\xsp) + (7\xs^{\prime 3} - 15\xsp)\cos(\xsp)\big]}{45(1 + \etaOTi)\xs^2\xs^{\prime 2}}, \qquad
 \Smn{s'c}0 = 0,
 \nn\\
 \Smn{s'c}1 &= -\frac{3\ii \big[(\xs^{\prime 4} - 6\xs^{\prime 2} - 15)\sin(\xsp) + (7\xs^{\prime 3}
  + 15\xsp)\cos(\xsp)\big]\etaOTi}{(1 + \etaOTi)\xs^2 x_0^3\xs^{\prime 2}}, \qquad
 \Smn{s't}0 =0, \qquad
 \Smn{s't}1 \sim \calO(1) , \qquad
  \Smn{s's}0 = 0,  \nn
 \eal
 \bal
 \Smn{s's}1 &= -\frac{\ee^{-\xs}\etaOTi}{(1 + \etaOTi)\xs^5\xs^{\prime 2}}\Big\{\big[(2\xs^2 + 6\xs + 6)\xs^{\prime 4} - (12\xs^2 + 36\xs + 36)\xs^{\prime 2} - 12\xs^3 - 42\xs^2 - 90\xs - 90\big]
 \sin(\xsp) \nn\\
 &\quad + 2\cos(\xsp)\xsp \big(7\xs^2\xs^{\prime 2} + 6\xs^3 + 21\xs\xs^{\prime 2} + 21\xs^2 + 21\xs^{\prime 2} + 45\xs + 45\big)\Big\}, \qquad
 \Smn{s'c'}0 = 0 , \nn\\
 \Smn{s'c'}1 &= - x_0^{\prime 2}\frac{2\big[(\xs^{\prime 4} - 45\xs^{\prime 2} + 105)\sin(\xsp) + (10\xs^{\prime 3} - 105\xsp)\cos(\xsp)\big]\etaOTi}{45(1 + \etaOTi)\xs^2\xs^{\prime 2}}, \qquad
 \Smn{s't'}0 = 0, \qquad
 \Smn{s't'}1 \sim \calO(1),
 \nn\\
 \Smn{s's'}0 &= 0 ,
 \nn\\
 \Smn{s's'}1 &=\frac{\etaOTi}{\xsp\xs^2(1+\etaOTi)}\Big[\big(-\sin(\xsp)\xs^{\prime 2} + 9\cos(\xsp)\xsp - 18\sin(\xsp)\big)\sinh(\xsp) - 3\xsp\cosh(\xsp)\big(\cos(\xsp)\xsp - 4\sin(\xsp)\big)\Big] .
 \eal
\end{widetext}


%

\end{document}